\documentclass[twocolumn]{emulateapj}
\usepackage{lscape}

\def\Min{\hbox{${}^{\prime}$\llap{.}}}
\def\Sec{\hbox{${}^{\prime\prime}$\llap{.}}}
\def\deg{\hbox{${}^\circ$}} \def\min{\hbox{${}^{\prime}$}}
\def\sec{\hbox{${}^{\prime\prime}$}}
\def\lae{\mathrel{<\kern-1.0em\lower0.9ex\hbox{$\sim$}}}
\def\gae{\mathrel{>\kern-1.0em\lower0.9ex\hbox{$\sim$}}}
\def\msun{M$_\odot$} 
\def\etal{et al.~} 
\def\nd{\nodata} 
\newcommand{\be}{\begin{equation}}
\newcommand{\ee}{\end{equation}}

\shorttitle{The {\it ACS} Virgo Cluster Survey. VI. Isophotal Analysis 
and  Surface Brightness Profiles
}
\shortauthors{Ferrarese \etal}

\begin{document}

\title{The {\it ACS} Virgo Cluster Survey. VI. \\
Isophotal Analysis and the
Structure of Early-Type Galaxies\altaffilmark{1}}

\author{Laura Ferrarese\altaffilmark{2}, Patrick
C\^ot\'e\altaffilmark{2}, Andr\'es Jord\'an\altaffilmark{3,4}, Eric
W. Peng\altaffilmark{2}, John P. Blakeslee\altaffilmark{5,6}, \\
Slawomir
Piatek\altaffilmark{7}, Simona Mei\altaffilmark{5}, David
Merritt\altaffilmark{8}, Milo\v s Milosavljevi\'c\altaffilmark{9,10},
John L. Tonry\altaffilmark{11},\\
 \& Michael J. West\altaffilmark{12}}

\altaffiltext{1}{Based on observations with the NASA/ESA {\it Hubble
Space Telescope} obtained at the Space Telescope Science Institute,
which is operated by the association of Universities for Research in
Astronomy, Inc., under NASA contract NAS 5-26555.}
\altaffiltext{2}{Herzberg Institute of Astrophysics, National Research
Council of Canada, 5071 West Saanich Road, Victoria, BC, V8X 4M6,
Canada; {\sf laura.ferrarese@nrc-cnrc.gc.ca, patrick.cote@nrc-cnrc.gc.ca, 
eric.peng@nrc-cnrc.gc.ca}} 
\altaffiltext{3}{European Southern Observatory,
Karl-Schwarzschild-Str. 2, 85748 Garching, Germany; {\sf ajordan@eso.org}}
\altaffiltext{4}{Astrophysics, Denys Wilkinson Building, University of 
Oxford, 1 Keble Road, Oxford, OX1 3RH, UK}
\altaffiltext{5}{Department of Physics and Astronomy, The Johns
Hopkins University, 3400 North Charles Street, Baltimore, MD
21218-2686; {\sf jpb@pha.jhu.edu, smei@pha.jhu.edu}}
\altaffiltext{6}{Department of Physics and Astronomy, PO Box 642814, 
Washington State University, Pullman, WA 99164}
\altaffiltext{7}{Department of Physics, New Jersey Institute of
  Technology, Newark, NJ 07102; {\sf piatek@physics.rutgers.edu}}
\altaffiltext{8}{Department of Physics, Rochester Institute of
Technology, 84 Lomb Memorial Drive, Rochester, NY 14623; {\sf merritt@cis.rit.edu}} 
\altaffiltext{9}{Theoretical Astrophysics,
California Institute of Technology, Mail Stop 130-33, Pasadena, CA
91125; {\sf milos@tapir.caltech.edu}} 
\altaffiltext{10}{Sherman M. Fairchild Fellow} 
\altaffiltext{11}{Institute for Astronomy, University of
Hawaii, 2680 Woodlawn Drive, Honolulu, HI 96822; {\sf jt@ifa.hawaii.edu}}
\altaffiltext{12}{Department of Physics and Astronomy, University of
Hawaii, Hilo, HI 96720; {\sf westm@hawaii.edu}}

\begin{abstract}

We present a detailed analysis of the morphology, isophotal parameters
and surface brightness profiles for 100 early-type members of the
Virgo Cluster, from dwarfs ($M_B = -15.1$ mag) to giants ($M_B =
-21.8$ mag). Each galaxy has been imaged in two filters, closely
resembling the Sloan $g$ and $z$ passbands, using the {\it Advanced
Camera for Surveys} on board the {\it Hubble Space Telescope}.

Dust and complex morphological structures are common. Dust is
detected in as many as 18, preferentially bright, galaxies. The
incidence rate in the 26 galaxies brighter than $B_T = 12.15$ mag, which
form a magnitude limited sample, is 42\%. The amount and distribution
of dust show no obvious correlations with galaxy morphology; dust
features range from faint wisps and patches on tens of parsec scales,
to regular, highly organized kpc-scale dust disks. Blue star
clusters are interspersed within the larger, clumpier dust disks,
while thin, dynamically cold stellar disks are seen in association
with the smaller, uniform nuclear dust disks.

Kiloparsec-scale stellar disks, bars, and nuclear stellar disks are
seen in 60\% of galaxies with intermediate luminosity ($-20 \lesssim
M_B \lesssim -17$). In at least one case (VCC 1938 = NGC 4638), the
large-scale stellar disk has a sharp inner edge, possibly produced
when disk instabilities led to the formation of a (now dissolved)
bar. This process might indeed be seen unfolding in one galaxy,
VCC 1537 (=NGC 4528). A spiral structure might be present in VCC 1199, an
elliptical companion of M49. In dwarf galaxies, spiral structures are
confirmed in VCC 856 and detected for the first time in VCC 1695.

Surface brightness profiles, ellipticities, major axis position
angles, and isophotal shapes are derived typically within 8 kpc from
the center for the brightest galaxies, and 1.5 kpc for the faintest
systems, with a resolution (FWHM) of 7 pc. For all but 10 of the
galaxies, the surface brightness profiles are well described by a
S\'ersic model with index $n$ which increases steadily from the
fainter to the brightest galaxies. In agreement with previous claims,
the inner profiles (typically within 100 pc of the center) of eight of
the 10 brightest galaxies, to which we will refer as ``core''
galaxies, are lower than expected based on an extrapolation of the
outer S\'ersic model, and are better described by a single power-law
function. Core galaxies are clearly distinct in having fainter central
surface brightness, $\mu_0$, and shallower logarithmic slope of the
inner surface brightness profile, $\gamma$, than expected based on the
extrapolation of the trend followed by the rest of the sample, for
which both $\mu_0$ and $\gamma$ increase steadily with galaxy
magnitude. Large-scale, global properties also set core galaxies
apart: the effective radius in particular is found to be almost one
order of magnitude larger than for only slightly less luminous
non-core galaxies.  Contrary to previous claims, we find no evidence
in support of a strong bimodal behavior of the inner profile slope,
$\gamma$; in particular the $\gamma$ distribution for galaxies which
do not show evidence of multiple morphological components (disks or
bars) is unimodal across the entire magnitude range (a factor 460 in
$B-$band luminosity) spanned by the ACSVCS galaxies. Although core
galaxies have shallow inner profiles, the shallowest profiles (lowest
$\gamma$ values) are found in faint dwarf systems. The widely adopted
separation of early-type galaxies between ``core" and ``power-law"
types, which had originally been prompted by the claim of a clear
bimodal distribution of $\gamma$ values, is therefore untenable based
on the present study.

Once core galaxies are removed, dwarf and bright ellipticals display a
continuum in their morphological parameters, contradicting some
previous beliefs that the two belong to structurally distinct
classes. However, dwarfs span a wider range in morphological
characteristics than brighter systems: their surface brightness
profiles vary from exponential to almost $r^{1/4}$ laws, they comprise
both nucleated and non-nucleated varieties, and several systems
display evidence of disks, spiral structures and recent star
formation. This is taken as evidence that dwarf galaxies, as currently
classified, form a heterogeneous class.

\end{abstract}


\section{Introduction}
\label{sec:introduction}
The innermost regions of galaxies have long played a pivotal role in
guiding our understanding of how galaxies form and evolve. Beginning
in the mid 1980's, ground-based studies started to pave the way
towards a characterization of the core (by ``core" we hereby refer
loosely to the innermost few hundred parsecs) structural parameters
(Lauer 1985; Kormendy 1985). However, the full extent of the diversity
and complexity of the core structure was not fully appreciated until
high-resolution {\it Hubble Space Telescope} (HST) images became
available (Crane et al.\ 1993; Jaffe et al.\ 1994; Ferrarese et al.\
1994; van den Bosch et al.\ 1994; Forbes, Franx \& Illingworth 1995;
Lauer et al.\ 1995, 2002, 2005; Verdoes Kleijn et al.\ 1999; Quillen,
Bower, \& Stritzinger 2000; Rest et al.\ 2001; Ravindranath et al.\
2001; Laine et al.\ 2003). These studies 
unveiled the existence of small, 100 pc scale, nuclear stellar and
gas/dust disks, which have become instrumental in probing the
central potential and the masses of central compact objects (Harms et
al.\ 1994; Ferrarese et al.\ 1996; van der Marel \& van den Bosch
1998).  They exposed nuclear bars and ``evacuated'' cores, possibly
scoured out by the coalescence of supermassive black holes. They
allowed the study of the properties of dust and nuclear star
clusters. In short, these observations dispelled the myth of elliptical
galaxies as easily characterizable, regular, dust-free stellar systems.

Despite a limited sample of just seven early-type galaxies (all but
one within 15 Mpc, observed with the pre-refurbishment HST {\it Faint
Object Camera} in one or more bands resembling Johnson's $U$, $B$ and
$V$ filters), the first HST study by Crane et al.\ (1993) established
that in none of the galaxies could the surface brightness profile be
characterized as isothermal. This result was subsequently confirmed by
Ferrarese et al.\ (1994) based on a somewhat larger and magnitude
limited sample of 14 early-type galaxies in the Virgo cluster,
observed in a single filter ($\sim$ Johnson's $V$) with the HST {\it
Planetary Camera 1}. Capitalizing on the completeness of their
sample, Ferrarese et al.\ (1994) noted that, for the brightest half of
their Virgo galaxies, the surface brightness profile could be
characterized as a broken power-law (these galaxies were referred to
as ``Type I''). For the rest of the sample (``Type II'' galaxies), the
profiles could be reasonably well described by a single power-law to
the smallest radii resolved by the instrument.

These issues were revisited in a series of papers by Lauer et al.\
(1995), Gebhardt et al.\ (1996), Byun et al.\ (1996), and Faber et
al.\ (1997). Armed with a larger sample of 45 to 61 galaxies
(depending on the paper), including a few spirals, these studies
significantly strengthened the conclusion that in no cases were cores
observed to have constant central surface density. Lauer et al.\
(1995) applied Hernquist's (1990, his equation 43) double power-law
model to the surface brightness profiles of their galaxies $-$ the
model, dubbed ``Nuker law'' by the authors, is similar to the broken
power-law proposed by Ferrarese et al.\ 1994, but includes an
additional parameter regulating the transition between the outer and
inner power-laws. Lauer et al.\ advocated the view that
early-type galaxies can be divided cleanly in two classes (which those
authors named ``core'' and ``power-law'' and correspond closely to the
Type I and Type II galaxies in Ferrarese et al.\ 1994) based on their
central surface brightness profiles. The working definition of a
``core'' galaxy in Lauer et al.\ (1995) is a galaxy for which the
logarithmic slope of the profile within the break radius (i.e., the
transition radius between the inner and outer power-law) is $\gamma <
0.3$. The authors also found that the distribution of $\gamma$ is
bimodal, with no galaxies in the range $0.3 < \gamma < 0.5$.

Although ground-breaking, the study of Lauer et al.\ (1995) suffered
from an ill-defined sample (although the galaxies span a wide range in
luminosities $-15.1 < M_V < -23.82$ and structural properties,  they
are sampled at vastly different spatial resolutions, ranging  in
distance between 3 and 280 Mpc, and were partly selected for
exhibiting features deemed interesting), and the fact that all
observations were made in one band ($\sim$ Johnson's $V$) with the
pre-refurbished HST. These shortcomings have been partly overcome by
Rest et al.\ (2001) and Lauer et al.\ (2005). Both of these
investigations capitalize on the use of the refurbished HST {\it Wide
Field and Planetary Camera 2} (WFPC2), but take rather different
approaches to sample selection. Rest et al.\ (2001) randomly targeted
an unbiased (though incomplete) sample of 67 galaxies selected from a
magnitude and space-limited sample of 130 objects within 54 Mpc. Each
galaxy was observed in a single band, similar to Johnson's $R$. The
Lauer et al.\ (2005) sample, on the other hand, is not easily
characterizable, being comprised of 77, mainly bright early-type
galaxies at distances between 10 and 100 Mpc. These galaxies were
observed as part of unrelated HST programs (and therefore with
different filter configurations and varying signal-to-noise ratios)
and downloaded from the HST public archive. Despite these drawbacks,
the Lauer et al.\ (2005) study was the first to make extensive use of
color information, with roughly half of the sample observed in two
filters, $\sim$ Johnson's $V$ and $I$.

Both studies confirmed the presence of dust in roughly half of the
galaxies. Rest et al.\ (2001) find embedded nuclear stellar disks in
51\% of their sample, while Lauer et al.\ (2005) suggest that ``all
power-law galaxies harbor disks; their visibility is simply due to
viewing angle'', at odds with previous claims in Lauer et al.\ (1995)
but in agreement with Ferrarese et al.\ (1994). Lauer et al.\ (2005)
reaffirm the strong dichotomy in the distribution of $\gamma$, with
core galaxies having $\gamma < 0.3$ and power-law galaxies having ``a
steep $\gamma > 0.5$ cusp [which] continues into the HST resolution
limit''.  This conclusion is based on Nuker model fits to the surface
brightness profiles, despite warnings by Rest et al.\ (2001) that the
dichotomous behavior of core and power-law galaxies observed by Lauer
et al.\ (1995) might be partly due to the fact that the adoption of
the Nuker model will bias the value of the inner slope for profiles
with slopes that vary smoothly with radius. By introducing a
non-parametric estimate of the logarithmic slope of the inner profile,
Rest et al.\ (2001) find that core and power-law galaxies form a
continuous sequence, with ``transition'' objects bridging the gap
between the two, a conclusion also reached by Ravindranath et al.\
(2001) based on HST/NICMOS data.

The stability of Nuker model fits has been challenged most recently by
Graham et al.\ (2003). The authors conclusively demonstrated that the
recovered Nuker model parameters are sensitive to the radial extent of
the data being fitted, owing to the simple fact that the kpc-scale
surface brightness profiles of early-type galaxies are not well
described by power-laws.  Graham et al.\ (2003) and Trujillo et al.\
(2004) show that  a S\'ersic model (S\'ersic 1968) provides a more
stable and accurate parametrization of the surface brightness profiles
for all but the brightest early-type galaxies, which are best
described by a S\'ersic model transitioning to a power-law  in the
inner regions. The authors define as ``core'' galaxies  displaying ``a
deficit of flux relative to the inward extrapolation of the outer
S\'ersic profile''. In a follow-up paper, Graham \& Guzman (2003) find
that the distribution of the inner profile slope is continuous, with
$\gamma$ increasing steadily from dwarf to giant ellipticals, only to
decrease again for the very brightest ($M_B < -20.5$ mag) systems.

This paper presents the next effort in the characterization of the
properties of early-type galaxies. Based on data obtained with the
{\it Wide Field Channel} (WFC) of the HST {\it Advanced Camera for
Surveys} (ACS), we discuss the morphology and structural parameters
for 100 confirmed early-type members of the Virgo cluster, each imaged
in the $g$ and $z-$bands. Compared to previous studies, the sample
presented in this paper is superior in terms of sample size,
completeness and dynamic range (a factor 460 in $B-$band luminosity),
depth (a factor three higher signal-to-noise ratio compared to a
typical galaxy in the most recent study by Lauer et al.\ 2005), color
baseline (a factor 1.5 greater wavelength range compared to Johnson's
$V$ and $I$, with all galaxies observed in two bands), spatial
resolution (a factor of $\sim$three improvement compared to studies
based on WFPC2 data, due both to the smaller distance of Virgo, and to
the sharper ACS/WFC point spread function), and field of view (a
factor 25 in area compared to studies based on WFPC2 data). This is
the sixth paper in a series discussing results from the ACS Virgo
Cluster Survey (ACSVCS). Previous papers have discussed the scientific
rationale and observational planning (C\^ot\'e et al.\ 2004, hereafter
Paper I), data reduction pipeline (Jord\'an et al.\ 2004a, hereafter
Paper II), the X-ray binary population in M87 (Jord\'an et al.\ 2004b),
the calibration of Surface Brightness Fluctuations (SBF) for distance
measurements (Mei et al.\ 2005a,b), the discovery of the population of
ultra-compact dwarfs (Hasegan et al.\ 2005), the properties of the
galaxy nuclei (C\^ot\'e et al.\ 2005, hereafter Paper VIII), the
color distribution and half-light radii of the globular clusters (Peng
et al.\ 2005a; Jord\'an et al.\ 2005), and the nature of diffuse star
clusters (Peng et al. 2005b).

The structure of this paper is as follows. The sample selection and
observational setup are described in \S 2. Data analysis, including
basic image reduction, correction for internal dust extinction,
recovery of the isophotal parameters, and parametrization of the
surface brightness profiles can be found in \S 3. The next section
reports on the morphological characteristics of the sample galaxies
(incidence of dust and other morphological peculiarities, such as
stellar disks, rings, bars and spiral structures, \S 4.1), the choice
of parametrization of the surface brightness profiles (\S 4.2), and
scaling relations linking the structural parameters (\S 4.3). A
general discussion is given in \S 5, where we reconsider the
definition of ``core'' galaxies and discuss their association with
supermassive black holes (SBHs, \S 5.1 and \S 5.2), address the
connection between dwarf and giant early-type galaxies (\S 5.3),
comment on the isophotal shapes (\S 5.4) and present color-color and
color-magnitude diagrams (\S 5.5). Notes on individual galaxies are
presented in an Appendix.

In this paper, we adopt a distance modulus to
Virgo of $31.09 \pm 0.03$ (random) $\pm 0.15$ (systematic) mag, which
corresponds to the mean Cepheid-calibrated surface brightness
fluctuation (SBF) distance as determined by Tonry et al.\ (2001),
corrected to reflect the latest calibration of the Cepheid PL relation
(Freedman et al.\ 2001; Blakeslee et al.\ 2002). The same mean
distance modulus to Virgo serves as the calibration for the SBF
distance scale adopted by Mei et al.\ (2005) for the ACSVCS
galaxies. The corresponding linear distance is $16.52 \pm 0.22$
(random) $\pm 1.14$ (systematic) Mpc. At this distance,
1\sec~corresponds to $80.1 \pm 1.1 \pm 5.53$ pc, and the ACS/WFC pixel
scale (0\Sec049) to $3.92 \pm 0.05 \pm 0.26$ pc.

\section{Sample Description and Observational Strategy}

We refer the reader to Paper I for a comprehensive description of the
sample and the scientific rationale that motivated the observational
setup. In essence, the ACSVCS comprises high resolution HST $g$ and
$z-$band images of 100 early-type galaxies, selected from the Virgo
Cluster Catalog (VCC) of Binggeli et al.\ (1985). Our initial sample
consisted of all galaxies 1) classified in the VCC as ``certain'' or
``probable'' members of Virgo, based on the observed morphology,
surface brightness, and (when available) radial velocity; 2) located
above +5\deg~in declination, thereby excluding galaxies in the
Southern Extension (Sandage et al.\ 1985); 3) brighter than $B_T = 16$
mag; and 4) with an E, S0, dE, dE,N, dS0 or dS0,N morphological
classification. This sample of 163 galaxies was further trimmed to a
total of 100 galaxies due to observing time limitations, by excluding
S0 or dwarf galaxies with uncertain or disturbed morphology,
significant dust contamination, lacking a visible bulge component, or
previously imaged with the WFPC2 (see Paper I for details). The final
ACSVCS sample spans a range of 460 in blue luminosity ($9.31 \leq B_T
\leq 15.97$). As an early-type galaxy catalog, it is magnitude limited
down to $B_T < 12.15$ mag (26 galaxies), and 54\% complete in the
$12.15 < B_T < 16.00$ magnitude range. If only E galaxies are
considered, the catalog is complete and magnitude limited over the
entire luminosity range. Morphologically, 25\% of the ACSVCS galaxies
are classified in the VCC as E, 33\% as S0, 7\% as E/S0, 5\% as dE,
24\% as dE,N, and 6\% as dS0. Basic properties of the galaxies,
ordered by increasing $B_T$ magnitude (decreasing luminosity), are
listed in Table 1. In addition to right ascension and declination
(both from NED), we list $B_T$ (from the VCC, uncorrected for
extinction), $J$, $H$, and $K_s$ isophotal magnitudes and errors from
2MASS (see below), foreground extinction $E(B-V)$ (from Schlegel et
al.\ 1998), heliocentric systemic velocity (from NED) and Hubble type
(from the VCC). Infrared photometry was retrieved from the 2MASS
All-Sky Extended Source Catalog for all but the brightest 15 of
galaxies, for which the 2MASS Large Galaxy Atlas was used instead
(Jarrett et al.\ 2003). In all cases, the magnitudes listed in Table 1
were measured within a $K_s=20~{\rm mag~arcsec}^{-2}$ elliptical
isophotal aperture.

All ACSVCS observations were carried out with the {\it Wide Field
Channel} (WFC) of the {\it Advanced Camera for Surveys} (ACS) between
2002, December 25 and 2003, December 28. The WFC consists of two
2048$\times$4096 pixel CCDs (WFC1 and WFC2), with a spatial scale of
0\Sec049/pixel. For each galaxy, the center was placed on the WFC1
chip at the reference pixel position (2096,200), and then offset away
from the gap separating the two chips by an additional 5\sec~to
10\sec. In practice, therefore, the center of each galaxy is $\sim$
130\sec~and $\sim$ 155\sec~from the nearest and farthest corners of
the CCD, respectively, and 15\sec~to 20\sec~from the gap separating
the WFC1 and WFC2 CCDs. Each galaxy was observed with an identical
strategy. Exposures in F475W (which resembles the SDSS $g$ filter)
totaled 750s, split in a pair of 375s exposures to aid the removal of
cosmic rays. Exposures in F850LP ($\sim$ SDSS $z$) totaled 1210s and
were divided in a pair of 560s exposure plus a shorter 90s exposure;
the latter was designed to allow recovery of the central regions of
galaxies whose nuclei might saturate in the longer 560s exposure. All
observations were performed in fine lock, giving a nominal pointing
stability of 5 mas (RMS) or better over 60-second intervals.

\section{Data Reduction and Analysis}

\subsection{Data Reduction}

\begin{figure}
\epsscale{0.6}
\plotone{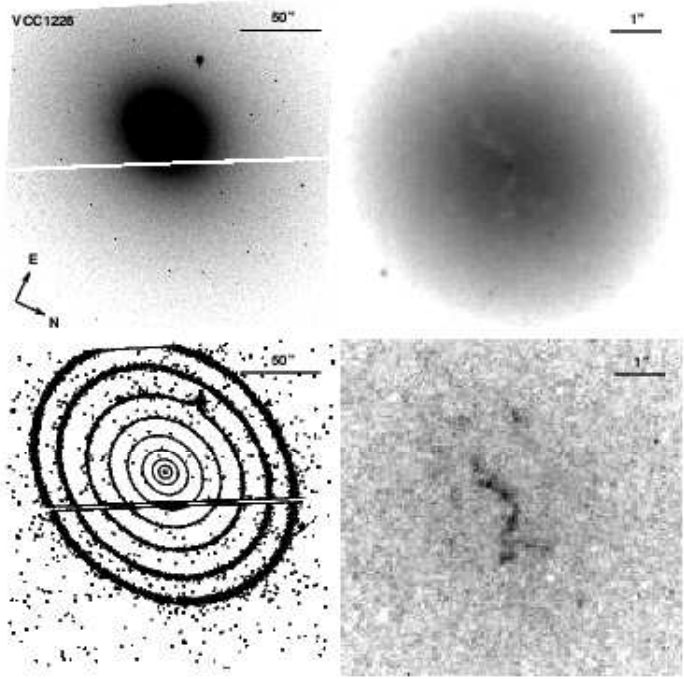}
\caption{The ACSVCS $g-$band image for VCC 1226
  (=M49; NGC 4472, Hubble Type E2/S01(2), see Table 1), the first
  ranked member of Virgo. The top panels
  show the ACS/WFC/F475W full frame (left panel) and a zoom towards
  the center (right panel). The bottom left panel shows a contour map of
  the $g-$band frame, with levels drawn at
  0.05,0.1,0.2,0.4,0.8,1.6,3.2,6.4,12.8,25.6 
  electrons s$^{-1}$ pixel$^{-1}$ (22.82 mag arcsec$^{-2}$ to 16.04  mag
  arcsec$^{-2}$, see equation 4).   Finally, the bottom right panel
  shows a central section of the $g-z$ color image, with red regions
  appearing as dark areas and blue regions as light. The linear scale
  is shown in each panel. All images have been background subtracted.
\label{ima1}}
\end{figure}

Details of the data reduction procedure are given in Paper II.  Images
were downloaded from the STScI data archive using the ``On the Fly
Reprocessing'' option, which processes the images with the task CALACS
as described in the ACS Data Handbook (Pavlovsky et al.\ 2004).
CALACS performs bias and dark subtraction, flat fielding, and 
overscan removal. Rotation and shifts between the five images of each
galaxy were calculated by matching coordinates for a list of objects
in common. Rotations were found to be negligible in all cases, while
shifts of up to 0.2 pixels were measured and applied to bring the
frames into alignment. Images taken with the same filter were
combined, rejecting cosmic rays in the process, using PYRAF's task
MULTIDRIZZLE. Distortion corrected images were also generated by
MULTIDRIZZLE using a Gaussian kernel: although a Lanczos3 kernel is
preferable for some applications (such as the SBF analysis, since it
reduces the correlation between output pixels), a Gaussian kernel is
more effective in repairing bad pixels, and therefore more suitable
for the analysis of the morphology and surface brightness profiles
presented here (Paper II). The final, combined and geometrically
corrected images have size 4256$\times$4256 pixels, with a pixel scale
of 0\Sec049. Because of geometric distortion, thin wedges along the
sides of the final image, as well as the 2\Sec5 gap separating the WF1
and WF2 chips, are not illuminated (e.g., see Figure \ref{ima1}). The
resulting field of view is approximately 202\sec $\times$ 202\sec~in
the shape of a rhomboid.

As described in Paper II, background estimates for the program
galaxies were obtained directly from the processed images as the mode
of the counts measured in two 5\sec~wide, 186\sec~long strips along
the top and bottom (in the CCD coordinate frame) edge of each
field. This procedure is sound only if the galaxies do not fill the
field of view (FOV) of the detector, a condition which is not
satisfied by the brightest ten galaxies  (Figure 2 of Paper II). For
these, the background in each filter was estimated from the relation,
which is followed quite accurately by the fainter 90\% of the sample,
between background count rates and the angle $\Phi_{Sun}$ from the Sun
to HST's V1 axis at the time of the observations (Figure 3 of Paper
II). We note that background estimates can be slightly biased for
galaxies which lie close to brighter companions, in particular VCC
1192 (NGC 4467) and VCC 1199, both of which project less than
5\min~from VCC 1226 (M49).

Images for the brightest two galaxies in the sample are shown in
Figure \ref{ima1}; images of all galaxies can be found on the ACSVCS
website ({\sf http://www.cadc.hia.nrc.gc.ca/community/ACSVCS/}). All images 
are corrected for geometric distortion and
have been background subtracted (see Figures 2 and 3 of Paper II);
the orientation differs from galaxy to galaxy (but is kept the same
for all figures pertaining to the same galaxy), and is shown by the
arrows in the upper left panel. The upper left panel in the Figure
shows the entire ACS FOV. The bottom left panel shows a contour
plot, drawn at the same levels given in the
caption. Finally, the panels on the right-hand side show a zoom around
the galaxy center (top panel), chosen to highlight particularly
interesting features, and a ratio image (bottom panel), produced by
dividing the $g$ and $z-$band background subtracted images. 

\subsection{Recovery of the Isophotal Parameters}

The IRAF task ELLIPSE was used to measure isophotal parameters for
each of the 100 ACSVCS galaxies, in both filters. ELLIPSE adopts the
methodology described in Jedrzejewski (1987): for each semi-major axis
length, the intensity $I(\phi)$ is azimuthally sampled along an
elliptical path described by an initial guess for the isophote's
center ($X,Y$), ellipticity ($\epsilon$), and semi-major axis position
angle ($\theta$). $I(\phi)$ is expanded into a Fourier series as:

\be
I(\phi) = I_0 + \sum_k[A_k{\rm sin}(k\phi) + B_k{\rm cos}(k\phi)].
\ee

The best fitting set of parameters $X, Y, \epsilon, \theta$ are those
that minimize the sum of the squares of the residuals between the data
and the ellipse when the expansion is truncated to the first two
moments (which completely describe an ellipse).  Higher order moments
($k \geq 3$) define deviations of the isophotes from ellipses. In
practice, moments beyond the fourth cannot be measured accurately;
third and fourth-order moments are calculated from the equation above
by fixing the first and second-order moments to their best-fit
values. The third-order moments ($A_3$ and $B_3$) represent isophotes
with three-fold deviations from ellipses (e.g., egg-shaped or
heart-shaped) while the fourth-order moments ($A_4$ and $B_4$)
represent four-fold deviations. Rhomboidal or diamond-shaped isophotes
translate into a non-zero $A_4$. For galaxies which are not distorted
by interactions, $B_4$ is the most meaningful moment: a positive $B_4$
indicates ``disky" isophotes (i.e., with semi-major axis
$B_4\times100$ percent longer than the best fitting ellipse), while a
negative $B_4$ indicates ``boxy" isophotes (i.e., with semi-major axis
$B_4\times100$ percent shorter than the best fitting ellipse;
Jedrzejewski 1987).

When running ELLIPSE on the program galaxies, we adopted the following
guidelines. To avoid having the isophotal solutions affected by bright
globular clusters, foreground stars, or bad pixels, points deviating
by more than three times the standard deviation of the mean intensity
along each trial isophote were excluded from the fit. Furthermore, any
bright galaxy in the FOV, any hot or dead pixel flagged during the
drizzling process, as well as any area affected by dust (see below)
were masked. The semi-major axis was increased logarithmically, i.e.,
each isophote was calculated at a semi-major axis 10\% longer than
that of the isophote preceding it. In general, all parameters were
allowed to vary freely, although in a few specific cases (see below),
it was found necessary to fix one or more of the parameters to achieve
convergence. The isophotal center was not allowed to wander by more
than 2 pixels between consecutive isophotes; in practice, when $X$ and
$Y$ were allowed to vary, the center was found to be stable.  The fit
was terminated when less than 50\% of the fitted ellipse resides
within the CCD boundaries; in practice, ELLIPSE fails to converge well
before this condition is met, since usually the small radial gradient
in the surface brightness profile which characterizes the outer
regions prevents convergence to be reached. When this happens, the
surface brightness profile at larger radii was recovered by running
ELLIPSE with $\theta$, $\epsilon$ and isophotal center fixed to the
mean values calculated for the outermost five fitted isophotes (which
span a factor $(1.1)^4=1.46$ in semi-major axis length).

It is important to stress that the procedure outlined above is
designed for predominantly elliptical isophotes. It will fail to
converge on isophotes which display large deviation from ellipses: in
such cases convergence can be achieved only by fixing the value of one
or more parameters (in what follows these cases will be clearly
identified). We therefore caution the reader that azimuthally averaged
parameters for edge-on disk systems (e.g. VCC 2095, VCC 1125) or
systems with strong bars (e.g. VCC759, VCC 654) might be uncertain.

Dust obscuration was dealt with in the following manner.  Galaxies
affected by dust were identified visually by inspecting both the
residual images obtained by subtracting from the $g-$band image a
smooth model galaxy described by the same isophotal parameters as
those derived for the original image; and the color image produced as
the ratio of the background-subtracted F475W and F850LP frames. As
discussed in Paper I, the longer baseline afforded by the F475W-F850LP
filter combination, when compared to the more commonly used
F555W$-$F814W ($V-I$) color, results in a 50\% increase in the
sensitivity to dust obscuration. Based on the $g-z$ color maps and
$g-$band residual maps, dust is firmly detected in 16 galaxies, and
likely present in two additional objects (Table 2). Under the
simplifying assumption that the dust responsible for the obscuration
is in the galaxy's foreground (i.e., the ``screen approximation"), at
each pixel location $(x,y)$, the absorption in the $g-$band is given
by:

$$
\; A(x,y)_g = - {{2.5} \over {A_z/A_g -1}} \times
$$

\be
\times \left[ \log{I(x,y)_{out,z}
\over I(x,y)_{out,g}} - \log{I(x,y)_{in,z} \over I(x,y)_{in,g}}\right],
\ee

\vskip .1in

\noindent where $I_{in}$ and $I_{out}$ are the intrinsic (unextincted)
and emerging (extincted) flux respectively; and $A_g$ and $A_z$ are
the absorption (in magnitudes) in the $g$ and $z-$band
respectively. Following Cardelli, Clayton \& Mathis (1989), and
assuming the spectral energy distribution of a G2 star, $A_g = 3.634
E(B - V )$ and $A_z = 1.485 E(B - V )$, therefore $A_g/A_z =
2.447$. It follows that $A_g$ (and therefore $A_z$) can be calculated
if $I_{in,z}/I_{in,g}$ (i.e., the intrinsic galaxy color) is known.
This is estimated by linearly interpolating across the dust affected
area (or extrapolating, if the dust affects the center) the 1-D
azimuthally averaged values of $I_{out,z}/ I_{out,g}$ measured in the
dust-free regions immediately surrounding it. A pixel is then flagged
as affected by dust if 1) $A(x,y)_g$ is positive; 2) $A(x,y)_g$ is
larger than the local standard deviation in the dust extinction map,
and 3) if more than two contiguous pixels are affected. The last two
conditions are necessary to avoid flagging noise spikes. In a typical
ACSVCS galaxy, this methodology is sensitive to $g-$band absorptions
$A_g$ of 0.01$-$0.04 magnitudes or larger within the innermost
1\sec. The case is significantly less favorable for galaxies with
extremely low central surface brightness (e.g., VCC 9, VCC 1857), for
which $A_g$ would have to be larger than a few tenths of magnitudes 
for dust to be detected.

Because the dust is likely distributed throughout the body of the
galaxy, rather than in the foreground, the extinction calculated as in
equation (2) represents a firm lower limit to the true extinction. If
the dust is distributed in a thin sheet in the galaxy mid-plane, for
instance, the extinction increases as

$$
\; A(x,y)_{mid-plane} = 
$$

\be
= -2.5\log\left[2\times\left(10^{-0.4A(x,y)_{foreground}} - 0.5\right)\right],
\ee

\vskip .1in

\noindent where $A(x,y)_{foreground}$ is given by equation (2). 

Knowledge of $A(x,y)$ allows the recovery of the intrinsic fluxes,
simply by multiplying the original images by an ``extinction image''
equal to $10^{-0.4A(x,y)}$. For most galaxies, this procedure
generates smooth, dust-free images, but fails spectacularly in a few
cases in which the unextincted color within the dust region is not
well represented by the color extrapolated from the surrounding areas,
as is the case when star formation is on-going within the dust
patches. A thorough discussion of these issues will be presented in a
future contribution, where dust masses will also be discussed.

Areas affected by dust were masked when running ELLIPSE. When the
dust affects less than 50\% of the pixels along individual isophotes
(which is sometimes the case when the dust is distributed in thin
filaments or small patches), a full isophotal solution can still be
recovered in the dust affected areas. For galaxies where more
than 50\% of an individual isophote is affected by dust (for instance
in the presence of dust disks), ELLIPSE was run on the image corrected
for dust as described above.

\begin{figure}
\epsscale{0.9}
\plottwo{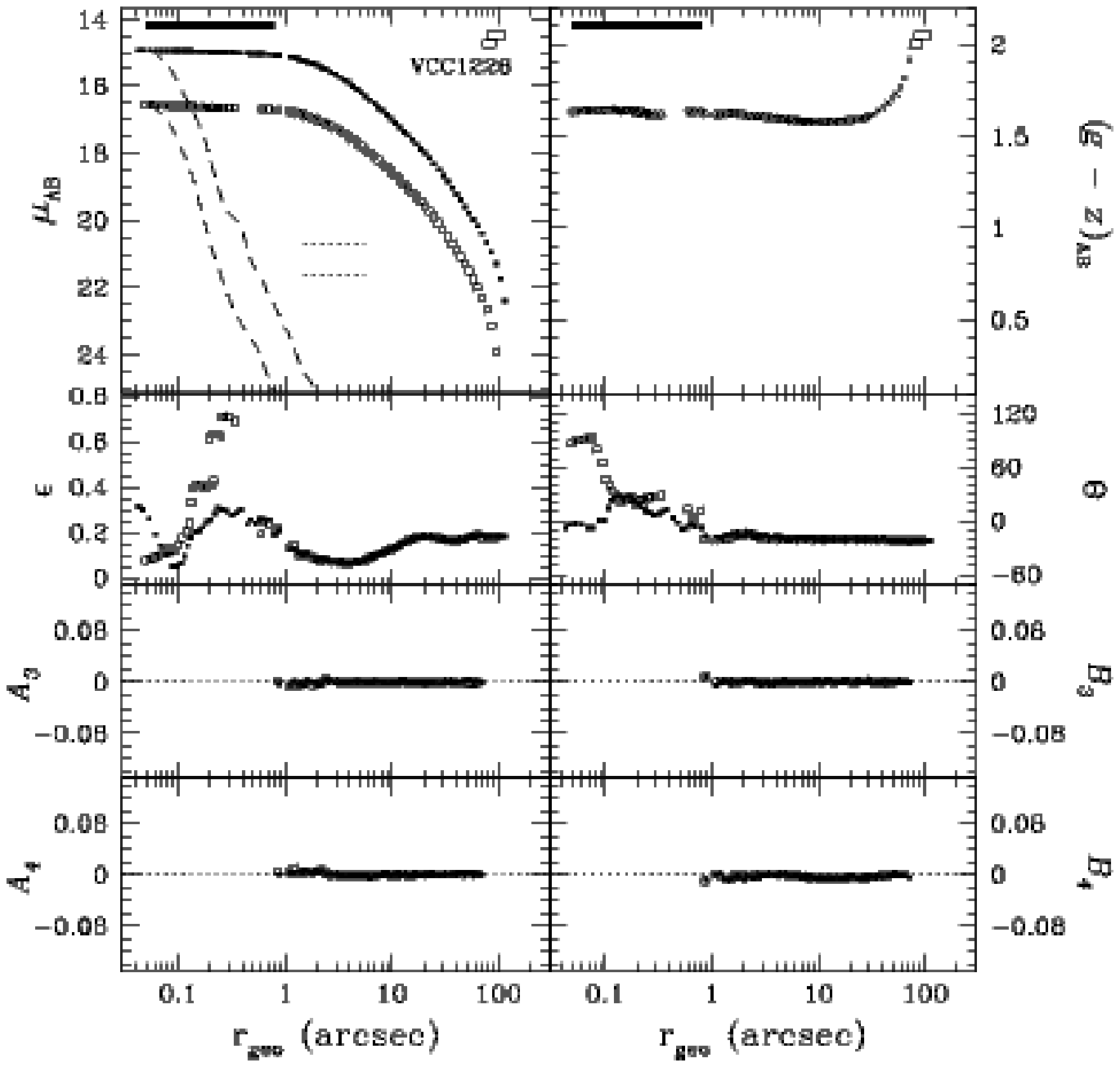}{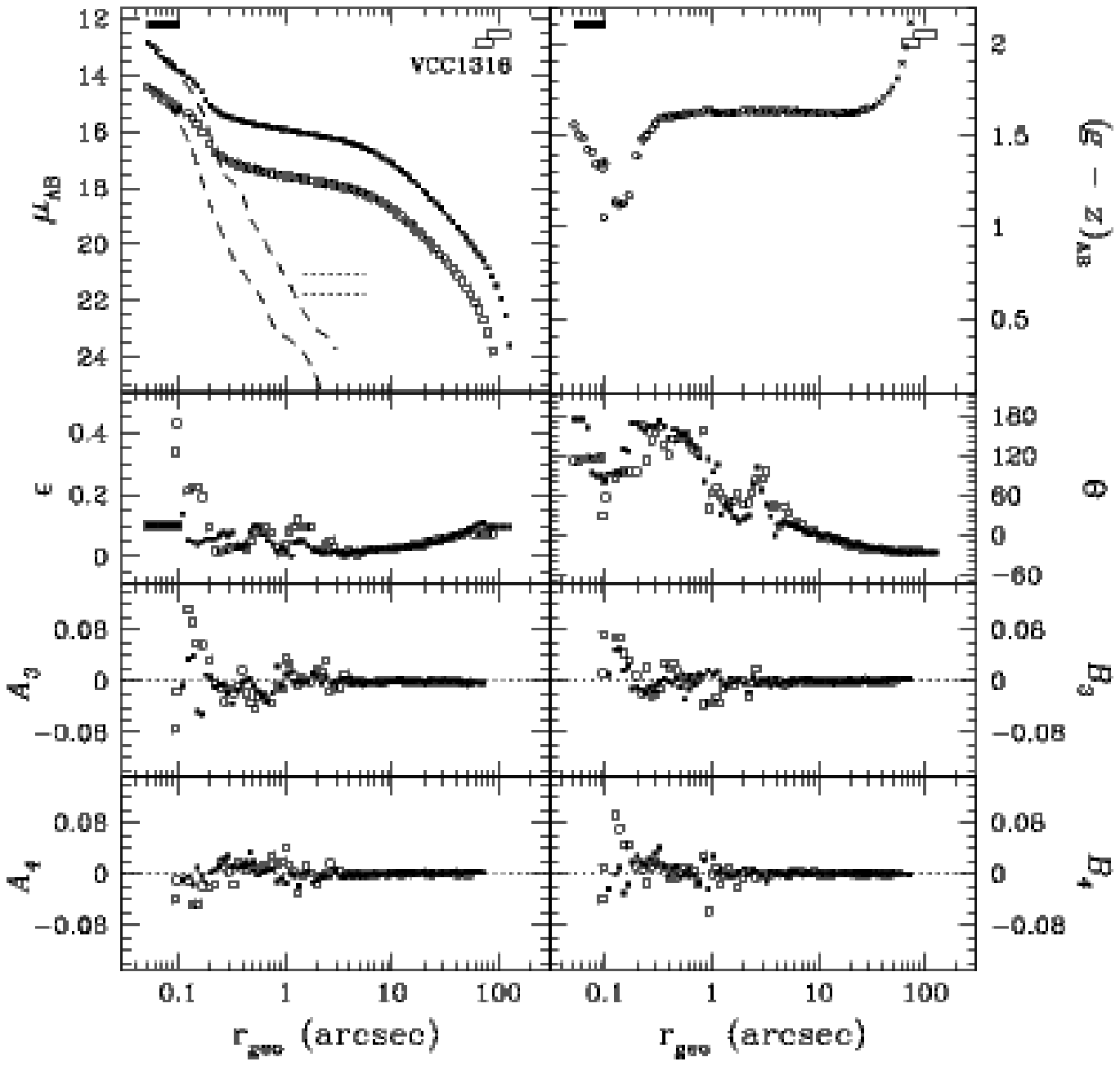}
\caption{{\it (Left Panel)} Isophotal parameters for VCC 1226 (=M49;
NGC 4472, Hubble Type E2/S01(2), see Table 1), plotted against the
`geometric mean' radius  $r_{geo} = a\times\sqrt{1-\epsilon}$, with
$a$ measured along the  semi-major axis of the galaxy. The panels show
surface brightness (in magnitude arcsec$^{-2}$) in both $g$ (open
squares) and $z-$ bands (filled squares), $g-z$ color profile (both
uncorrected for extinction), ellipticity, position angle (in degrees,
measured from North to East) and parameters measuring deviations of
the isophotes from pure ellipses.  In all panels with the exception of
the top right, open and closed symbols refer  to $g$ and $z-$band data
respectively.  Data are only plotted between 0\Sec049 (= 1 WFC pixel)
and the radius  at which the galaxy counts fall below 10\% of the sky;
sky surface brightnesses are shown by the two short dashed horizontal
lines (the fainter sky level is in the $g-$band). Crosses are used in
plotting the color profile when the sum of the (background subtracted)
counts in the two filters is lower than the sum of the background
levels. Finally, a full horizontal bar in the top-left panel
identifies a region where the analysis was performed on images
corrected for dust obscuration, while two open horizontal bars
identify regions where the fit was performed with fixed values of
$\epsilon$ and $\theta$ (the topmost bar is for the $z$-band data, the
bottom bar for the $g$-band data). Higher order coefficients are not
shown in these regions (see text for further details).  {\it (Right
panel)} The same as for the left panel, but for VCC 1316 (=M87;
NGC 4486, Hubble Type E0). 
\label{prof1}}
\end{figure}

The complete results of the ELLIPSE fitting for the brightest two
galaxies are shown in Figure \ref{prof1} for both filters. Plots
showing the results for all other galaxies can be found on the ACSVCS
website ({\sf
 http://www.cadc.hia.nrc.gc.ca/community/ACSVCS/}). Surface brightness
$\mu$, ellipticity $\epsilon$, position angle $\theta$ (measured in
degrees from North to East) of the major axis and deviation of the
isophotes from pure ellipses, as described by the A3, A4, B3 and B4
coefficients (equation 1), are plotted as a function of the `geometric
mean' radius $r_{geo}$.  This is defined as $a\times
\sqrt{1-\epsilon(a)}$, $a$ being the radius measured along the
isophotal semi-major axis. The data are plotted up to the geometric
mean radius at which the surface brightness falls below 10\% of the
sky surface brightness, shown by the two short horizontal lines (the
upper in the $z-$band and the lower in the $g-$band) in the upper left
panel of each figure. Measurements in the $g$ and $z-$band are plotted
using open and closed circles, respectively. A horizontal solid bar in
the upper left panel marks the region within which the isophotal
parameters are derived from images corrected for dust, using the
procedure detailed above. Isophotal shapes cannot be reliably
recovered from the corrected images, and therefore the A3, A4, B3 and
B4 coefficients are not plotted within this range. These coefficients
are also not plotted when the isophotal fit is carried out by fixing
the ellipticity and position angle as described above (these regions
are identified in the Figures by horizontal open bars).

The transformation to the AB photometric system was performed as 
described in Paper II:

\begin{equation}
g_{AB} = -2.5\log f(F475W) + 26.068
\end{equation}

\begin{equation}
z_{AB} = -2.5\log f(F850LP) + 24.862,
\end{equation}

\noindent where $f(F475W)$ and $f(F850W)$ refer to the integrated
fluxes in units of electrons/second (Sirianni et al.\ 2005). Note that
the surface brightness profile and color profile shown in
Figure \ref{prof1} are not corrected for foreground
extinction. In the following discussion, it will be made clear whether
correction for foreground extinction is performed; in such case, we
adopt the differential $E(B-V)$ reddening values listed in Table 1
(from Schlegel et al.\ 1998), and the absorption coefficients listed
immediately following equation 2.

\begin{figure}
\epsscale{0.65}
\plotone{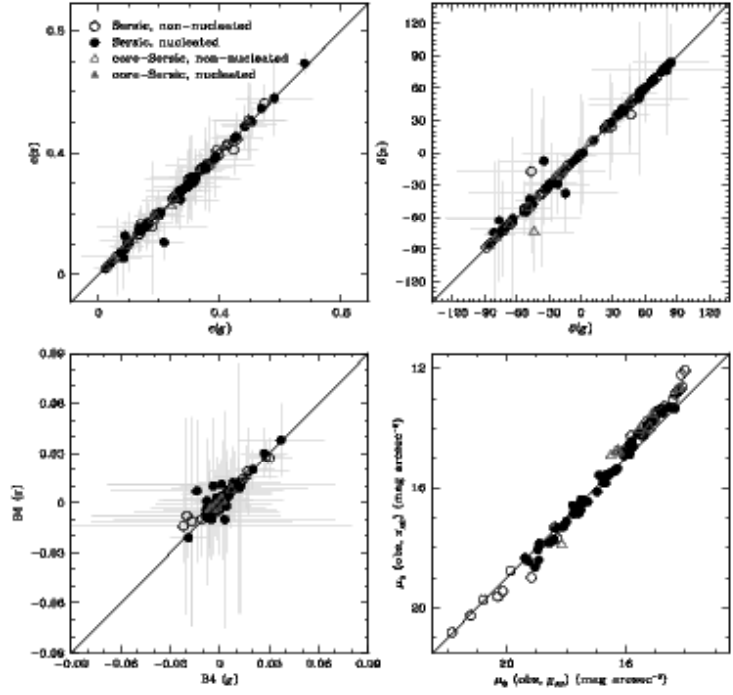}
\caption{Comparison of the mean ellipticity, $\epsilon$, position
  angle, $\theta$, B4 coefficients, and extinction corrected central
  surface brightness, $\mu_0$, measured in the $g$ and $z-$bands.  The
  mean values are calculated between 1\sec~and one effective radius,
  excluding regions affected by dust or nuclei. Standard deviations
  from the mean values are plotted in yellow to avoid over-crowding
  the plot -- such standard deviations are over-estimates of the
  intrinsic errors in the mean, since they are affected by real radial
  variation in $\epsilon$, $\theta$ and B4.   Filled and open symbols
  represent nucleated and non-nucleated galaxies
  respectively. Galaxies best fit by a S\'ersic profile are shown in
  black; galaxies best fit by a core-S\'ersic profile are shown in red
  (see \S 3.3).
\label{comparison2}}
\end{figure}

The mean ellipticity, position angle and B4 coefficients recovered by
ELLIPSE in the $g$ and $z-$band are compared in Figure
\ref{comparison2}. The means are calculated between 1\sec~(thus
excluding a possible nuclear component) and one effective radius $r_e$
(estimated as described in the next section), excluding areas
affected by dust. The parameters recovered in the two bands compare
favorably, attesting to the robustness of the fitting procedure. A
comparison of central surface brightnesses (here and for the rest of
the paper, central surface brightnesses are measured at 0\Sec049),
corrected for foreground extinction as described above, is also shown
in the Figure. The solid line represents a $g-z$ color of 1.0; the
tendency of brighter galaxies to be redder than fainter ones is real
and will be discussed in \S 5.5.

Tables containing the results of the isophotal analysis for each
galaxy will be made available electronically on the ACSVCS website:
{\sf  http://www.cadc.hia.nrc.gc.ca/community/ACSVCS/}.

\subsubsection{A Background Check for the Brightest Galaxies}

\begin{figure}
\epsscale{1.0}
\plottwo{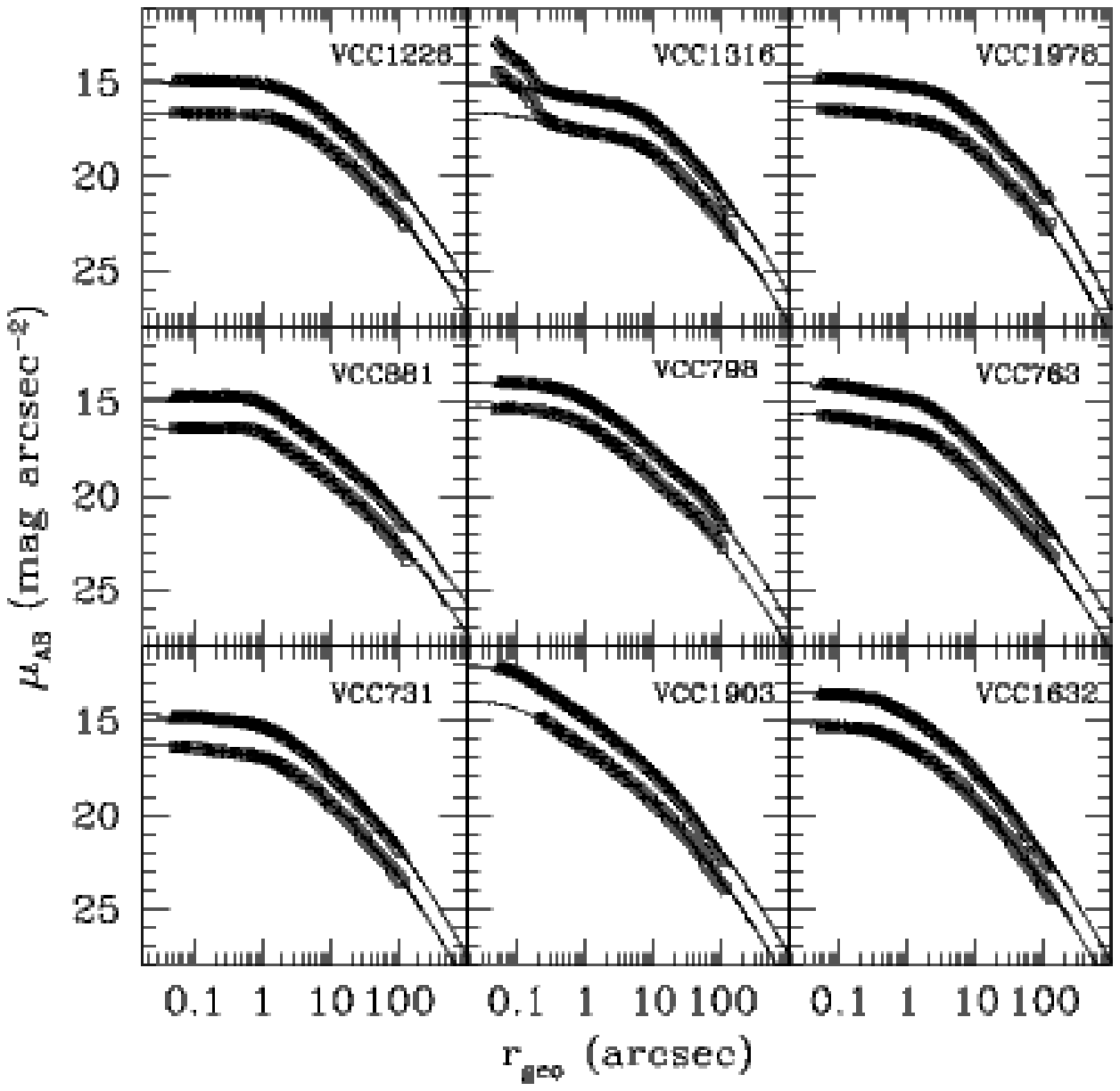}{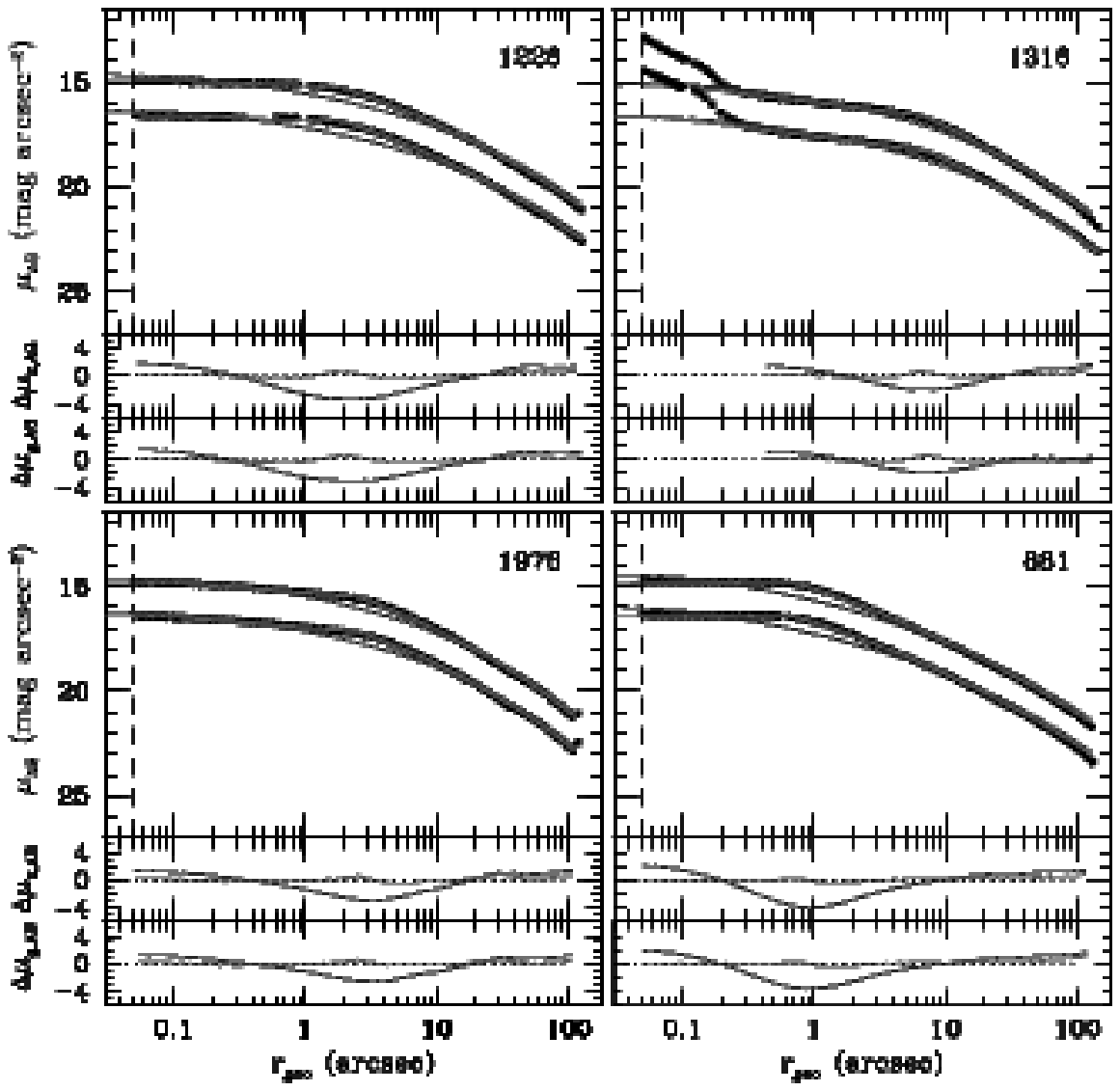}
\caption{(Left) Composite surface brightness profiles, obtained by combining
 the ACSVCS $g$ and $z-$band profiles (shown as open symbols) with
 ground-based $B-$band profiles drawn from the literature, and plotted
 as a function of the geometric mean radius. The solid curve
 represents the best-fit core-S\'ersic model to the final composite
 profile (shown as solid symbols). For each galaxy, the brighter of
 the two profiles is measured in the $z-$band.
\label{comp1}}
\end{figure}

As mentioned earlier, background determination for the ten brightest
galaxies is particularly challenging since these galaxies completely
fill the ACS/WFC FOV. A comparison of our measured profile with
wide-field $B-$band surface brightness profiles available from the
literature for nine of the ten brightest galaxies (Caon et
al.\ 1990, 1994; Michard 1985; no attempt to compare profiles for
VCC 1535 (NGC 4526) was made, because of the severe dust obscuration in this
galaxy) shows generally excellent agreement. In a few cases, however,
some small discrepancies exist at large radii ($r >$
50\sec$-$100\sec). To avoid biases associated with an incorrect
background determination, as well as to increase the radial range of
the data, the ACS $g$ and $z$ surface brightness profiles for these
galaxies were combined with the ground-based $B-$band profiles. To do
so, all three profiles were first fit with spline polynomials, and
the mean $B-g$ and $B-z$ colors over the range 10\sec~to 15\sec~
were determined. Constant offsets were then applied to the $B-$band
profiles, to create pseudo $g$ and $z$ profiles; the fixed color
offset seems warranted given that these galaxies show little or no
evidence for strong radial color gradients within $\sim 0.5r_e$. The
final composite $g$ and $z$ profiles for each galaxy then consist of
the original ACS data inside 30\sec, and the transformed $B$-band
profiles outside of this radius. Figure \ref{comp1} shows the final
brightness profiles for the nine galaxies. The fitted curve in each
case represents the best-fit S\'ersic or core-S\'ersic models, described in \S 3.3.

\subsection{Parametrization of the Surface Brightness Profile}
\label{sec:parmodels}

\subsubsection{Galaxy Profiles}

We opt to fit the surface brightness profiles of the ACSVCS galaxies
using both a S\'ersic (1968; see Graham \& Driver 2005 for a review)
and a core-S\'ersic model (as in Trujillo et al.\ 2004, see also
Graham et al.\ 2003). This section is designed to provide a brief
exposition of other models available in the literature, in particular
the ``Nuker" model (Lauer et al.\ 1995), and to justify our choice.

The first HST studies of early-type galaxies (Crane et al.\ 1993;
Ferrarese et al.\ 1994; Lauer et al.\ 1995) revealed surface
brightness profiles which remained steep all the way to the smallest
resolved radius, and could be closely approximated, in the inner
regions, as power-laws, $I(r) \propto r^{\gamma}$. It was immediately
apparent that in order to fit the HST profiles, one had to move beyond
traditional parametrizations, such as exponential laws, de Vaucouleurs
or Jaffe profiles (de Vaucouleurs 1948; Jaffe 1983). To reproduce the
profiles of 14 bright elliptical galaxies in Virgo, observed with the
(pre-refurbishment) Planetary Camera on board HST, Ferrarese et al.\
(1994) introduced a double power-law model described by four
parameters: the inner and outer power-law slopes, the break radius
which marks the transition between the two, and the intensity at the
break radius. Examining a larger sample of galaxies, again observed
with the pre-refurbished HST, Lauer et al.\ (1995) proposed a modified
five parameter model (the ``Nuker'' model):

\begin{equation}
I(r) = I_b2^{(\beta-\gamma)/\alpha} \biggl ({r \over r_b} \biggl
)^{-\gamma} \biggl [1 + \biggr({r \over r_b} \biggl )^{\alpha} \biggr
]^{ (\gamma - \beta)/\alpha}. 
\end{equation}

Like the model of Ferrarese et al.\ (1994), the Nuker model is a blend
of inner and outer power-laws (of slopes $\gamma$ and $\beta$
respectively), but has the added flexibility that the sharpness of the
transition between the two, which occurs at the break radius $r_b$,
can be controlled by the parameter $\alpha$.

Graham et al.\ (2003) point out that since the Nuker model was
specifically designed to describe only the innermost part of the
profile (none of the profiles studied by Lauer et al.\ (1995) extends
beyond 10\sec, corresponding to $\sim$ 800 pc at the distance of a
typical galaxy in their sample), the model fails to reproduce the
kpc-scale  profiles of bulges and early-type galaxies, which deviate
significantly from power-laws. More seriously, Graham et al.\ showed
that the fitted Nuker parameters are sensitive to the extent of the
radial range which is being fit; the degree of sensitivity is enough
to compromise the physical interpretation of  scaling relations based
on such parameters. Finally, the integral of a Nuker model for $0 \leq
r \leq \infty$ has no guarantee of converging, preventing the
calculation of fundamental quantities such as effective radii and
total magnitudes. To remedy these shortcomings, Graham et al.\
proposed a new model, composed of  an outer S\'ersic (1968) model,
given by

\begin{equation}
I(r) = I_e \exp \left\{ -b_n \biggl [ \biggl ({r \over r_e} \biggl
)^{1/n} - 1 \biggr ] \right\},
\end{equation}

\noindent and an inner, power-law profile. The ``core-S\'ersic'' model
is therefore described as

$$
I(r) = I^{\prime} \biggl [1 + \biggl ( {r_b \over r} \biggr )
^{\alpha} \biggl ]^{\gamma / \alpha}
$$

\be
\times  \exp \biggl [-b_n \biggl (
{r^{\alpha} + r_b^{\alpha} \over r_e^{\alpha}} \biggr ) ^{1/(\alpha
n)} \biggr ].
\ee

\vskip .1in 
\noindent where $I^{\prime}$ is related to the intensity, $I_b$, at the break 
radius $r_b$, as:

\begin{equation}
I^{\prime} = I_b2^{-\gamma / \alpha} \exp \biggl [b_n \biggl
(2^{1/\alpha}r_b/r_e \biggl )^{1/n} \biggr ].
\end{equation}

As in the Nuker model, $r_b$ separates the inner power-law of
logarithmic slope
$-\gamma$ from the outer part of the profile, which in this case is
described by a S\'ersic model with effective radius $r_e$. The
parameter $\alpha$ controls the sharpness of the transition between
the inner and outer profile, with higher values corresponding to
sharper transitions.

For a S\'ersic model (or for the S\'ersic component of a core-S\'ersic
model) with $1 \lesssim n \lesssim 10$ , the effective radius $r_e$
contains roughly half the integrated light of the model if $b_n
\approx 1.992n-0.3271$ (e.g., Capaccioli 1989; Caon et al.\ 1993;
Graham \& Driver 2005). Effectively, therefore, the number of free
parameters in a S\'ersic model is three, and in a core-S\'ersic model
six. The parameter $n$ controls the overall shape of a S\'ersic
profile, with low $n$ values producing curved profiles with
logarithmic slopes which are shallow in the inner regions, and steep
in the outer parts, while high $n$ values produce extended profiles
with less overall curvature (i.e. steep inner profiles and shallow
outer profiles).  For $n=1$, a S\'ersic model becomes a pure
exponential, which is known to provide a good fit to the surface
brightness profiles of some dwarf ellipticals, while for $n=4$ it
reduces to a classical de Vaucouleurs profile, which is a good
approximation of the large-scale profiles of most bright ellipticals.

Trujillo et~al.\ (2004) found it convenient to restrict the
core-S\'ersic model to the limit $\alpha \rightarrow \infty$,
corresponding to an infinitely sharp transition between the S\'ersic
and power-law profiles at the break radius $r_b$:

$$
I(r) = I_b \biggl [(r_b/r)^{\gamma}u(r_b-r) +
$$

\begin{equation}
+ e^{b_n(r_b/r_e)^{1/n}}e^{-b_n(r/r_e)^{1/n}}u(r-r_b) \biggr ];
\end{equation}

\vskip .1in
\noindent where $u(x-z)$ is the Heaviside step function and all other
parameters have the same meaning as in equation 8.  For $\alpha
\rightarrow \infty$, the number of free parameters in a core-S\'ersic
model is reduced from six to five, the same number of parameters as in
the Nuker model. The intensities (and corresponding surface
brightnesses) at the effective radius and at the break radius are
related as:

\begin{equation}
I(r_e) = I_b\exp\left\{b_n \biggl [ \biggl ({r_b \over r_e} \biggl
)^{1/n} - 1 \biggr ] \right\}
\end{equation}

\begin{equation}
\mu_e = \mu_b - 2.5b_n \biggl [\biggl ({r_b \over r_e} \biggl )^{1/n}
- 1 \biggl ] \log e.
\end{equation}

The S\'ersic and core-S\'ersic models offer significant advantages
compared to a Nuker model. First and foremost, they provide as good a
description of the inner (100 pc scale) profiles, and a significantly
better description when the profiles are extended to kpc-scale (Graham
et al.\ 2003; Trujillo et al.\ 2004) . The S\'ersic profile might have
wide applicability: for instance, Merritt et al.\ (2005) find it to
provide a good description of the mass profiles of ${\rm \Lambda}$CDM
halos. Unlike the case of the Nuker model, the parameters of a
S\'ersic and core-S\'ersic model are  robust against the extent of the
radial range covered by the data (as long as the profile is sampled to
radii of order $r_e$, i.e. large enough to allow to secure the
curvature, Graham et al.\ 2003), and both models have finite integrals
for $r \rightarrow \infty$. Finally, while the Nuker model was
designed specifically as a phenomenological description of the
observed projected profile, the S\'ersic model has a simple physical
interpretation: the roughly Boltzmann energy distribution
corresponding to a S\'ersic model is what is expected for systems
which are well mixed as a consequence of violent relaxation or merging
(Hjorth \& Madsen 1995).

Based on these considerations, we opt to fit S\'ersic and
core-S\'ersic models for the ACSVCS galaxies, with only one
caveat. Many of the ACSVCS galaxies host distinct stellar nuclei at
their centers (Paper VIII); in these cases, we allowed the inclusion
of a nuclear component $I_{nuc}(r)$ in the models. This component was
modeled as a King (1966) profile, which is described by a lowered
isothermal distribution function of the form

\begin{equation}
f(E) = \exp(-\beta E)-1,
\end{equation}

\noindent where $E$ is the energy. There is no analytic expression for
the density corresponding to this distribution function, but the
function $I_{nuc}(r)$ is readily obtained by integrating Poisson's
equation. The King (1966) models are a one parameter family usually
characterized by the concentration parameter $c=\log(r_t/r_c)$, where
$r_t$ is the tidal radius (i.e., the radius beyond which the density
is zero) and $r_c$ is the core radius (i.e., is the radius at which
the density falls to about half its central value). In addition there
are two scale parameters, the overall flux $I_k$ and a spatial scale
parameter, which we take to be the half-light radius $r_h$. The core
and half-light radii are uniquely related, as their ratio is a
function of $c$ (e.g., McLaughlin et al.\ 2000). Thus, a King model
adds three parameters, $c,~r_h$, and $I_k$, to the fit of the surface
brightness profiles of the ACSVCS galaxies.

Finally, we point out that in the interest of providing a homogeneous
analysis, single S\'ersic and core-S\'ersic models (with the possible
addition of a nuclear King model) were applied to all galaxies,
including those showing evidence of multiple morphological components
(e.g., stellar disks or bars, see \S4.2). For these galaxies our
approach is clearly unsatisfactory, and 2-D multi-component fits
should be used instead for a rigorous analysis (e.g., Peng et
al.\ 2002). {\it We therefore caution the reader against attaching too
much weight to the profile fits for galaxies flagged as having
multiple components; in the following discussion and figures, we will
single out these galaxies to remind the reader of the limitation of
our single 1-D model fits.}

\subsubsection{Parameter fitting}

We used a $\chi^2$ minimization scheme to find the S\'ersic and
core-S\'ersic (the latter with $\alpha$ set to infinity as in Trujillo
et al.\ 2004) models which best fit the $g$ and $z-$band intensity
profiles of each galaxy. When the presence of a nucleus could be
established, either based on the inspection of the images, or of the
surface brightness profiles (see Paper VIII), a central King profile
was also included in the fitting. Because of difficulties in
estimating  reliable errors in the intensity measurements, we adopt
the usual procedure (e.g. Byun et~al.\ 1996) of assigning equal
fractional weights to all points in the profile (i.e. the amount by
which the fitted profile is allowed to deviate at each point is
proportional to the intensity at that point, see equation 14).  The
assumption of equal fractional weights is not unjustified given the
way the profiles are constructed: the lower signal-to-noise ratio
associated with the lower intensity of the profile at large radii is
offset somewhat by the larger number of points sampled along the outer
isophotes. As will be shown shortly, the profile fits under this
assumption appear satisfactory, however, the lack of reliable error
estimates in the intensity measurements prevents us from deriving
meaningful errors in any of the fitted parameters.  Finally, to avoid
parameter mismatches due to differing initial conditions in the
minimizations, the same starting inner radius was used when fitting
the model in both filters.

All minimizations were carried out using the {\tt Minuit} package in
CERN's program library.  The minimization procedure used consists of a
Simplex minimization algorithm (Nelder \& Mead 1965) followed by a
variable metric method with inexact line search (MIGRAD) to refine the
minima found by the Simplex algorithm. The models are convolved with
the ACS point spread function (PSF) before being compared to the
observed profile. PSFs for F850LP and F475W were obtained from images
of the globular cluster 47Tuc, processed in exactly the same way as
the ACSVCS images analyzed here. The PSF varies across the field of
view and, at a fixed location, it depends on the star's centroid within
the pixel; therefore to each galaxy we associate the PSF corresponding
to the exact (sub-pixel) location of the galaxy center, measured as
the light centroid within the inner 5 pixels (see Paper II and Paper
VIII for additional details).

Although the best-fit parameters need not be the same in the two
filters (e.g., the presence of color gradients would lead to a
physically meaningful difference in a parameter related to the profile
slope), to first order such differences should be small. However,
external factors (for instance, the fact that it is sometimes possible
to fit the surface brightness over a larger radial extent in one
filter than in the other) might cause the minimization routine to
converge to significantly different solutions in the two filters, a
situation which is not physically justified.

To circumvent this problem, the fitting procedure was divided in two
separate steps. First, we fit models {\it simultaneously} to the $g$
and $z$ profiles, $I(g)$ and $I(z)$. In doing so, initial guesses for
the parameters are chosen based on the observed profiles; all
parameters are constrained to be identical in both filters, with the
exception of the intensity parameters, which are kept independent.
For instance, in the case of a S\'ersic fit with a nuclear component
included, we obtain the parameters by minimizing

$$
\chi^2_{sim}[I_e(z),I_e(g),r_e,n,c,r_h,I_k(z), 
I_k(g)] = 
$$

$$
= \sum_i \{[I_i(g)-S[r_i|I_e(g),r_e,n,c,r_h,I_k(g)]]/ \epsilon 
I_i(g)\}^2
$$

\be
\; + \sum_i \{[I_i(z)-S[r_i|I_e(z),r_e,n,c,r_h,I_k(z)]]/ 
\epsilon I_i(z)\}^2,
\ee

\noindent where the sum is performed over all measured values of $I$,
$S$ is the PSF-convolved S\'ersic plus King model profile and
$\epsilon$ is an arbitrary factor (set to 0.001) chosen to effectively
assign fractionally small (and equal) errors to all points. Second,
the best parameters obtained from the simultaneous fit are used as
initial guesses for individual fits to the $g$ and $z$ intensity
profiles. In this case, for each band we obtain the best fit
parameters by minimizing (using again the S\'ersic fit with a nuclear
component as an example):

$$
\chi^2_{Sersic}(I_e,r_e,n,c,r_h,I_k) = 
$$

\be
= \sum_i
[(I_i-S(r_i|I_e,r_e,n,c,r_h,I_k)/\epsilon I_i]^2.
\ee

\vskip .1in
Because the nuclei of galaxies brighter than $B_T = 13.5$ mag are
generally rather extended and do not clearly stand out against the
bright galaxy background, fits for these galaxies were performed by
constraining the S\'ersic index $n$, the half-light radii and the
concentration parameter of the King model to be the same in the two
filters (see Paper VIII for further details).

The parameters of the S\'ersic or core-S\'ersic models best describing
the surface brightness profiles of the ACSVCS galaxies are listed in
Table 3. Global properties derived from these profiles are listed in
Table 4, and will be discussed in more detail in later sections. The
best fit S\'ersic (blue) and core-S\'ersic (red) profiles for the
brightest four galaxies are shown over-plotted to the $g$ and $z-$band
surface brightness profiles in Figure \ref{fit1}. Fits
for all other galaxies can be found on the ACSVCS website ({\sf
 http://www.cadc.hia.nrc.gc.ca/community/ACSVCS/}). 
Solid curves show the final fits, including a nuclear component when
present, while dashed and dotted curves show the individual galaxy and
nuclear components respectively. Residuals are shown in the lower
panels. With the best-fit parameters in hand, we calculated the total
$g$ and $z-$ band magnitudes and effective radii (defined as
containing half of the total light of the galaxy\footnotemark)
\footnotetext{In the case of a core-S\'ersic model, and with the
definition of $b_n$ given in the text, $r_e$ in equations 8 and 10
represents the radius within which half of the total light of the
outer S\'ersic model, rather than of the complete model, is contained
(see Trujillo et al. 2004). In this case, the $r_e$ values in Tables 3
and 4 are not equal.} by integrating the model to $r =
\infty$\footnotemark.  \footnotetext{We note that with this
definition, the total magnitudes of galaxies which are tidally
truncated might be overestimated.}

The model profiles (before and after PSF convolution) will be made 
available on the project website. 
The results of the profile fitting will be discussed in \S
4.2. Before doing so, we will digress momentarily to describe, in
broad strokes, the galaxies' morphologies (\S 4.1).

\section{Results}
\label{sec:results}

\subsection{Galaxy Morphologies}

The ACSVCS galaxies display a staggering variety of morphologies. A
detailed description of the morphological features detected in each
galaxy is given in the Appendix, while Table 2 summarizes the
incidence of dust, stellar disks, bars, stellar nuclei and other
morphological anomalies. The properties of the stellar nuclei   are
the subject of Paper VIII, and will be discussed only  briefly here.

\subsubsection{Dust Features}

Dust is firmly detected in 16 galaxies, and likely present in two
additional systems (Table 2). Forty-two percent (11 out of 26)
galaxies brighter than $B_T = 12.15$ mag (which, as mentioned earlier,
form a magnitude-limited sample), have dust features. This is
consistent with previous studies which targeted preferentially bright
elliptical galaxies (e.g., Goudfrooij et al.\ 1994; van Dokkum \&
Franx 1994; de Koff et al.\ 2000; Tran et al.\ 2001). Dust is also
detected in seven of the galaxies fainter than $B_T = 12.15$. Although
only one of these (VCC 1779) is classified in the VCC as a dwarf
(giving a nominal detection rate of $1/35 = 3$\%), strong selection
effects are at play, since the generally low surface brightness of
dwarf galaxies prevents us from detecting low levels of dust
obscuration. Three bright early-type Virgo galaxies excluded from the
ACSVCS are known to have dust based on ground based observations (VCC
2066 = NGC 4694, VCC 801 = NGC 4383 and VCC 758 = NGC 4370; Sandage \&
Bedke 1994; Caon, Capaccioli \& D'Onofrio 1994). A firm lower limit on
the incidence of dust in the 65 E+S0 galaxies in the ACSVCS, plus VCC
2066, VCC 801 and VCC 758 is therefore $(17+3)/(65+3) = 29$\%.

Dust features can be broadly grouped in four categories: 

\begin{description}

\item{C1.\phantom{x}Faint wisps and patches, often crossing the center (VCC
1226=M49, VCC 1316=M87, VCC 881=M86, VCC 1632=M89, VCC 355=NGC 4262,
VCC 1619=NGC 4550, VCC 1779 and, possibly, VCC 1422, VCC 798=M85);}

\item{C2.\phantom{x}Large, kiloparsec-scale, high extinction irregular dust
lanes (VCC 763=M84, VCC 571);}

\item{C3.\phantom{x}Large (over 700 pc, or 8\Sec7, across), patchy dust disks
(VCC 1535=NGC 4526, VCC 1154=NGC 4459, VCC 1030=NGC 4435, VCC
1250=NGC 4476);}

\item{C4.\phantom{x}Small (less than 300 pc, or 3\Sec7 across), thin,
featureless, regular dust disks (VCC 759=NGC 4371, VCC 685=NGC 4350,
VCC 1327=NGC 4486A).}

\end{description}

Evidence of recent/on-going star formation, in the form of bright blue
associations, is always associated with the large, clumpy dust disks
(category 3). Star formation within cold, dusty gas disks has been
discussed by Tan \& Blackman (2005); indeed all three of the smaller,
regular disks (category 4) are seen in conjunction with regular,
smooth blue stellar disks/rings, perhaps suggesting that these disks
might represent a later evolutionary phase of the larger, clumpier
dust disks. Star formation is not seen in connection with the
irregular dust lanes, which may therefore represent an earlier
evolutionary phase of the large dust disks, when the dust has not yet
had the time to settle into a stable configuration and star formation
has yet to begin. This would suggest an evolutionary scenario in which
dust features evolve from category C2 to C4 through C3. Since the
unsettled, irregular morphology of the dust lanes (C2) implies an
external origin (Goudfrooij et al.\ 1994l; van Dokkum \& Franx 1995; de
Koff et al.\ 2000; Tran et al.\ 2001; Temi et al.\ 2004), this would
argue for an external origin even for the regular, smaller dust disks
(C4). Alternatively, the nuclear dust disks might be the result of the
bar-induced accumulation of gas and dust towards the center, and might
be the precursors of the blue, nuclear stellar disks seen in several
of the ACSVCS galaxies (\S 4.1.2). The connection between the
disks/lanes, and the thinner, less massive, more tenuous dust
filaments, none of which is accompanied by star formation, is not
immediately apparent. These issues will be discussed in more detail in
a future contribution.

Large-scale, highly organized dust disks are not found in the faintest
half of the sample; had they been present in the galaxies omitted from
the ACSVCS, they would certainly have been detected based on ground
based images. This fact aside, there appears to be no correlation
between either the amount of dust obscuration, or the morphology of
the dust features, and the magnitude or morphology of the host galaxy.
For instance, VCC 1226 (M49) and VCC 1316 (M87, Figure \ref{ima1}),
the two brightest galaxies in Virgo, show only faint dust wisps and
patches, while VCC 1250 (NGC 4476), which is 3.6 mag fainter, sports a
clumpy, nuclear dust disk 1.8 kpc ($\sim$ 22\sec) across. 
VCC 571, 1.8 magnitudes fainter still, shows only thin
dust filaments emanating from the center. The
most massive dust features, well organized disks $\sim 1.4$ kpc
(17\Sec5) across, are seen in VCC 1535 (NGC 4526) and VCC 1154 (NGC
4459), two bright galaxies of
similar magnitude. These disks are remarkably similar in size,
extinction, and overall structure. However, VCC 1154 (NGC 4459) shows
very regular isophotes, while the isophotes of
VCC 1535 (NGC 4526) are highly disturbed: the galaxy becomes extremely
boxy at a distance between 10\sec~and 20\sec~from the center, switches
to a diamond shape further out, and becomes very disky at a distance
of 70\sec. VCC 759 (NGC 4371) and VCC 355 (NGC
4262), two of the galaxies which host a dominant stellar bar, show a
small regular dust disk and thin dust filaments respectively, however, 
no dust is detected in the
barred galaxies VCC 2092 (NGC 4754) and VCC 654 (NGC 4340).

\subsubsection{Stellar Disks}

Estimating the incidence and strength of disk structures in early-type
galaxies is a complex problem owing to the effects of inclination on
the measured surface brightness profile, ellipticity and isophotal
shape. Rix \& White (1990) estimate that for inclinations angles $i$
(with $i$ measured between the plane of the disk and the plane of the
sky, so that $i = 90$\deg~for edge-on systems)  larger (i.e., more
edge on disks) than $\sim$ 55\deg, a disk containing as little as 10\%
of the light of the bulge component would cause the isophotes to
depart from a pure elliptical shape. At smaller inclinations (more
face-on disks), the projected isophotes of even luminous disks are
almost perfectly elliptical, and disks can be identified only through
the perturbations they impart on the surface brightness profile of the
accompanying bulge. Unfortunately, such deviations can be recognized
only if the intrinsic profile of the bulge is known {\it a priori}, an
invalid hypothesis given the considerable scatter in the shapes of the
profiles displayed by galaxies of comparable magnitude (\S 4.2). It
follows that a morphological analysis would not detect the vast
majority of  bulge-embedded disks with $i \lesssim$ 55\deg~.

Keeping these biases in mind, disks are detected in two basic
varieties within the ACSVCS sample: fully embedded, 100-pc scale inner
disks hosted in galaxies that are bulge-dominated on large-scales; and
kpc-scale outer disks that dominate the light distribution in the
galaxy outskirts. In some cases (e.g., VCC 1692=NGC 4570, VCC 685=NGC
4350, see Table 2), the two classes overlap, i.e., a ``nested-disk''
structure composed of two morphologically distinct inner and outer
disks is detected $-$ similar morphologies have been reported in the
past by, among others, van den Bosch et al.\ (1994), Lauer et al.\
(1995), Scorza et al.\ (1998) and Erwin et al. (2003). Because the
disks are almost always close to edge-on, they impart a characteristic
signature to the surface brightness profile. For instance, Figure
\ref{vcc1692} shows a cut through the photometric major axis of VCC
1692. At large radii, the profile is dominated by the outer,
exponential, disk. In the inner region, a blue stellar ring with
radius 150 pc, but less than 7 pc wide, surrounds the nucleus; 
the outer boundary of the ring produces a clear
discontinuity in the profile at 1\Sec8 from the center. The structure
of the inner region of VCC 1692 has been addressed by Van den Bosch \&
Emsellem (1998) in the context of secular bar evolution.

Outer disks have been detected in 16 of the ACSVCS galaxies; in
two additional cases (VCC 798=M85 and VCC 1231=NGC 4473, the 5th and
11th ranked early-type members of Virgo) the presence of a moderately
inclined disk is suggested by the positive B4 coefficients. In VCC 1938 (NGC 4638) 
the outer disk, which is seen edge-on, is rather abruptly
truncated approximately 300 pc (4\Sec0) from the center; within this
region is nested a boxy, elongated bulge. Figure \ref{vcc1938} shows a
cut through the photometric major axis for VCC 1938, clearly showing
the inner truncation  of the disk. In addition, the North-West side 
of the disk (shown on the left
side of the profile, left panel) shows a discontinuity at 1.1 kpc
(14\sec) from the center. This discontinuity, which is not immediately
apparent in the South-East side of the profile, might correspond to
density fluctuations within the disk, as could result from spiral
structure. ``Inner Truncated Disks'' (ITD), such as the one observed
in VCC 1938, have surfaced regularly in the literature since Freeman's
(1970) observation that in a significant number of spiral galaxies the
inward extrapolation of the exponential surface brightness profile of
the outer disk would exceed the observed central surface brightness
(Freeman's Type II disks). The incidence of ITDs, detected on the
basis of a 1-D or 2-D bulge/disk decomposition of the surface
brightness profiles, is of order 25\% in spiral galaxies (Baggett,
Baggett \& Anderson 1998, 1996; Anderson, Baggett \& Baggett 2004),
and as high as 65\% in S0 galaxies (Seifert \& Scorza 1996). The
general consensus is that ITDs are the result of secular evolution
induced when the outer disk becomes bar unstable (Baggett et al.\
1996; Emsellem et al.\ 1996; van den Bosch \& Emsellem 1998; Scorza \&
van den Bosch 1998); indeed, in the case of VCC 1938, an edge-on bar
could be disguised as the boxy bulge detected within the inner cutoff
of the edge-on outer ITD. In addition, the bar-driven influx of
(presumably enriched) gas is identified as the mechanism by which the
inner disks seen in nested-disk structures are built (Erwin 2004;
Emsellem et al.\ 1996; Scorza et al.\ 1998; van den Bosch \& Emsellem
1998; Scorza \& van den Bosch 1998). The large line strength values
detected at the centers of S0 galaxies support this hypothesis
(Fisher, Franx \& Illingworth 1996; van den Bosch \& Emsellem 1998);
in VCC 1125 (NGC 4452) the resolved nucleus and
the inner stellar disk are bluer than the surrounding bulge,
indicating that both might have formed at a later time by accumulation
of gas towards the center. The small, nuclear dust disks surrounded by
blue stellar disks and rings seen in VCC 685 (NGC 4350) and VCC 1327
(NGC 4486A) could represent
earlier stages of the inner, blue stellar disks, when gas and dust
driven towards the center by the bar has not yet been completely
converted into stars.

The bar-driven accumulation of mass in the center, whether it results
in the formation of a disk, bulge or the feeding of a central
supermassive black hole, would ultimately have a stabilizing influence
through the creating of a strong inner Lindblad resonance, thereby
leading to dissolution of the bar (e.g., Norman, Sellwood \& Hasan
1996). In the next section, we will argue that VCC 1537 (NGC 4528), in
which a bar structure is clearly detected, could
represent a transition stage in which an ITD, seen at  moderate
inclinations, is currently being formed.

VCC 2095 (NGC 4762) and VCC 1125 (NGC 4452) display a morphology 
intermediate between IDTs and
nested-disks. In both galaxies, the edge-on outer disk continues
uninterrupted to the center of the galaxy, but its thickness appears
variable, giving the impression of a ``nested-disk'' structure (Figure
\ref{vcc1125}).  Cuts along the major axis (Figure \ref{vcc2095}) show
multiple ``humps''; a detailed kinematical study is necessary to
decide whether the structure is due to nested-disks, a single, warped
disk, or a disk plus a ``lens'' (Freeman 1975; Kormendy 1984).  Either
a lens or an ITD could be responsible for the morphology of  VCC 575
(NGC 4318, left panel of Figure \ref{vcc2095}), which shows a bright
bulge confined within a sharply bounded, low surface brightness
region, about 7\sec~across (280 pc radius), beyond which a disky
component (positive B4, higher ellipticity) sets in.

Finally, we mention in passing the admittedly speculative possibility
that the inner ``dip'' detected in the surface brightness profile of
VCC 881 (Figure~\ref{vcc881}; Carollo et al. 1997; Lauer et al. 2002;
Lauer et al. 2005) might be associated with the inner truncation of an
outer, almost face-on disk, which would  otherwise remain undetected
based on the isophotal analysis. VCC881 (NGC 4406) is kinematically
peculiar in that the main galaxy exhibits minor axis rotation, while
the core is kinematically decoupled (Bender 1988; Wagner et
al. 1988). Whether secular evolution of a (now dissolved) bar formed
within a disk structure could explain the observed kinematics is a
possibility worth investigating; based on its stellar velocity
dispersion, the galaxy hosts a $\sim 4\times10^8$ \msun~supermassive
black hole at its center (Ferrarese \& Ford 2005), the (bar-induced?)
growth of which could be partly responsible for the bar dissolution.

\subsubsection{Stellar Bars and Rings}

Stellar bars have been detected in a few early-type galaxies (Scorza
et al.\ 1998; Emsellem, Goudfrooij \& Ferruit 2003; Busarello et al.\
1996; Erwin 2004); and indeed they might be connected with the
truncated and nested-disk structures discussed in the previous
section.  Unfortunately, investigating the extent of such a connection
is complicated by the fact that bars are most easily detected in
face-on configurations which are least conducive to disk detection.

Stellar bars are seen in at least seven, and perhaps as many as nine
of the ACSVCS galaxies (Table 2); although the presence of a disk is
not immediately apparent in any of them (as noted above, observational
biases work against the detection of a disk in these objects), their
profiles are not dissimilar from those displayed by galaxies with
large-scale stellar disks. With two exceptions (VCC 1537=NGC 4528 and
VCC 654=NGC 4340), the bars are, however, of significantly
larger size (over 1 kpc) than, and are therefore likely unrelated to,
the ones needed to explain the ITDs in VCC 1938 (NGC 4638) and,
possibly, VCC 575 (NGC 4318).

VCC 1537 (NGC 4528), however, is a prime candidate for a transition
object between regular S0s and ITDs. The isophotal parameters show a
dramatic change in ellipticity, A4 and B4 coefficients (and, to a
lesser extent, major axis position angle), between 480 and 960 pc
(6\sec~and 12\sec); inspection of the images
clearly reveals the presence of a bar, misaligned by $\sim
60$\deg~relative to the major axis of the main body of the galaxy. 
The surface brightness profile
outside 470 pc  from the center is roughly exponential (Figure
\ref{vcc1537}), the ellipticity is larger compared to the value
measured in the inner region, and the B4
coefficient is slightly positive, indicating the presence of a
large-scale disk. Immediately inside 470 pc of the center,  the
surface brightness falls below the extrapolation of the exponential
describing the large-scale profile.  This morphology is consistent
with the presence of an ITD, inclined approximately 60\deg~to the
plane of the sky and truncated at an inner radius of 470 pc,
corresponding to the outer Lindblad resonance within which the bar is
contained.

The other bars detected in the ACSVCS galaxies vary widely in terms of
strength, size and the structure of the surrounding region. Beside VCC
1537 (NGC 4528), the most spectacular cases are VCC 654 (NGC 4340) and
VCC 759 (NGC 4371).  VCC 654 is listed as a double-barred galaxy in
Erwin (2004). The outer (primary) bar extends for 5.5 kpc (68\sec),
displaying sharp edges where the bar meets an outer stellar ring. The
latter has major axis approaching 10 kpc (125\sec), and minor axis
misaligned, by approximately 17\deg, with respect to the bar. The
inner (secondary) bar extends for 0.6 kpc (8\sec), and is responsible
for the ``stretching'' of the isophotes and the dramatic change in the
major axis position angle registered beyond a few arcsec. The inner
bar is bounded by a nuclear
stellar ring, extending for 1.5 kpc (19\sec) along the major
axis. Nuclear and outer rings appear aligned, and the inner bar is
roughly aligned with the minor axis of the nuclear ring. Inner and
outer bars, however, are slightly misaligned, in other words the outer
bar does not appear to be a simple extension of the inner bar.

Like VCC 654, VCC 759 displays a large stellar ring, 13 kpc (160\sec)
across along the major axis, and a large scale bar. 
The inner region is characterized by at least two
kpc-scale stellar rings; however there appears to be no evidence for
an inner bar. With VCC 355 (NGC 4262), VCC 759 is one of two barred
galaxies to host dust: a small, 120 pc (1\Sec5), dust disk surrounds the
center of the galaxy. Dust disk and stellar rings appear to be
concentric, coaxial, and share the same ellipticity. In VCC 2092 (NGC
4754) and VCC 355 only the
bar, but no ring, is visible. Weaker bars are likely present in VCC
778 (NGC 4377) and VCC 1883 (NGC 4612); in both cases, the bar is
misaligned with respect to the
main body of the galaxy. In two more galaxies (VCC 369=NGC 4267, and
VCC 1283=NGC 4479), although a bar is not immediately
apparent by a simple visual inspection of the images, its presence is
inferred from the rapid changes in ellipticity, higher order
coefficients and, in particular, major axis position angle exhibited
by the isophotes. These are characteristics of the barred galaxies
described above (see also Erwin 2004); in particular a radial
variation in position angle is unlikely the result of an edge-on disk.

\subsubsection{Other Morphological Peculiarities}

In the next few paragraphs, we will briefly touch upon a few
additional galaxies that display interesting morphologies.

Four of the ACSVCS galaxies project close to bright companions. VCC
1327 and VCC 1297 (NGC 4486A and NGC 4486B respectively) project
7\Min5 and 7\Min3 respectively from M87, while VCC 1192 (NGC 4469) and
VCC 1199 project 4\Min2 and 4\Min5 respectively from M49.  For all
except VCC 1327 (NGC 4486A) there is clear evidence that the galaxies
have been tidally stripped. Indeed, NGC 4486B has long been held as
one of the textbook cases of compact ellipticals which result from
tidal truncation (Rood 1965; Faber 1973); the surface brightness
profile declines rapidly beyond a few
arcseconds. With VCC 1192 and VCC 1199, and compared to galaxies of
similar $B_T$, NGC 4486B has brighter central surface brightness,
smaller effective radius, and a surface brightness profiles which is
best fit by a larger S\'ersic index $n$ (see \S 3.3.1, \S 4.2 and
Table 3). The case is not as clear-cut for VCC 1327: the surface
brightness profile, effective radius and central surface brightness of
this galaxy are consistent with those of galaxies of similar magnitude.

We confirm the existence of a face-on spiral pattern in VCC 856 (IC
3328), a galaxy formally classified as a dE1,N. The pattern was first
noticed by Jerjen et al.\ (2000). Since then, evidence of spiral
patterns or disk/bar structures have been found in a handful of Virgo,
Fornax and Coma dwarf ellipticals (Jerjen et al.\ 2001; De Rijcke et
al.\ 2003; Graham et al.\ 2003; Barazza et al.\ 2002); these objects
have been interpreted as late-type galaxies in the process of being
transformed to an early-type morphology by harassment within the
cluster environment. In addition to VCC 856, for which the pattern is
clearly visible both in the $g$ and $z-$band images, tantalizing
evidence for spiral structure is found in VCC 1695 (IC 3586, classified
as dS0:) and VCC 1199 (classified as E2). Although the structure is
not immediately apparent from inspection of the single images, it does
appear in the residual images, obtained by subtracting the best
isophotal model from the original images. The $g-$band residual images
of both galaxies can be found on the ACSVCS webpage ({\sf
 http://www.cadc.hia.nrc.gc.ca/community/ACSVCS/}). In the case of VCC
1199, which, as noted above, is a close companion of M49, it is
possible that the spiral arms might have been triggered by the
interaction between the two galaxies (Toomre 1969; Goldreich \&
Tremaine 1979)

Finally, four galaxies, VCC 21 (IC 3025), VCC 1779
(IC 3612), VCC 1499 (IC 3492),
and VCC 1512, all classified as dS0 galaxies,
should instead be reclassified as dwarf irregular/dwarf elliptical
transition objects, based on the presence of loose associations of
bright blue star clusters, and the irregular isophotal shapes.

\subsection{Surface Brightness Profiles}

Figure \ref{fig07} shows representative profiles for six `regular'
(i.e., displaying elliptical, regular isophotes, and no obvious
stellar disks or bars) galaxies in our survey, ordered from top to
bottom according to their total $B-$band magnitude. The diversity in
the light distribution is immediately apparent: the extent and slopes
of the profiles differ greatly. In addition, some of the profiles are
dominated by the presence of a nuclear component (Paper VIII). The
profiles for all galaxies, to which the contribution of the nuclear
component has been subtracted, are plotted in Figure \ref{allprofiles}
as a function of $r/r_e$, with $r_e$ the effective radius. When
plotted in this fashion, some common traits between the profiles start
to become apparent: in general, as the galaxy becomes fainter, the
curvature in the profiles becomes more pronounced and the profile's
slope within $\sim$ 1\sec decreases. The brightest ten or so galaxies
deviate from this trend and seem to define a class of their own, with
extended profiles which flatten towards the center in a manner which
is uncharacteristic of the remaining galaxies in the sample. In the
following we will quantify these statements in a more rigorous
fashion, and discuss the implications in \S 5.

\subsubsection{S\'ersic or Core-S\'ersic?}

To justify the choice of parametrization of the surface brightness
profile, for each galaxy we calculated the $\chi^2$ of the best
S\'ersic and core-S\'ersic models, following Equation (15). In an
absolute statistical sense, such $\chi^2$ values are not easily
interpretable since realistic error estimates in the surface
brightness data are not simply quantifiable because of uncertainties
in the sky values, fitting procedures, and, when applicable,
correction for dust obscuration. However, comparing $\chi^2$ values
for different galaxies allows us to assess the {\it relative} goodness
of fit for the two models. In what follows, we will assign to each
galaxy the largest of the $g$ and $z-$band $\chi^2$ (calculated in a
radial range which excludes the nuclear star cluster if present), and
convert this into a reduced $\chi^2$, $\chi^2_r$, by dividing by the
number of fitted data points minus the number of fitted parameters
(three and five for the S\'ersic and core-S\'ersic models
respectively).

Figure \ref{chiplot} plots $\chi^2_r$ of the best fit S\'ersic model
against the ratio of the $\chi^2_r$ of the best fit S\'ersic and
core-S\'ersic models. The figure is divided in two panels: on the
left, we plot all `regular' galaxies, which appear to be well
described by a single morphological component (nucleus
not-withstanding), while on the right we plot all galaxies with
multiple morphological components (e.g., bulge + stellar disks, bars
or spiral structures). Several points are noteworthy:

\begin{enumerate}

\item{The $\chi^2_r$ for a S\'ersic model is never significantly lower
than for a core-S\'ersic model, which is not surprising since a
core-S\'ersic model can always be reduced to, and hence perform as
well as, a pure S\'ersic model by choosing the right combination of 
$r_b$ and $\gamma$.}

\item{Although several of the galaxies with multiple morphological
components (nucleus not-withstanding) are, somewhat surprisingly,
reasonably well fit by a S\'ersic, or core-S\'ersic, model (i.e.,
$\chi^2_r$ (S\'ersic) $\lesssim$ a few), a larger majority has high
$\chi^2_r$, compared to regular galaxies. This is of course due to the
fact that the presence of morphologically distinct features creates
irregularities in the 1-D profile which the models are not designed to
reproduce. In all these cases, it seems sensible to choose the simpler
S\'ersic over the core-S\'ersic model as a first-order description of
the profile.}

\item{While for most regular galaxies a S\'ersic fit provides a good
description of the data, as demonstrated by the low value of the
corresponding $\chi^2_r$, there are a handful of exceptions. In
particular, there are galaxies for which both S\'ersic and
core-S\'ersic models perform equally ``badly'', i.e., the $\chi^2_r$
for the two models are large and similar [$\chi^2_r$ (S\'ersic) $\gg$
1 and $\chi^2_r$ (S\'ersic)/$\chi^2_r$ (core-S\'ersic) $\leq 2$]. For
these cases, the simpler S\'ersic model is again chosen over the
core-S\'ersic model.}

\item{For a few galaxies, a core-S\'ersic model provides a
significantly better description of the data than a S\'ersic
model. This is the case for eight of the 10 brightest Virgo galaxies
(in the case of VCC 1535=NGC 4526, the profile is seriously
contaminated by dust and a reliable fit cannot be carried
out). Examples are VCC 1226 (M49), VCC 1978 (M60) and VCC 881 (M86):
for all, the slope of the inner profile is shallower than the one
given by a S\'ersic model constrained by the profile at large radii. 
This result is robust against the extent of the
radial range to which the model is fit; in particular it is not
affected by the inclusion of the ground-based data in the surface
brightness profile (\S 3.2.1). In the left-hand panel of Figure
\ref{chiplot}, these galaxies are plotted as red triangles and are all
found in the right upper half of the diagram, i.e., $\chi^2_r$
(S\'ersic) / $\chi^2_r$ (core-S\'ersic) $\gtrsim$ 10, and $\chi^2_r$
(S\'ersic) $> 200$: their $\chi^2_r$ then supports the conclusion that
a core-S\'ersic model is preferable to a pure S\'ersic model.}

\item{For two additional galaxies, a core-S\'ersic fit also seems to
provide a better description of the data than a S\'ersic fit. These
are VCC 1250 (NGC 4476) and VCC 1512. In the left panel of Figure 
\ref{chiplot}, both galaxies
(plotted again as red triangles) are found near
$\chi^2_r$(S\'ersic)/$\chi^2_r$ (core-S\'ersic) $\sim 10$ and
$\chi^2_r$ (S\'ersic) $\sim 100$, and in both cases a S\'ersic model
fails to reproduce the abrupt transition observed between the outer
and inner profile. Nevertheless, the choice of a core-S\'ersic model
should be considered marginal: VCC 1250 harbors a 22\sec~dust disk and
a stellar nucleus; although the correction for dust extinction
performs well (\S 3.2), both the effective radius and the break radius
of the best fit core-S\'ersic model are within the dust-affected area,
therefore the fits should be interpreted with caution. VCC 1512 is
best classified as a dwarf irregular/dwarf elliptical transition
object, therefore its properties cannot be simply related to those of
the early-type galaxies in the sample. For the three other dIrr/dE in
the sample, VCC 21, VCC 1779 and VCC 1499, a core-S\'ersic model is
well suited at following the small gradient of the profile towards the
center; however the difference between core-S\'ersic and S\'ersic fits
is not as pronounced as in the case of VCC 1512. }

\end{enumerate}

In conclusion, the choice of a core-S\'ersic model is amply justified
only for the eight bright galaxies discussed in 4), and should be
considered tentative for VCC 1250 (NGC 4476) and VCC 1512.  Figure
\ref{comparison1} shows a comparison of the  best fit S\'ersic index
$n$, effective radius $r_e$, break radius $r_b$ (for the galaxies best
fit by a core-S\'ersic model), and inner slope $\gamma$ measured in
the $g$ and $z-$bands. In the case of the 90 galaxies best fit by a
S\'ersic profile, $\gamma$ is defined as the logarithmic slope of the
model profile at 0\Sec1:

\be
\gamma_{0.1} = -{{\rm d} \log I \over  {\rm d} \log r} =  {b_n \over n} \left[0.1 \over 
r_e {\rm (arcsec)}\right]^{1/n}.
\ee

Note that with this definition, $\gamma$ is not affected by the
presence of a nuclear component and represents the inner slope of the
profile prior to PSF convolution. The consistency of the parameters
recovered by the independent fits in the two bands testifies as to the
robustness of the fitting procedure. Finally, $g$ and $z-$band
magnitudes are compared to $B_T^0$ from the VCC in Figure
\ref{magcomp}. Notice that the $g$ and $B-$band are similar but not
identical, furthermore, the zero point uncertainty associated with the
VCC $B-$band magnitudes is of order 0.25 mag (Binggeli et al.\
1985). Both effects can lead to a small systematic $g-B$ color
difference. The tendency of fainter galaxies to be bluer than brighter
ones is a known trend (\S 5.5).

\subsection{Scaling Relations}
 
The fact that the morphological and kinematical properties of
elliptical galaxies, bulges, galaxy nuclei and disks are markedly
different is the reflection of the clearly distinct physical
mechanisms regulating the formation of these systems. Here we will
focus on scaling relations involving only morphological parameters;
the study of the fundamental plane and other relations including
kinematical properties will be left to future papers.

Figure \ref{all1} presents a collection of correlations between fitted
parameters in the $g$ (plotted below the diagonal line running from
the lower left to the upper right corner) and $z-$band (plotted above
the same diagonal line). Along the diagonal are shown histograms for
all of the parameters: $g$ and $z-$band data are plotted using a solid
and dashed line respectively; in both cases, the value of the
parameter is plotted along the x-axis, and the number of galaxies on
the y-axis (for which units are not shown). Some particularly relevant
correlations, discussed in more detail below, are reproduced in Figure
\ref{fcorr2}, where galaxies are color coded according to the
parametrization of the surface brightness profile, and the
presence/absence of multiple morphological components and/or a
nucleus. The following points are worth noticing:

\begin{enumerate}

\item{The central surface brightness $\mu_0$ (panels {\it ae} and {\it
  ea} of Figure \ref{all1}) correlates well with the total galactic
  magnitude. $\mu_0$ is measured directly from the images as the
  intensity recorded at the central pixel, and therefore includes the
  contribution of a nuclear component, as well as PSF smearing. The
  basic trend, however, is unchanged if $\mu_0$ is substituted with
  the surface brightness $\mu_{0.05}$ measured at 0\Sec049 (1 pixel)
  from the model best fitting the galaxy surface brightness profile
  (equations 7 or 10), before PSF convolution and excluding a nuclear
  contribution, if present (Figure \ref{fcorr2}). The clear trend of
  central surface brightness becoming brighter with decreasing
  magnitude (increasing luminosity) is broken only by the brightest
  half a dozen galaxies, for which the trend is reversed. The
  brightest galaxies do not follow the trend defined by the rest of
  the sample even if to the measured  $\mu_0$ one substitutes the
  value obtained from the inner extrapolation of the S\'ersic model
  best fitting the outer parts of the profile. Substituting $\mu_0$ or
  $\mu_{0.05}$ with the surface brightness at one effective radius,
  $\mu_e$, leads to a similar correlation but increases the scatter
  (panels {\it af} and {\it fa} of Figure \ref{all1}, and Figure
  \ref{fcorr2}). This is due to the fact that, both for a S\'ersic and
  core-S\'ersic model, $\mu_e$ is related to $\mu_{0.05}$ through a
  term depending on $r_e$, and the latter shows larger scatter as a
  function of galaxy magnitude (see point 4, panels {\it ad} and {\it
  da} of Figures \ref{all1}, and~\ref{fcorr2}).}

\item{There appears to be a sharp upper limit to the value of the
  central surface brightness achieved by early-type galaxies, as is
  apparent from the histogram showing the distribution of $\mu_0$
  (panel {\it ee} of Figure \ref{all1}). This cutoff occurs at $\mu_0
  = 14.0$ mag ($\mu_{0.05} = 13.5$ mag) in the $g-$band, and $\mu_0
  = 12.1$ mag ($\mu_{0.05} = 11.9$ mag) in the
  $z-$band\footnotemark. \footnotetext{It should be remembered that
  while $\mu_0$ is measured directly from the images, $\mu_{0.05}$ is
  computed from the galaxy model before PSF convolution, therefore
  $\mu_{0.05}$ can be brighter than $\mu_0$).} The brightest galaxies
  in the sample fall below this upper cutoff.}

\item{The S\'ersic index $n$ (which, in the case of galaxies best fit
  by a core-S\'ersic model, is equal to the index of the S\'ersic model
  that best describes the outer profile) increases monotonically as
  the galaxies become brighter (panels {\it ac} and {\it ca} of
  Figure \ref{all1}, and Figure \ref{fcorr2}). In particular, the
  shape of the outer profile for the brightest galaxies is consistent
  with that expected by simply extrapolating the trend defined by the
  fainter systems: in contrast with the claim of Prugniel \& Simien
  (1997), there is no clear indication in our sample for a leveling
  off in the value of $n$ at the bright end. Prugniel \& Simien
  (1997), using S\'ersic fits to ground-based data for galaxies with
  $-22 < M_B \leq -16$ mag, found $n\sim$ const $\sim4$ for $M_B \leq
  -20$ mag. The constancy of $n$ for the brightest galaxies is likely a
  bias introduced by the fact that these galaxies have cores which
  are shallower than expected based on a S\'ersic fit to the outer
  parts: a S\'ersic fit to the entire profile will therefore be
  forced to have larger curvature (lower $n$) than it would were the
  inner region to be excluded from the fit. Indeed, S\'ersic fits to
  the ACSVCS data for the brightest galaxies produce values of $n$ 
  which are approximately constant and clustered around $n \sim 4$.}

\item{The effective radius $r_e$, which, with $n$ and $\mu_0$, is one
  of three independent parameter in a S\'ersic fit, shows a weak
  correlation with $M$ (panels {\it ad} and {\it da} of Figure
  \ref{all1}, and Figure \ref{fcorr2}). The brightest galaxies ($M_B
  \lesssim -20.2$ mag) are distinct from the rest of the sample,
  producing a clearly bimodal distribution in the $r_e$ histogram
  (panel {\it dd} of Figure \ref{all1}). The mean effective radius for
  galaxies with $M_B \gtrsim -20.2$ mag is 13\Sec6 $\pm$ 6\Sec6
  ($1.1\pm0.5$ kpc), while for brighter galaxies $r_e$(mean) =
  169\sec~$\pm$103\sec~($13.5\pm8.2$ kpc) $-$ an abrupt change in the
  trend of $r_e$ {\it vs.} magnitude at $M_B \sim -20$ mag was also
  noted by Binggeli, Sandage \& Tarenghi (1984, their Figure
  7). Although the ACSVCS sample is complete for $M_B \leq -18.94$
  mag, the bimodality observed in the $r_e$ values is likely due to
  the small number of galaxies straddling the critical magnitude range
  $-20.5 < M_B < -20$; indeed a discontinuity in $r_e$ was not noted
  by either Binggeli, Sandage \& Tarenghi nor Graham \& Guzman (2003),
  using larger samples of galaxies.  We note that the large $r_e$
  values measured for the brightest galaxies are not an artifact of
  the choice of parametrization for the surface brightness profile:
  while forcing a S\'ersic fit to the brightest galaxies (as done by
  Binggeli et al.) decreases $r_e$ by 30\% to 40\% relative to
  the value obtained using a core-S\'ersic fit (as a consequence of
  the bias described in point 3 above), this is not enough to bring
  the brightest galaxies in agreement with the rest of the sample.}
 
\item{The logarithmic slope of the inner brightness profile, $\gamma$,
  becomes steeper as the galaxy luminosity increases (panels {\it ab}
  and {\it ba} of Figure \ref{all1}, and Figure \ref{fcorr2} $-$ where
  $\gamma$ is defined as the slope of the profile {\it underlying} a
  nuclear component, when present, and prior to PSF convolution, see
  equation 16). The scatter in the relation between $\gamma$ and $M$
  is quite large, with galaxies with steep central cusps ($\gamma
  \sim 0.6$) found over the entire magnitude range (once the
  brightest galaxies are removed). Galaxies brighter than $M_B \sim
  -20.2$ mag (best fit by a core-S\'ersic profile), are clearly distinct
  from the rest of the sample in having inner slopes shallower than
  expected based on the extrapolation of the trend shown by the
  fainter systems (in agreement with the correlation between $\mu_0$,
  $\mu_{0.05}$ and $M$ noted in point 1). }

\end{enumerate}

Points 1, 2 and 5 leave no doubt that the brightest galaxies ($M_B
\lesssim -20.5$) are set apart from the rest of the sample insofar as
innermost regions (a few hundred parsecs) are concerned. However, the
clear dichotomy in the distribution of effective radii (see point 4
above) reveals a physical disconnect in the overall structure as
well. We will discuss the implications of these findings in the next
section. Fits to some of the scaling relations shown in Figure
\ref{fcorr2} are given below, where ``S'' and ``CS'' refer to the
samples of S\'ersic and Core-S\'ersic galaxies respectively.

\be
\mu_{0.05}(z) = (1.65\pm0.09) (M_B+18) + (15.7\pm0.13)   \hspace{0.3in} {\rm S}
\ee

\be
\mu_{0.05}(z) = -(1.06\pm0.38) (M_B+18) + (11.0\pm1.3)   \hspace{0.3in} {\rm CS}
\ee

\be
\mu_{0.05}(g) =  (1.51\pm0.09) (M_B+18) + (17.1\pm0.13)   \hspace{0.3in} {\rm S}
\ee

\be
\mu_{0.05}(g) = -(0.97\pm0.38) (M_B+18) + (12.8\pm1.2)   \hspace{0.3in} {\rm CS}
\ee

\be
\mu_{r_e}(z) =  (0.76\pm0.10) (M_B+18) + (20.5\pm0.1)   \hspace{0.5in} {\rm S}
\ee

\be
\mu_{r_e}(z) =  (0.31\pm0.48) (M_B+18) + (23.2\pm1.3)   \hspace{0.5in} {\rm CS}
\ee

\be
\mu_{r_e}(g) =  (0.63\pm0.10) (M_B+18) + (21.8\pm0.13)   \hspace{0.5in} {\rm S}
\ee

\be
\mu_{r_e}(g) =  (0.14\pm0.43) (M_B+18) + (24.1\pm1.2)  \hspace{0.5in} {\rm CS}
\ee

\be
\log n(g) = -(0.10\pm0.01) (M_B+18) + (0.39\pm0.02)   \hspace{0.3in} {\rm All}
\ee

\be
\log r_e(g) = -(0.055\pm0.023) (M_B+18) + (1.14\pm0.03)   \hspace{0.3in} {\rm S}
\ee

\be
\log r_e(g) = -(0.22\pm0.13) (M_B+18) + (1.50\pm0.37)   \hspace{0.3in} {\rm CS}
\ee

\section{Discussion and Conclusions}
\label{sec:discussion}

\subsection{Are Early-Type Galaxies divided in ``Core'' and
  ``Power-Law'' types?}

Although the first systematic surveys of the cores of elliptical
galaxies date back to the late '70s - early '80s (King 1978; Schweizer
1979; Kormendy 1985; Lauer 1985), it was only with HST that core
properties could be studied in detail. For instance, in the best
ground-based seeing conditions, the cores of a few giant ellipticals
appeared isothermal (Kormendy 1985); even before refurbishment, HST
revealed this not to be the case (Crane et al.\ 1993). The first HST
studies also seemed to indicate a dichotomy in the core properties of
bright elliptical galaxies. Ferrarese et al.\ (1994) divided a
magnitude limited sample of fourteen galaxies in the Virgo cluster
into two broad classes: the fainter galaxies appeared to have
power-law profiles, which showed no indication of flattening to the
smallest radius resolved by the instrument, while the brightest third
of the sample showed resolved (non isothermal) cores, producing a
clear break in the surface brightness profile. Lauer et al.\ (1995)
strengthened these results using a larger (although not as
homogeneous) sample of galaxies. Lauer et al.\ (1995) and Faber et
al.\ (1997) introduced the currently adopted definition of a ``core''
galaxy as one for which the logarithmic slope of the surface
brightness profile (measured using a Nuker fit) is $\gamma < 0.3$ as
one approaches the resolution limit of the instrument (0\Sec1 for
their study). In agreement with Crane et al.\ (1993) and Ferrarese et
al.\ (1994), Lauer et al.\ also found no cases in which $\gamma =
0$. Furthermore, Lauer et al.\ found no galaxies for which $0.3 <
\gamma < 0.5$, and refer to galaxies for which $\gamma > 0.5$ as
``power-law''. The separation between core and power-law galaxies in
Lauer et al.'s sample is dramatic, with no overlap between the two
classes; a more recent study by the same authors (Lauer et al.\ 2005)
confirms this stark bimodality. This conclusion was partly challenged
by Rest et al.\ (2001), who found that 10 galaxies (15\% of their
sample) had profiles with $0.3 < \gamma < 0.5$, the region devoid of
galaxies in Lauer et al.\ (1995) sample. Rest et al.\ note that a
Nuker model, by virtue of the fact that it approximates the profile
with two power-laws, will bias the value of $\gamma$ if the slope of
the surface brightness profile varies smoothly with radius.

Figure \ref{hist2} shows the $\gamma$ distribution for the ACSVCS
galaxies. The logarithmic slope in the figure is calculated from the
best fitting model at $r=$ 0\Sec1; it is equal to the slope of the
inner power-law for galaxies best fit by a core-S\'ersic profile, and
follows equation 16 for S\'ersic galaxies. We stress that the
definition of $\gamma$ adopted in this paper is significantly
different from that commonly found in the literature. The slope of a
S\'ersic model changes continuously with radius, therefore it will
generally be shallower at the center than the slope obtained if a
power-law fit (as in the Nuker model) is enforced (a point already made
by Rest et al.\ 2001). More importantly, nuclei are excluded in this
study, i.e., $\gamma$ is defined as the slope of the underlying host galaxy,
unbiased by the nuclear component, if present. Previous studies often
blocked out the nucleus when fitting the galaxy profile, however,
unrecognized nuclei (especially in bright galaxies) would have biased
$\gamma$ towards larger values.

The $\gamma$ distribution for the entire ACSVCS sample (plotted as a
thick solid line in Figure \ref{hist2}) is strongly skewed; however,
the only hint of bimodality is a modest, and not statistically
significant, excess of galaxies with $\gamma \sim 0.5$. Moreover, the
physical cause driving the shape of the $\gamma$ distribution shown in
Figure \ref{hist2} is dramatically different from the one proposed by
Lauer et al.\ (1995) and subsequent papers by the same group. In these
earlier works, $\gamma$ was found to increase going from bright to
faint galaxies:  the low-$\gamma$ peak of the distribution was
occupied exclusively by the brightest ``core'' galaxies, while the
high-$\gamma$ tail was defined by the faintest ``power-law'' galaxies
(some of which were as faint as the faintest ACSVCS galaxies). This
result was challenged by Graham \& Guzman (2003) and is indeed at odds
with our findings: {\it it is the faintest, not the brightest galaxies
that have the shallower inner profiles}.

The left-side panel of Figure \ref{hist2} shows the $\gamma$
distribution for three samples: all bright galaxies best fit by a
core-S\'ersic profile, and all galaxies brighter and fainter than $B_T
= 13.7$ mag (corresponding to the dividing point between elliptical
and dwarf galaxies in the VCC). The high $\gamma$ peak is populated
almost exclusively by bright ($B_T < 13.7$ mag) S\'ersic galaxies,
although their distribution extends to values of $\gamma$ as low as
0.1, i.e., the range previously believed to be populated exclusively
by core galaxies. Galaxies best fit by a core-S\'ersic profile have
generally lower values of $\gamma$ ($\lesssim 0.4$) but their
distribution overlaps with that of the previous sample. The histogram
bin at the lowest $\gamma$ value is populated preferentially not by
the core galaxies, but by the faintest galaxies in the survey ($B_T >
13.7$).  We note that the slight peak at $\gamma \sim 0.5$ disappears,
producing a monotonic decrease in the distribution as $\gamma$
increases, if only ``regular'' galaxies (i.e., not showing evidence of
multiple morphological components - bars, stellar disk, rings) are
considered, as shown in the right panel of Figure \ref{hist2} (dashed
line). In other words, the $\gamma$ distribution for the population of
E+dE galaxies is skewed, but most certainly unimodal.

The above results indicate that there is no justification, based on
the shape of the $\gamma$ histogram, for dividing early-type galaxies
into ``core'' and ``power-law'' types as defined by Lauer et
al.\ (1995). Such a distinction was born from the observation of a
clearly bimodal distribution of inner slopes. Our study shows instead
that the distribution of the slopes of the inner profiles for E+dE
galaxies  is unimodal although strongly skewed towards small
$\gamma$. The addition of galaxies with multiple morphological
components (mainly disks) enhances the high-$\gamma$ end of the
histogram, but not enough to produce a clearly bimodal distribution.

In view of the above discussion, the definition of ``core'' galaxies
as in Lauer et al.\ (1995) appears untenable: according to this
definition, most of the ACSVCS dwarfs (e.g., VCC 1886, VCC 1993, VCC 21,
etc.) should be classified as core galaxies. Lauer et al.\ claim that
cores are the result of the evolution of binary black holes, a clear
impossibility for the dwarf galaxies -- insisting on such definition
would therefore demand different physical explanations for the
formation of ``cores'' in dwarfs and bright ellipticals. As stressed
by Graham et al.\ (2003) and several papers by the same group, adoption
of a S\'ersic instead of a Nuker model to characterize the surface
brightness profiles (a choice justifiable on the simple ground that
certainly on large scales, and often on small scales, profiles are not
well approximated by power-laws) allows one to interpret the
flattening of the inner profiles in faint systems as a natural
extension of the trend seen in brighter counterparts, a trend which is
broken only by the very brightest ellipticals with $M_B \lesssim
-20.5$. A revision along the lines proposed by Graham et al.\ (2003),
who define a galaxy as having a ``core'' if the surface brightness 
in the inner region is lower than the inward extrapolation of the
S\'ersic profile best fitting the outer parts, is more consistent with
our findings. With this definition, the brightest $\sim10$ galaxies
(those best fit by a core-S\'ersic model and which, ironically, are the
only ones to have power-law inner profiles) are ``core'' galaxies,
while the faintest galaxies (which follow a S\'ersic profile all the
way to the center) are not, in spite of having shallower inner
profiles.

\subsection{Core Galaxies, Mass Deficit and Supermassive Black Holes}

The brightest $\sim 10$ galaxies are clearly distinct from the rest of
the sample insofar as the properties of the inner 300-400 pc region
are concerned. These galaxies lack a distinct stellar nucleus (see
also Paper VIII), and have central surface brightness and inner
profile slopes which are lower than expected based on the
extrapolation of scaling relations defined by fainter systems (upper
and lower left panels of Figure \ref{fcorr2}). Although this is in
agreement with previous studies (e.g., Graham \& Guzman 2003; Jerjen
\& Binggeli 1997), such studies also found that if the inner few
hundred parsecs are neglected, there is a continuity in the structural
properties of bright and faint ellipticals, likely suggesting a common
formation scenario. This statement is not entirely supported by the
ACSVCS data; i.e., we find that core galaxies can be distinguished
from the rest of the sample also on the basis of some large-scale
structural properties. In particular, although the S\'ersic index $n$
for core galaxies lies along the relation established by the rest of
the sample (lower right panel of Figure \ref{fcorr2}, see also Graham
et al.\ 2003; Graham \& Guzman 2003; Jerjen \& Binggeli 1997),  the
effective radii $r_e$  are significantly larger than expected based on
the trend displayed by fainter systems (upper right panel of Figure
\ref{fcorr2} and panel $dd$ of Figure \ref{all1}, {\it cf.} Figure 7
of Binggeli, Sandage \& Tarenghi 1984). The significance of these
results will be addressed in a forthcoming paper, in which the ACSVCS
sample will be augmented with a similarly defined sample of early-type
galaxies in the Fornax cluster. Based on the present study, however,
our interim conclusion is that {\it the distinguishing features of core
galaxies are not relegated exclusively to the innermost regions, but
extend to include the overall galactic structure.}

Finally, we will comment briefly on the connection between cores and
supermassive black holes. The two have long been linked together
(Lauer et al.\ 1995; Faber et al.\ 1997; Rest et al.\ 2001).  N-body
simulations by Milosavljevi\'c \& Merritt (2001) followed the
evolution of the SBHs binary which forms as a consequence of the
merger of two galaxies with inner power-law profiles. They find that
the binary decays through the gravitational slingshot of stars in the
inner region: in a gas-free environment the process can eject an
amount of mass comparable to the combined mass of the two SBHs, with
the result that the merger product will form a core. Estimates of the
``missing mass'' in core galaxies have been carried out by
Milosavljevi\'c et al.\ (2002), Ravindranath et al.\ (2002) and Graham
(2004). We follow Graham (2004) by defining the mass deficit within
the break radius as the difference between the mass contained within
the extrapolation of the S\'ersic model which best fits the profiles
at larger radii, and the mass enclosed by the power-law which
corresponds to the best fitting core-S\'ersic model to the actual
profile. Figure \ref{deficit} shows the result of this exercise; SBH
masses have been calculated from the $M_{SBH} - \sigma$ relation of
Ferrarese \& Ford (2005) using published values of $\sigma$, the
stellar velocity dispersion\footnotemark.\footnotetext{References can
be found in the HyperLeda database, {\sf
http://leda.univ-lyon1.fr/hypercat}, Prugniel et al.\ (1997).} Mass
deficits have been calculated as in Graham et al.\ (2004), assuming
that the $R-$band mass to light ratio $M/L$ scales with luminosity as
in van der Marel (1991), and the $R$ and $z-$band magnitudes and
colors for the Sun and giant ellipticals are (Fukugita et al.\ 1995):
$M_{R,\odot} = 4.28$, $M_{z,\odot} = 4.51$, $(R-z)_{gal} =
0.54$\footnotemark.  \footnotetext{Note that $R$ magnitudes are
defined in the Vega photometric system, while $z$ magnitudes are
defined in the AB system.} Excluding VCC 798, for which  the S\'ersic
index and mass deficit might be biased by the presence of a
large-scale stellar disk, the mean ratio between mass deficit and SBH
mass for our sample is $2.4 \pm 1.8$, in agreement with Graham et al.\
(2004, $2.1 \pm 1.1$), implying that the galaxies have undergone no
more than a few major mergers since the last event which involved gas
dissipation.

\subsection{One Sequence of Early-Types, from Dwarfs to Giants?}

The question of how ``regular'' and ``dwarf'' ellipticals are related,
if indeed they are, has been discussed in the literature since the mid
1980s. That a distinction might exist was originally prompted by the
observation that dwarf, low luminosity ($M_B \gtrsim - 18$ mag,
Binggeli et al. 1985) ellipticals are not well described by the same
de Vaucouleurs $R^{1/4}$ law as brighter ellipticals, their profiles
being more closely exponential (Binggeli, Sandage \& Tarenghi 1984;
Lin \& Faber 1983). A structural dichotomy between regular and dwarf
ellipticals was further stressed by Kormendy (1985), who claimed a
markedly different correlation between core radius, central surface
brightness and total magnitude for the two classes. In particular,
according to the latter study, while the central surface brightness of
dwarf ellipticals becomes brighter as the total luminosity increases,
in bright ellipticals the two quantities appear anti-correlated,
prompting the conclusion that ``dwarf elliptical galaxies are very
different from the sequence of giant ellipticals. [...] dwarf
spheroidals and ordinary ellipticals were formed in very different
ways.'' Indeed, the term ``spheroidal'' (rather than elliptical), when
referring to elliptical galaxies fainter than $M_B \sim -18$ mag was
introduced partly to reinforce the physical distinction between the
two classes (e.g., Kormendy 1985; Djorgovski 1992). More moderate
views were taken, for instance, by Caldwell (1983), Binggeli, Sandage
\& Tarenghi (1984) and Binggeli (1994). In particular, a parametric
representation of the surface brightness profile carried out in the
first two contributions, led to a continuous sequence extending from
$M_B \sim -22$ to $M_B \sim -12$, thus uniting ``regular'' ellipticals
and ``dwarf'' systems under one single $M_B - \mu_0$ scaling
relation. Binggeli (1994) cautioned that the striking dichotomy
observed by Kormendy (1985) could partly be due to the lack, in
Kormendy's sample, of galaxies in the $-20 < M_B < -17$ range,
corresponding precisely to the transition region between the two
families.

In the past ten years, the view that dwarf ellipticals are,
morphologically, the extension of brighter systems has gained further
support, and the question has shifted instead to whether the very
brightest ellipticals (the core galaxies discussed in \S 5.1 and
\S5.2, with $M_B \lesssim -20.5$ mag) are distinguished from the rest
of the elliptical population. For instance, Young \& Currie (1994)
found that when the surface brightness profile of dwarf ellipticals is
characterized as a S\'ersic model, the S\'ersic index $n$ strongly
correlates with the total luminosity: the profile evolves from
exponential ($n \sim 1$) to an almost de Vaucouleurs profile ($n \sim
4$) as $M_B$ decreases (i.e., the luminosity increases). Jerjen \&
Binggeli (1997) showed that the trend extends to bright ellipticals,
translating into a well defined correlation between $n$ and $M_B$ in
the range $M_B \sim -22$ to $M_B \sim -14$. The ensuing correlation
between $M_B$ and $\mu_0$ (with $\mu_0$ increasing with luminosity) is
violated only by galaxies brighter than $M_B \sim -20.5$ mag (what we
defined  in \S 5.2 as core galaxies), but no break is seen
corresponding to the ``canonical'' $M_B = -18$ magnitude which
supposedly divides ellipticals from dwarfs. Jerjen \& Binggeli
conclude that ``the diffuse, low-surface brightness dwarf galaxies are
the true low-luminosity extension of the classical giant
ellipticals''. In the most recent contribution to the subject, Graham
et al.\ (2003) and Graham \& Guzman (2003) confirm that, from a
structural point of view, 1) there is no dichotomy in the $M_B -
\mu_0$ scaling relations at $M_B \sim -18$, indeed there is nothing
significant altogether associated with this magnitude; 2) dwarf
ellipticals are the low-luminosity extension of, and likely share the
same formation mechanisms as bright ellipticals; and 3) the ``break''
in the $M_B - \mu_0$ correlation observed at $M_B \sim -20.5$ mag is
the result of the core evolution undergone by the brighter systems
after formation.

These studies have established that the structural dichotomy  between
dwarf and regular ellipticals as advocated by Kormendy (1985) was
likely the result of observational biases.  Nevertheless, different
clues indicate a high degree of complexity within the population of
dwarf galaxies. Conselice et al.\ (2001) and Drinkwater et al.\ (2001)
found that, in dense environments, the spatial and kinematical
distribution of dwarf ellipticals is broader than that of giant
ellipticals, and more closely resembles that of spirals. Dynamically,
dwarfs do not form a homogeneous family, with at least some systems
being rotationally supported (Mateo 1998; Pedraz et al.\ 2002; Geha et
al.\ 2003; van Zee, Skillman \& Haynes 2004). Finally, the stellar age
and metallicity differ significantly from dwarf to dwarf (Poggianti et
al.\ 2001), suggesting that not all dwarfs share a common origin and
evolutionary history. There is some support in  our data for dwarfs
forming a complex and diverse population. In particular:

\begin{enumerate}

\item{The fainter galaxies ($-15 \lesssim M_B \lesssim -17$) display
$-$ more so than their bright counterparts $-$  a broad variety of
surface brightness profiles, from  purely exponential (e.g., VCC 1948,
$M_B = -16.0$) to almost  $R^{1/4}$ law (e.g., VCC 1993, $M_B =
-15.8$).}

\item{Once the nuclear component has been removed, the central surface
brightness estimated for the faintest galaxies spans a broad range (a
factor $\sim 200$ at a fixed galaxy magnitude, Figure \ref{fcorr2}),
more so than for the brighter systems.}

\item{Dwarf galaxies with $M_B > -17.5$ comprise both nucleated and
non-nucleated varieties (Figure \ref{fcorr2}). The brightest galaxies
($M_B < -20$ mag) do not show any evidence of stellar nuclei. Some of
the galaxies with intermediate luminosity are nucleated, although they
all are consistent with a nucleus being present, but falling below the
level of detectability due to the high central surface brightness of
these galaxies (Paper VIII). The dwarf class is therefore
heterogeneous insofar as the presence of a nuclear component. Brighter
galaxies might not be (they could all be nucleated), although the very
brightest certainly are not (none has a stellar nucleus).}

\item{Morphologically, dwarf ellipticals form a rather heterogeneous
 class, with some systems showing evidence of spiral structures, disks
 and recent star formation.}

\item{Although dwarf and giant ellipticals show comparable amount of
 scatter in a color-magnitude diagram (\S 5.5), Mei et al.\ (2005)
 noted that dwarf ellipticals show larger scatter in the $z-$band
 Surface Brightness Fluctuation (SBF) magnitudes $\bar{m}_{850}$. This may
 indicate a larger spread in ages/metallicity for the dwarfs compared
 to bright ellipticals, in agreement with Poggianti et al.\ (2001).}

\end{enumerate}

The origin of dwarf galaxies is still highly controversial, and a
detailed review is beyond the scope of this paper. Briefly, dwarfs
could have formed early during the life of a protocluster (e.g., White
\& Frenk 1991; Dekel \& Silk 1986); they could be the result of
passive evolution of compact blue galaxies (e.g., Babul \& Rees 1992);
or they could be the remnants of stripped dwarf spirals which are
accreted onto a cluster (e.g., Moore, Lake \& Katz 1998; Conselice et
al.\ 2001; Mastropietro et al.\ 2005). Here, we will comment briefly on
the latter scenario. Figure ~\ref{shape} shows the S\'ersic index $n$
plotted against $g-$band central surface brightness and total $B-$band
magnitude for the ACSVCS galaxies and the sample of spiral bulges
analyzed by Graham (2001). The latter study presented S\'ersic fits to
the bulge components of 86 undisturbed face-on spiral galaxies,
originally observed by de Jong \& van der Kruit (1994). Low
luminosity ellipticals and bulges are clearly distinct: at a given
profile shape ($n$), bulges are brighter than ellipticals by at least
two $B$-band magnitudes, and have brighter central surface
brightness. A least square fit to the bulge population shown in
Figure \ref{shape} gives:

\be
\log n(z) = -(0.073\pm0.01)(M_B+18) + (0.021\pm0.03)
\ee

\be
\log n(z) = -(0.10\pm0.01)(\mu_0-18) + (0.052\pm0.022)
\ee

\vskip .1in
For comparison, the equivalent equations for early-type galaxies
best described by S\'ersic profiles are as given by equation 25 and:

\be
\log n(z) = -(0.070\pm0.066)(\mu_0-18) + (0.35\pm0.01)
\ee

\vskip .1in
Figure \ref{shape} can provide some observational constraints on
galaxy harassment theories, if indeed dwarf ellipticals are the result
of tidal stripping of spirals accreted onto clusters (e.g., Conselice
et al.\ 2001 and references therein). For instance, Moore et al.\ (1998)
predict that the surface brightness profile of harassed spirals should
steepen (their Figure 4), i.e., the harassed bulge should move up
(larger $n$) and to the right (brighter $\mu_0$) in the left panel of
Figure \ref{shape}, i.e., towards the general area occupied by the ACSVCS
galaxies (this argument assumes, of course, that present day bulges
are representative of bulges at $z \sim 0.4$, i.e., at the epoch when
stripping has yet to take place). An additional possibility to
reconcile the early-type and bulge populations in Figure \ref{shape}
is to invoke merging: unequal mass mergers would only slightly impact
the magnitude of the product, but could have profound effects on its
density profile. For instance, N-body simulations by Eliche-Moral et
al.\ (2005) show that unequal mass mergers of spiral galaxies would
lead to a merger product with structural parameters which might
resemble those of the dwarfs in the ACSVCS sample. The arrow in the
right panel of Figure \ref{shape} shows where the exponential ($n =1$)
bulge of a spiral would ``move'' after merging with a satellite of
half its mass (and, we assume, luminosity), according to these
simulations. Dissipation during merging would also move bulges in the
same general direction: by steepening the density profile it would
lead to an increase in luminosity, $n$, and central density.

\subsection{Isophotal Shapes}

It has long been known that the shapes of ellipticals show good
correlations with a variety of other properties, such as optical
luminosity (Nieto \& Bender 1989; Nieto et al.\ 1991), optical central
surface brightness (Nieto et al.\ 1991), X-ray and radio fluxes
(Bender et al.\ 1987, 1989), and stellar kinematics (Bender 1988). The
extent of these correlations, which apply to the core properties as
well (Ferrarese et al.\ 1994; van den Bosch et al.\ 1994; Lauer et
al.\ 1995; Rest et al.\ 2001; Lauer et al.\ 2005), has prompted the
use of the term ``dichotomy'', with optically bright ellipticals being
preferentially round, boxy, pressure supported and often showing
nuclear activity, while fainter systems are more elongated, disky,
rotationally supported and show little or no radio and X-ray emission.

Figure \ref{corr4} shows the intensity weighted mean B4 coefficient
and ellipticity for the ACSVCS galaxies, calculated in two radial
ranges: $r < 0.1r_e$ (left panels) and $0.1r_e < r < r_e$ (right
panels). While on large scales all galaxies are consistent with having
regular isophotes ($B4 \sim 0$), in the innermost regions there is a
trend for $B4$ to increase as galaxies get fainter. In agreement with
previous works, this trend is a result of the fact that the core
galaxies (i.e., the brightest galaxies in the sample) are
preferentially boxy, while galaxies immediately following in the
luminosity function often display stellar disks, producing elongated
isophotes. Dwarf systems have isophotes which are consistent, within
the errors, with being elliptical. With the exception that no high
ellipticity objects are found in core galaxies, there is, however, no
clear trend of ellipticity with luminosity: galaxies with $M_B > -20$
span the entire ellipticity range, regardless of luminosity.

The issue of the intrinsic shape of elliptical galaxies has been
addressed in many works (Sandage, Freeman \& Stokes 1970; Lambas,
Maddox \& Loveday 1992; van den Bergh 1994; Rix \& White 1990, 1992;
J\o rgensen \& Franx 1994; Michard 1994). Figure \ref{ell} shows the
distribution of minor to major axes ratios, $b/a$, for the entire
sample of ACSVCS galaxies, and for samples divided according to
magnitude, S\'ersic index $n$, and morphological classification.  The
thick solid line in the upper right panel of the Figure shows the
distribution expected for a population of oblate galaxies with
intrinsic axial ratio $b/a = 0.2$ $-$ selecting a different intrinsic
$b/a$ has the effect of truncating the distribution at the new $b/a$,
without changing the overall shape. The observed distribution is in
excellent qualitative agreement with the distribution derived by
Lambas, Maddox \& Loveday (1992) for a sample of over 2000 ellipticals
(upper panel of their Figure 2). Lambas et al.\ note that a pure
oblate or prolate population cannot account for the observed
distribution, and conclude that a population of mildly triaxial
galaxies (with major to intermediate axial ratio of 0.8) matches the
observations best.

Different axial ratio distributions are expected if galaxies belonging
to different sub-samples (giant and dwarf ellipticals, or galaxies
described by de Vaucouleurs and exponential profiles) have different
distributions of intrinsic shapes. Although larger samples would be
desirable, statistically different distributions are seen only when
the sample is divided between E+dE and S0+dS0 galaxies. In this case,
a Kolmogorov-Smirnov test shows that the ellipticity distributions
differ at the  1.5$\sigma$ level, with a deficiency of round S0+dS0
galaxies. This is a likely an observational bias due to the fact that
bulge-dominated face-on bulge+disk systems are not easily recognizable
(e.g. J\o rgensen \& Franx 1994). The ellipticity distribution for
samples divided between bright ($M_B < -18.5$) and faint ($M_B >
-17.5$) galaxies, E and dE and galaxies with nearly de Vaucouleurs
profiles ($n > 3$) and those with nearly exponential profiles ($n <
2$) are consistent well within the 1$\sigma$ level.

\subsection{The Color-Magnitude and Color-Color Relations of
 Early-Type Galaxies in Virgo} 

Because the colors and magnitudes of a stellar population are affected
by its age, metallicity, as well as its mass, these can be constrained
by a study of color-magnitude diagrams. The $(g-z) - g$ color
magnitude relation for the ACSVCS galaxies is shown in
Figure \ref{cmd1}. Magnitudes in the $g$ and $z-$bands (Table 4)  are
calculated by integrating to infinity the best fitting S\'ersic or
core-S\'ersic models (thus excluding a nuclear component, if present);
while $g-z$ colors (last column of Table 4) are calculated directly
from the images, by integrating the azimuthally averaged profiles
between 1\sec~(to avoid a nuclear component) and either one effective
radius, or the radius at which the surface brightness profile, in
either band, is less than one magnitude above the sky, whichever
radius is the smallest. The error associated to the $g-z$ color
accounts only for the error in the sky background values, added in
quadrature. The left panel of Figure \ref{cmd1} shows the data with
error-bars, while in the right panel are over-plotted single burst
stellar population models from Bruzual \& Charlot (2003), assuming a
Salpeter IMF and lower/upper mass cutoffs of 0.1/100\msun. The models
are calculated for metallicities between [Fe/H] $= -2.25$ (yellow)
and [Fe/H] = +0.56 (red), ages between 1 and 15 Gyr (shown as solid
dots along each iso-metallicity track, with older galaxies becoming
redder), and mass between $10^8$ and $10^{12}$ \msun~(with the least
massive models having fainter total magnitude). The well defined
relation seen in the Figure, with fainter galaxies becoming
progressively bluer than brighter ones, has been discussed widely in
the literature (e.g., Baum 1959; Bower, Lucey \& Ellis 1992; Blakeslee
et al.\ 2003; Bernardi et al.\ 2005; Hilker et al.\ 1999; Secker et
al.\ 1997). It is immediately apparent from Figure \ref{cmd1} that the
relation is non-linear: dwarf galaxies are bluer than expected based
on a linear extrapolation of the relation defined by the bright
systems. Error-weighted least square fits to galaxies brighter and
fainter that $M_g = -18$ give, respectively:

\be
(g-z) = (1.41 \pm 0.02) - (0.045 \pm 0.005) (M_g+18) 
\ee

\vskip .1in
\noindent and

\be
(g-z) = (1.30 \pm 0.03) - (0.070  \pm 0.020) (M_g+18)
\ee

\vskip .1in
These results do not change significantly if galaxies with multiple
morphological components (shown as blue points) are removed from the
sample. The two relations given in equations 31 and 32 are shown in
Figure \ref{cmd1} as dashed and dotted lines, respectively. A
satisfactory fit to the entire sample requires the use of a quadratic
relation:

$$
(g-z) = (1.373 \pm 0.003) - (0.086 \pm 0.003) (M_g+18)  -
$$

\be
- (0.0077 \pm 0.0011) (M_g+18)^2 
\ee

\vskip .1in

\noindent (see Jord\'an et al.\ 2005); this is shown as a solid line
in Figure \ref{cmd1}. A more subtle point is whether the intrinsic
scatter in the relation increases as one moves from brighter to
fainter systems, as one would expect if, for instance, dwarf galaxies
span a wider range in ages and/or metallicities than brighter
galaxies. The reduced $\chi^2$ of the quadratic fit given in equation
33 is $\chi^2_r = 3.7$, indicating that part of the observed scatter
is indeed intrinsic. However, the $\chi^2_r$ is virtually identical
whether the sample is restricted to faint ($M_g > -18$) or bright
($M_g < -18$) systems; in other words, the apparently larger scatter
shown by faint galaxies in Figure \ref{cmd1} is consistent with being
due to the larger measurement errors in the $g-z$ colors for these
systems. The same conclusion applies if $\chi^2_r$ is measured from
the linear fits given in equations 31 and 32.

Six galaxies stand out as being significant outliers in Figure
\ref{cmd1} and were excluded from the fits discussed above. In
particular, four galaxies  are located above the color distribution
shown in Figure \ref{cmd1}, i.e., they have significantly redder
colors than expected given their magnitude. In order of decreasing
luminosity, these galaxies are VCC 1327 (NGC 4486A), VCC 1297 (NGC
4486B), VCC 1192 (NGC 4467), and VCC 1199. VCC 1327 projects close to a
bright foreground star and hosts a nuclear dust disk; its color and
magnitude are more uncertain than is the case for other galaxies. In
addition, the galaxy lies just over 7\min~to the North-West of M87,
and might be tidally truncated. The other three galaxies are also
found in close proximity of brighter companions (see Appendix and \S
4.1.4), and were indeed flagged as being tidally stripped from
inspection of the images and brightness profiles. It is therefore
likely that for all four galaxies, the red colors are consistent with
the magnitude of the galaxy before tidal stripping took place. On the
blue side of the main relation there are two significant outliers, VCC
1499 ($g-z = 0.51$; $M_g =-16.16$), and VCC 798 (M85, $g-z = 1.38$;
$M_g =-22.03$). VCC 1499 is classified as a dIrr/dE transition object
from inspection of the images\footnotemark. VCC 798 likely hosts a
large-scale stellar disk, and is peculiar in having a significant
excess of diffuse star clusters, relative to other galaxies in the
sample (Peng et al. 2005b); it also displays strong
isophotal twists.

\footnotetext{Four more galaxies $-$ VCC 21, VCC 1488, VCC 1857 and
 VCC 1779 $-$ which also closely resemble VCC 1499 (low surface
 brightness, elongated system with a small gradient in the surface
 brightness profile) are also found to lie in a group at $g-z \sim
 0.85$, $M_g \sim -16.3$. }

The brightest core galaxies are well characterized as 12 Gyr old
stellar systems with baryonic mass $\sim 10^{12}$ \msun~and solar
metallicity ([Fe/H] $\sim$ +0.09). As expected, the faintest galaxies
are less massive, in the neighborhood of $10^9$ \msun; their bluer
colors can be interpreted as either (or both) a consequence of younger
age or lower metallicity. Figure \ref{cmd2} shows an optical-NIR
color-color diagram; although the scatter is large, most systems are
consistent with stellar ages between 8 and 15 Gyrs, and metallicities
in the range $-0.64 <$ [Fe/H] $< 0.09$, with metallicity being the
primary driver for the colors becoming bluer with decreasing galaxy
luminosity.  This conclusion is consistent with the observed lack of
redshift evolution of  the slope of the color-magnitude relation for
early-type galaxies  (Kodama \& Arimoto 1997; Kodama et al.\ 1998;
Blakeslee et al.\ 2003)

\vskip .2in
\noindent{\bf{Appendix A: Notes on Individual Galaxies}}

This appendix presents a collection of notes on individual galaxies.  Some
of the features discussed, especially for the brightest galaxies, had
already been known based on previous studies (see for instance, studies by
and references in Crane et al.\ 1993; Jaffe et al.\ 1994; Lauer et al.\ 1995,
2002, 2005; Verdoes Kleijn et al.\ 2005; Carollo et al.\ 1997; van den Bosch \&
Emsellem 1998). The purpose of this section is simply meant to highlight
features which are immediately apparent by visually inspecting the $g$, $z$ and
$g-z$ images, as well as the isophotal parameters.

\noindent {\bf 1. VCC 1226 (M49, NGC 4472)}. A thin, boomerang-shaped
dust lane crosses the galaxy's center at a position angle of roughly
--45$^{\circ}$,  extending $\approx$ 1\Sec5 on either side. Faint dust
patches are seen within 3\Sec5 of the center. There is no indication
of a nucleus from the dust-corrected images. The isophotes show no
significant deviations from ellipses.

\noindent {\bf 2. VCC 1316 (M87, NGC 4486)}. The inner $0\Sec25$ of
the $g-$band image is saturated. The optical synchrotron jet emanating
from the unresolved, non-thermal nucleus, is clearly visible and was
masked prior to the isophotal analysis.  Dust filaments radiate from
the center outwards, extending out to 13$\arcsec$ from the
nucleus. There are two very red objects close to the galaxy center:
one directly south of the nucleus, at a distance of $\approx$
6$\arcsec$, the other southeast of the nucleus at a distance of
11\farcs5. The former appears to be a  globular cluster, and the
latter a faint background galaxy. The isophotes appear very regular.

\noindent {\bf 3. VCC 1978 (M60, NGC 4649)}. This galaxy shows very
regular isophotes, with no evidence of a morphologically distinct
nuclear component.

\noindent {\bf 4. VCC 881 (M86, NGC 4406)}. NGC 4406 is a well known
example of a galaxy moving through a cluster at supersonic velocity
(e.g., White et al.\ 1991; Rangarajan et al.\ 1995; Stickel et al.\
2003; Elmegreen et al.\ 2000); on-going dust stripping is believed to
be taking place as a consequence. The ACSVCS residual images indeed
show several dust filaments radiating from the center, mainly in the
South-East quadrant, all the way to the edge of the chip.  The
isophotes are regular, a fact reflected in the constancy of the
ellipticity and position angle, and in the fact that all higher order
coefficients are consistent with zero. However, in the inner 0\Sec4,
the surface brightness profile decreases inward, a fact first noticed
by Carollo et al.\ (1997). At first sight, this behavior is consistent
with what is expected from a small dust disk, however, the color image
shows no color difference between this region and the surrounding
areas. The other explanation is that the stellar density actually
decreases within this region.  However, the ``depleted'' area appears,
in projection, quite round, rounder than the galaxy isophotes, and
might even be slightly mis-centered (extending more to the South-East).

\noindent {\bf 5. VCC 798 (M85, NGC 4382)}. The isophotes are very
boxy within 1\sec; the boxiness persists all the way to the center,
where there almost seems to be a ``butterfly" pattern,  extending
about 0\Sec3 on either side and aligned with the major axis of the
galaxy. The isophotes become distorted beyond $\sim 50$\sec. The
residual image shows the possible presence of two small dust patches
about 4\sec~South-West of the center. There is no evidence of a
nucleus.

\noindent {\bf 6. VCC 763 (M84, NGC 4374)}. Like M87 (=3C 374), M84
(=3C 272.1) is a well known radio source. Two parallel dust lanes, one
crossing (but not centered on) the galaxy's center, the other just
North of the first, run in the East-West direction. The larger
(northern) dust lane extends for approximately 14\sec, while the lane
crossing the center extends for about 5\sec. Both lanes are surrounded
by fainter dust wisps and filaments. The isophotes are very regular.

\noindent {\bf 7. VCC 731 (NGC 4365)}. The galaxy is a member of the
W-cloud, at a distance of $\sim 23$ Mpc (Mei et al.\ in preparation).
A resolved, slightly elongated nucleus is clearly visible in the color
image, as well as in the individual frames. The nucleus is bluer than
the surrounding galaxy and clearly affects the surface brightness
profiles, in particular in the $g-$band.  The isophotes appear quite
regular.

\noindent {\bf 8. VCC 1535 (NGC 4526)}. This galaxy hosts a large dust
disk,  very reminiscent of the one seen, at lower inclination, in VCC
1154. The disk extends 15\Sec6 on either side of the galaxy's center,
and is aligned with the major axis of the galaxy (in the
East-SouthEast direction). The plane of the disk makes an angle of
approximately 15\deg~to the line of sight, the Northern side being the
closest. There are several blue knots within the disk, probably
indicative of star formation.  Because the underlying stellar
population within the disk is likely different from that
characterizing the main body of the galaxy, the dust-correction for
this galaxy fails, and the isophotes can only be recovered outside a
semi-major axis of 7\sec. At a radius of approximately 10\sec~the
isophotes become visibly boxy; the boxiness is rather extreme at a
radius of 20\sec. Beyond, the isophotes assume a diamond shape
appearance (as a consequence the A4 coefficient becomes non-zero), and
finally become disky at a radius of approximately 70\sec.

\noindent {\bf 9. VCC 1903 (M59, NGC 4621)}. The $g-$band images are
saturated within the inner 0\Sec25.  Based on the $z-$band image alone,
it is hard to judge the presence of a morphologically distinct nucleus;
however, an edge-on, thin, faint stellar disk,
extending at least 10\sec~from the center is clearly present. 
The presence of the disk is reflected in the B4 coefficient.

\noindent {\bf 10. VCC 1632 (M89, NGC 4552)}. Two thin dust filaments
radiate from the center in the  West-NorthWest direction, out to a
distance of about 6\sec. Small dust patches are visible within
11\sec. A small dust patch, less than 1\sec~across, surrounds (but it
is not centered on) the center of the galaxy. There is a very red
object to the South-West of the center, at a distance of about
10\sec. The isophotes are very regular, there is no indication of a
nucleus.

\noindent {\bf 11. VCC 1231 (NGC 4473)}. Rather elongated system; the
color image shows no evidence of dust, however, there is a small (less
than 1\sec) thin rectangular feature crossing the center of the
galaxy. This feature is aligned with the major axis and bluer than the
surrounding regions; it is probably responsible for the negative B4
coefficient in the very innermost region. Further out, the galaxy
becomes disky, and remains so out to about 10\sec. There doesn't
appear to be a morphologically distinct nucleus.

\noindent {\bf 12. VCC 2095 (NGC 4762)}. This galaxy has a very thin,
edge-on disk, extending all the way to the edge of the CCDs. Within
the inner 15\sec~ on either side of the nucleus, the disk appears to
become thinner and fainter as the radius decreases; it reaches maximum
surface brightness between 15\sec~and 40\sec~from the center, and it
is bluer than the surrounding galaxy.

\noindent {\bf 13. VCC 1154 (NGC 4459)}. This galaxy harbors a
spectacular dust disk, similar to the one seen, at higher inclination,
in VCC 1535. The disk is about 17\sec~across, and its plane is
inclined by $\sim$ 45\deg~to the line of sight. There are several blue
clumps in the disk, probably indicative of star formation. Several
wisps and filaments extend beyond the edge of the disk. As for the
case of VCC 1535, the correction for dust fails,  and the profile can
only be recovered beyond 7\Sec5 from the center. Unlike the case of
VCC 1535, the isophotes look regular.

\noindent {\bf 14. VCC 1062 (NGC 4442)}. This galaxy appears quite
regular within the inner 15\sec. Beyond, the isophotes become boxy,
and a faint elongated outer envelope dominates the galaxy's
appearance. This is slightly tilted (towards South) relative to the
inner regions, causing the isophotes to twist slightly. The images
show no strong evidence of a morphologically distinct nucleus.

\noindent {\bf 15. VCC 2092 (NGC 4754)}. This galaxy has two
morphologically distinct components: the main body, which appears
quite regular, and a misaligned (by about 45\deg) bar, extending out
to 25\sec~from the center. A faint nucleus is likely present.

\noindent {\bf 16. VCC 369 (NGC 4267)}. This galaxy is quite round in
the innermost  6\sec; beyond it becomes more elongated, with perhaps a
slight twist in the major axis position angle, to revert to round
again beyond 50\sec. It is possible that the change in the isophotal
shape is due to a stellar bar similar (but much fainter) to the one
seen in VCC 2092. A faint nucleus is likely present.

\noindent {\bf 17. VCC 759 (NGC 4371)}. This galaxy has a small dust
disk at the center, only about 1\Sec6 across. The disk is inclined by
about 26\deg~to the line of sight, and its major axis seems to
coincide with the major axis of the galaxy (which is aligned in the
East-West direction). The structure beyond is not easily
interpretable. There seems to be a thin (less than 1\sec~across) blue
nuclear stellar ring centered on the galaxy's center, with a radius
just below 11\sec.  The inclination and orientation of the stellar
ring are the same as those of the nuclear dust disk. Within this ring,
there seems to be at least one additional blue arc. Just beyond the
stellar ring, there seems to be a second, this time redder, stellar
ring, about 6\sec~ across. Farther out, at a distance of about
34\sec, a ring rests at the end of a kpc-scale bar. The presence of
the rings makes the A4 and B4 coefficients non zero. The position
angle, ellipticity and the surface brightness profile are also
irregular. There doesn't appear to be a stellar nucleus, although the
presence of the dust makes this statement hard to quantify.

\noindent {\bf 18. VCC 1692 (NGC 4570)}. This galaxy shows a thin,
circum-nuclear stellar ring, centered on the nucleus (which appears
resolved), extending just less than 2\sec~ on either side and aligned
with the main body of the galaxy. Although almost edge-on, the
structure indeed appears to be a ring rather than a disk; an inner gap
is apparent from the images.  The ring is bluer than the surrounding
galaxy. The galaxy is very elongated, with a thin, edge-on disk which
appears to become slightly thicker towards the center, where it joins
the main bulge.

\noindent {\bf 19. VCC 1030 (NGC 4435)}. There is a small dust disk,
about 8\sec~across, centered on the galaxy's center. The disk is
inclined $\sim$ 16\deg~to the line of sight. Although similar, this
disk is smaller and less structured than those in VCC 1154 and VCC
1532. Several regions within the disk are quite blue, probably due to
star formation. This prevents an accurate dust correction, and the
isophotes are not recovered within the inner 2\sec. Because of the
dust, it is also impossible to assess the presence of  a nucleus. The
isophotes are visibly boxy in the inner $\sim$10\sec, switching to
disky further out.

\noindent {\bf 20. VCC 2000 (NGC 4660)}. The center is saturated in
the $g-$band image. The galaxy is very elongated, but becomes boxy at
a distance of about 6\sec~ (along the major axis) from the center. At
the corresponding distance along the minor axis (4\Sec5), to the
South, there is a small, faint, very blue structure, resembling a
nucleus with two spiral arms. The ``arm" facing the nucleus of VCC
2000 is more extended and looser than the arm facing away. The entire
structure is about 2\Sec5 across.

\noindent {\bf 21. VCC 685 (NGC 4350)}. This is a very elongated system, with a
large-scale edge-on stellar disk typical of S0 galaxies. The presence
of the disk affects the B4 coefficient, which becomes positive beyond
a few arcsec. There is a nuclear, nearly edge-on dust disk, about
4\sec~in diameter. The dust disk is surrounded by a blue stellar disk,
extending perhaps 1\Sec5 beyond the edge of the dust disk. Dust,
stellar, and outer disks are all aligned. In correcting for dust
absorption, the color of the region underlying the dust disk has been
assumed equal to the color of the surrounding stellar disk, although
the correction is still not entirely successful. There is no obvious
evidence for the presence of a nucleus from the color images or the
surface brightness profiles (although this is not a secure conclusion,
because of the presence of the dust). The vast majority of globular
clusters are found in the equatorial plane of the galaxy, and are
therefore likely associated with the large-scale disk.

\noindent {\bf 22. VCC 1664 (NGC 4564)}. There appears to be a faint
nucleus in the color image, although its presence is not obvious in
the individual images. The galaxy becomes quite disky beyond 10\sec.

\noindent {\bf 23. VCC 654 (NGC 4340)}. The galaxy is quite round
within the inner 1\Sec7, beyond which it becomes stretched in the
NortEast-SouthWest direction, due to the presence of an inner bar
surrounded by a nuclear stellar ring.  Beyond 6\sec, the galaxy is
elongated in the East-West direction. The very innermost region
(within 0\Sec5) is also elongated, in the same direction as the outer
parts. From the color image, the innermost structure appears bluer
than the main body of the galaxy.  An outer stellar ring with
semi-major axis of about 56\sec~bounds an outer bar; a small irregular
galaxy lies along this ring about 40\sec~to the North-East of the
center; it is possible that this is a disrupted satellite.

\noindent {\bf 24. VCC 944 (NGC 4417)}. This galaxy has a thin edge-on
blue stellar disk, aligned with the main body. There appears to be a
faint nucleus, both from the color image and from the $g-$band surface
brightness profile.

\noindent {\bf 25. VCC 1938 (NGC 4638)}. The galaxy is visibly disky
beyond a few arcsec. A thin,  edge-on stellar disk seems to be
embedded in the outer, thick disk of the galaxy. The latter does not
extend all the way in, stopping at a distance of about 4\Sec5 from the
center. There is a faint indication of two tightly wound dust spiral
arms extending from the the disk, at a distance of 13\sec~from the
center, although there is no feature associated with these structures
in the color image. The latter do show, however, an inner, red disk
about 12\sec~in radius. There is no obvious dust absorption associated
with this disk in the individual images, and therefore this is
probably a stellar feature. The color images also indicate the
presence of a compact nucleus. In the isophotal analysis, the center
seems to wonder, and was therefore held fixed. A fainter galaxy
(possibly a nucleated dwarf elliptical) is seen about 100\sec~to the
east, while a very red object is present about 17\sec~East of the
nucleus.

\noindent {\bf 26. VCC 1279 (NGC 4478)}. This galaxy is generally well
behaved on large scales. There is no hint of the existence of a
large-scale disk, but the isophotes become more elongated in the very
innermost region. The color image betrays the existence of a small
(less than 1\sec~in radius) blue stellar disk, which is seen almost
edge-on and oriented along the major axis of the galaxy. The presence
of a disk is also reflected in the B4 coefficient. From the surface
brightness profile, it is possible that a compact nucleus exists,
affecting the profile within the inner 1\sec or so, although the
feature seen could also be attributable to the disk.

\noindent {\bf 27. VCC 1720 (NGC 4578)}. Well behaved  elliptical,
with regular isophotes. Both the color image and the surface
brightness profiles show evidence of a stellar nucleus.

\noindent {\bf 28. VCC 355 (NGC 4262)}. Two dust filaments radiate
away from the center in a spiral pattern, reaching out to a distance
of about 4\sec. Other than that, the galaxy appears regular within
this region. Beyond, the isophotes twist (from a position angle of
about -30\deg~to approximately +20\deg), and the isophotes become
quite disky. Inspection of the images reveal the presence of a bar,
aligned at a position angle of +10\deg and extending about 27\sec~from
side to side. The bar appears slightly misaligned with respect to the
main body of the galaxy at the same radius. The end of the bar has a
``flattened" appearance, resembling a hammer, although there is no
obvious indication of a ring spiral arms protruding from it. There is
no obvious color feature associated with the bar. There might be a
hint of a compact stellar nucleus from the surface brightness profiles.

\noindent {\bf 29. VCC 1619 (NGC 4550)}. This galaxy is intermediate,
as far as dust morphology is concerned, between the ``dust filaments"
types (e.g., M87 and NGC 4472) and the ``large-scale dust disk" types
(e.g., VCC 1154). Dust filaments radiate outwards from the center, but
they have larger optical depth compared to the ones seen in M87 or VCC
355, for instance. Although the filaments form a disk structure, this
is not quite as organized as the disks seen in VCC 1154 or VCC
1532. The main dust filaments/lanes are confined within a 10\sec~from
the center, but dust patches can be seen further out, up to a distance
of 25\sec. The dust correction performs well, indicating that the
galaxy's color within the dust region is similar to that of the
neighboring areas. The surface brightness profile indicates the
presence of a large nucleus.

\noindent {\bf 30. VCC 1883 (NGC 4612)}. This galaxy resembles VCC
355, in that is shows a dramatic isophotal twist. In this case,
however, there is no evidence of dust. The presence of a bar is not
quite as obvious as it is in the case of VCC 355, but the galaxy
isophotes become ``stretched" at a distance of about 17\sec~from the
center, as would be expected if a boxy, elongated component were
superimposed to the main body of the galaxy.  The ``ends" of the bar
are visible, albeit barely, in the individual images.  The bar is
misaligned with the main body of the galaxy, producing a 40\deg~change
in the position angle of the major axis. A stellar nucleus causes an
inflection in the innermost 0\Sec2~of the surface brightness profile.

\noindent {\bf 31. VCC 1242 (NGC 4474)}. This galaxy resembles VCC 685
but for the presence of dust. Very elongated system, with a
large-scale edge-on stellar disk typical of S0 galaxies. The presence
of the disk affects the B4 coefficient, which becomes positive beyond
a few arcsec, and the galaxy appears bluer in the equatorial plane. A
nucleus is visible in both of the individual images, and is reflected
in the shape of the surface brightness profiles in the inner
0\Sec3. The galaxy has a significantly larger Globular Cluster (GC) 
system compared to galaxies in the same magnitude range.

\noindent {\bf 32. VCC 784 (NGC 4379)}. The isophotes become slightly
disky at a distance of approximately 10\sec~from the center, as shown
by the B4 coefficient and ellipticity change between 10 and 20\sec.
Unlike the case of VCC 1883 and VCC 355, there is no convincing
evidence that a bar might be present, although the fact that the
position angle changes (just as it does for VCC 1883 and VCC 355 where
the bar sets in) is suggestive. The images and surface brightness
profile are clearly affected by the presence of a nucleus in the inner
0\Sec2.

\noindent {\bf 33. VCC 1537 (NGC 4528)}. This galaxy is in the same
class as VCC 1883 and VCC 355 (and perhaps VCC 784 and VCC 778), but
shows even more extreme properties.  The bar is very prominent, and is
misaligned relative to the major axis of the main body of the galaxy
by about of 60\deg, with the bar almost aligned in the East-West
direction. Ellipticity, position angle, A4 and B4 coefficients change
abruptly where the bar sets in. The surface brightness in the main
body of the galaxy changes abruptly at the radius corresponding to the
edge of the bar; being fainter within the region swept by the bar than
outside. There is no clear evidence for a nucleus, although the surface
brightness profile is quite steep.

\noindent {\bf 34. VCC 778 (NGC 4377)}. There are three small spiral
galaxies at approximately 17\sec, 19\sec~and 31\sec~from the center of
VCC 788; the largest galaxy is approximately 11\sec~across. The three
galaxies are aligned approximately in the East-West direction; there
is no obvious isophotal disturbance in VCC 778 at the radius where
these galaxies are, and therefore the three spirals are probably part
of a background group. All three galaxies were masked in fitting the
isophotal parameters. This galaxy shows the same features seen in VCC
784: at radii between approximately 3\Sec5 and 8\sec, the isophotes
twist, the ellipticity, the B4 and, especially, the A4 coefficients
jump, as due to the presence of a misaligned bar (the misalignment
here is about 20\deg), producing ``diamond" shaped isophotes. The bar
is not directly visible in the individual frames, but produces a
discontinuity in the surface brightness profile. 

\noindent {\bf 35. VCC 1321 (NGC 4489)}. This galaxy shows a dramatic
(almost 90\deg) isophotal twist which takes place abruptly at
approximately 5\sec~from the center. However, unlike the case of VCC
1537, the twist is not accompanied by a change in the higher order
coefficients, in fact the isophotes remain very regular all the way
out. There is therefore no evidence that this twist is due to a bar.

\noindent {\bf 36. VCC 828 (NGC 4387)}. The galaxy is visibly boxy
beyond 1\sec, but otherwise it doesn't show evidence of structure. A
nucleus is clearly visible from the individual frames and the surface
brightness profiles.

\noindent {\bf 37. VCC 1250 (NGC 4476)}. This galaxy harbors a
nuclear, clumpy, dust disk, about 11\sec~ in radius. The clumps are
clearly organized in a spiral structure. The plane of the disk is
inclined by about 20\deg~to the line of sight; the North-West side of
the disk is the closest.  This disk is similar, but clumpier and
smaller than the ones seen VCC 1154 or VCC 1532. The correction for
dust performs better than in the case of VCC 1154 or VCC 1532,
indicating that the change in the stellar population embedded in the
disk (relative to the one surrounding it), is not as extreme as in
those galaxies. Several blue clumps, organized in an almost complete
arc structure, are however clearly visible within the disk, and the
isophotal parameter within the inner 10\sec~should be interpreted with
caution. There is clear indication of the presence of a
nucleus from both the individual and the color image, and from the
surface brightness profiles. The color of the nucleus cannot be
assessed because of the presence of dust. The nucleus appears to be
offset by just over 1\sec~ in the North direction relative to the
center of the dust disk (the disk's major axis is aligned in the
North-NorthEast direction). The isophotal center is not very stable
even in the outer regions, varying by as much as 0\Sec1 between
subsequent isophotes; but it does seem to be consistent with the
location of the nucleus. The isophotes seem quite regular, at least in
the outer region.

\noindent {\bf 38. VCC 1630 (NGC 4551)}. From the individual images,
in the very innermost region, it is barely possible to make out the
presence of a faint stellar disk, possibly flaring out at the
edges. This is reflected in the abrupt change of the A4, B4
coefficients in the inner 0\Sec1. There is a clear resolved nucleus.

\noindent {\bf 39. VCC 1146 (NGC 4458)}. The innermost isophotes
appear quite disky, as reflected in the negative B4 coefficient in the
inner 0\Sec2. The ellipticity increases steadily towards the
center. The galaxies hosts a large, bright nucleus.

\noindent {\bf 40. VCC 1025 (NGC 4434)}. This galaxy is quite round,
although it becomes visibly more elongated in a narrow radial range
around $\sim$ 8\sec. There are no features seen in the color
image. There is a hint of the presence of a large nucleus from a
``break" in the surface brightness profiles around 1\sec.

\noindent {\bf 41. VCC 1303 (NGC 4483)}. This galaxy appears to have
two smooth spiral arms, first noticeable extending from the central
bulge at a distance of approximately 7\sec~from the galaxy's
center. The arms extend in the East-West direction at first (the major
axis of the galaxy is at a position angle of 60\deg), then bend
towards North in a clockwise fashion. The position angle, B4 and A4
coefficients all change outwards of 10\sec~because of the arms. There
are no features visible in the color image.

\noindent {\bf 42. VCC 1913 (NGC 4623)}. This galaxy is somewhat
similar to VCC 1630 in that it shows (more clearly in this case) the
presence of a thick, edge-on nuclear stellar disk, about 1\sec~in
radius. From the color image, the disk appears slightly bluer than the
host galaxy. A nucleus is quite apparent in the individual
images, and seems to affect the surface brightness profile in the
inner 0\Sec3. At larger scales, the galaxy appears to have a
``structure within the structure" morphology: within the inner 7\sec,
the light is dominated by a boxy bulge. Beyond, there appears to be a
disky (or at least highly elongated) ``halo" or disk. The transition
between the two structures is quite abrupt and can be seen as a
discontinuity in the surface brightness profile in the individual
images. All of the structures share the same major axis.

\noindent {\bf 43. VCC 1327 (NGC 4486A)}. The galaxy is tidally
truncated due to interactions with M87, which projects only 7\Min5
away. Small, faint galaxy with a bright foreground star nearby. An
edge-one dust disk is clearly visible.

\noindent {\bf 44. VCC 1125 (NGC 4452)}. This galaxy is an edge-on
S0. There seem to be four separate components: a resolved 
nucleus, an inner, thin stellar disk, confined within 2\sec~from the
center, a ``middle" thicker, stellar disk, extending out to distances
of about 17\sec~from the center, and an outer, thinner stellar disk,
which is highly flattened. An inner bulge seems to be almost an
extension of the middle disk. The smaller inner disk is bluer in color
than the outer disks; there is no appreciable color difference between
the middle and the outer disk and the bulge of the galaxy. The
presence of the nucleus is very clear from the surface brightness
profile.

\noindent {\bf 45. VCC 1475 (NGC 4515)}. The galaxy shows a 
nucleus, surrounded by an edge-on blue disk, which
flares out slightly at the end. The disk extends for a total of about
4\sec. The nucleus is visible in the surface brightness profile as
well.

\noindent {\bf 46. VCC 1178 (NGC 4464)}. A compact nucleus is
clearly visible in the color image, but barely discernible in the
surface brightness profiles. The isophotes appear regular.

\noindent {\bf 47. VCC 1283 (NGC 4479)}. A clear nucleus is
visible from the color image and the surface brightness profile. The
isophotes appear quite regular, although there seems to be a faint bar
extending out to about 30\sec.

\noindent {\bf 48. VCC 1261 (NGC 4482)}. This galaxy hosts a 
resolved nucleus. The isophotes appear regular. Several faint
background galaxies are seen within 50\sec.

\noindent {\bf 49. VCC 698 (NGC 4352)}. The galaxy is quite disky
within 1\sec, becoming boxy further out, and then disky again outside
10\sec.  The inner disk (which is blue) and a resolved nucleus
are visible in the color image; the nucleus is noticeable in the
surface brightness profile as well.

\noindent {\bf 50. VCC 1422}. This galaxy has a bright, resolved
nucleus, clearly dominating the surface brightness profile. The
isophotes appear regular. The color image shows possible evidence of a
large filamentary dust structure about 26\sec~South-West of the
nucleus, distributed azimuthally within a 90\deg~radius.

\noindent {\bf 51. VCC 2048}. This galaxy has a  bright nucleus,
dominating the surface brightness profile and clearly visible in the
color image. The nucleus is surrounded by a thin edge-on stellar disk,
as a consequence of which  the isophotes are very disky.

\noindent {\bf 52. VCC 1871}. Unlike other galaxies in the same
magnitude range, this is quite round, perhaps because of inclination
effects. The isophotes are regular. A bright resolved nucleus is
clearly visible in the surface brightness profiles and color image.

\noindent {\bf 53. VCC 9}. This galaxy has fainter surface brightness
than galaxies in the same magnitude range, and a flat core. There are
two bright clusters within 2\sec~from the center, but nothing at the
center of the isophotes. 

\noindent {\bf 54. VCC 575 (NGC 4318)}. This galaxy shows a clear
structure in the stellar population. At a distance of about 4\sec~from
the center, a stellar disk or wide ring sets in; the ring is brighter
than the region within it, causing a kink in the surface brightness
profile. It is possible that the ring is actually the result of a
spiral structure, although the spiral ``arms" do not appear to extend
all the way to the center of the galaxy. The ring might be similar to
the structure seen edge-on in other galaxies, for instance VCC 1938,
where an outer ``disk'' is present but does not extend all the way to
the center. Although there is no structure associated with the
ring/truncated disk in the color image, this does show a thin, edge-on
red disk extending approximately 2\Sec5 across the nucleus.

\noindent {\bf 55. VCC 1910}. The surface brightness profile is
dominated by the stellar nucleus, which is  brighter
than any globular cluster within the field. The isophotes are
regular. There is a very red globular cluster just over 2\sec~West of
the nucleus.

\noindent {\bf 56. VCC 1049}. From the brightness profile there
doesn't seem to be evidence of a separate nuclear component, although
the profile is quite steep and the galaxy becomes significantly bluer
towards the center. The isophotes are fairly regular. There are two
small, faint irregular/spiral galaxies, one 20\sec~to the South-East,
the other 19\sec~ to the North, probably both in the background.

\noindent {\bf 57. VCC 856}. This galaxy shows a clear face-on spiral
pattern. The arms are quite wide, smooth and regular. The center of
the galaxy has a bright, resolved nucleus. There is no evidence
of a bulge component.  This galaxy would be better classified as a
dwarf spiral.

\noindent {\bf 58. VCC 140}. A nucleus is clearly visible both in
the color image and in the surface brightness profiles. The isophotes
are regular. Several faint background galaxies are seen within the
field.

\noindent {\bf 59. VCC 1355}. The galaxy has a bright, resolved
nucleus. The isophotes are quite round and regular.

\noindent {\bf 60. VCC 1087}. The nucleus is marginally resolved, the
isophotes appear regular.

\noindent {\bf 61. VCC 1297 (NGC 4486B)}. Lauer et al.\ 1996 identified
a double nucleus, separated by 12 pc (0\Sec15), based on deconvolved
WFPC2 images of this galaxy. The two nuclei are not immediately
apparent in the ACS images, although the galaxy is quite disky within
the inner 100 pc, with this trend persisting all the way to the
center, and the profile is slightly asymmetric in the inner 40 pc.
Beyond about 10\sec, the surface brightness profile is clearly
truncated by interaction with M87, which projects only 7\Min3
away. 

\noindent {\bf 62. VCC 1861}. Like other galaxies in this magnitude
range (with the exception of VCC 1297) this galaxy has a bright, 
resolved nucleus superimposed to a flat core. The isophotes are quite
round and regular.

\noindent {\bf 63. VCC 543}. A resolved nucleus dominates the
surface brightness profile.  This galaxy is quite elongated, but the
isophotes are regular.

\noindent {\bf 64. VCC 1431}. Round galaxy with a resolved
nucleus. There are three red, faint irregular galaxies between 9\sec~
and 12\sec~to the North-East; probably background. The isophotes are
regular.

\noindent {\bf 65. VCC 1528}. This galaxy has a bright, very compact
nucleus. The isophotes  are regular.

\noindent {\bf 66. VCC 1695}. This galaxy has a bright
nucleus. The isophotes  are regular, there is no evidence of dust,
although the residual images show evidence of a large-scale spiral
structure. In the color image, there seems to be an elongated, large,
red, faint feature about 20\sec~to the East, although this is not
visible in the single images.

\noindent {\bf 67. VCC 1833}. Unlike galaxies in a similar magnitude
range, this galaxy does not appear to have a distinct nucleus. The
isophotes are regular.

\noindent {\bf 68. VCC 437}. This galaxy has a bright, resolved
nucleus. The isophotes are regular.  A background galaxy cluster is
visible in the frame.

\noindent {\bf 69. VCC 2019}. This galaxy has a bright nucleus, and
low surface brightness. Several bright clusters are seen, and a
small but resolved object is detected about 12\sec~to the North.

\noindent {\bf 70. VCC 33}. This galaxy has a bright nucleus
dominating the surface brightness profile.  The globular cluster
population is small. An extremely red object (not obviously extended)
is seen about 4\sec~East of the nucleus.

\noindent {\bf 71. VCC 200}. This galaxy has a bright nucleus
dominating the surface brightness profile.

\noindent {\bf 72. VCC 571}. The center of this galaxy is crossed by a
dust lane running in the North-West direction (approximately along the
major axis of the galaxy). The dust lane extends out to about
8\sec~(becoming wider and more diffuse with increasing distance from
the center) in the South-East direction, and just over 1\sec~in the
North-West direction. More wisps are seem further out, especially on
the North-West side. The correction for dust performs well. Several
blue star clusters are seen near the center (although not along the
dust lane itself), probably indicative of recent star formation.

\noindent {\bf 73. VCC 21}. Elongated system, may be better classified
as a dwarf irregular/dwarf elliptical transition object.  There are
several bright clusters near, but none exactly at, the center.

\noindent {\bf 74. VCC 1488}. Elongated system. There is a small
spiral galaxy, about 6\sec~across, 8\sec~to the South; this galaxy has
been masked in fitting the isophotes. An unconstrained fit to the
isophotes runs into problems because of the low gradient in the
surface brightness profile, but between 0\Sec5 and 5\sec~the isophotes
are centered very accurately on the nucleus. ELLIPSE was therefore run
over a more extended radial range by fixing the position of the center
to that of the nucleus.

\noindent {\bf 75. VCC 1779}. This galaxy might be best classified as
a dIrr/dE transition object. Dust filaments are seen radiating from
the center outwards within the inner 6\sec. From a visual inspection
of the image, there appears to be very few objects which could be
identified as GCs; three of these are near the center,  although none is at
the nuclear position. A small spiral galaxy is seen about 17\sec~to
the South-West.

\noindent {\bf 76. VCC 1895}. Highly elongated, faint elliptical. The
nucleus is clearly visible and may be slightly elongated in the same
direction as the major axis of the galaxy.

\noindent {\bf 77. VCC 1499}. This galaxy should be classified as a
dIrr/dE. There are several bright clusters and blue associations close
to the center, although there doesn't appear to be a proper nucleus,
and there is no convincing evidence of dust. The center of the
isophotes drifts to the East as the radius is increased.

\noindent {\bf 78. VCC 1545}. This galaxy has a bright, resolved
nucleus. The galaxy is fairly round, the isophotes are regular. The
galaxy has a fairly substantial population of bright globular clusters.

\noindent {\bf 79. VCC 1192 (NGC 4467)}. This galaxy is a close
companion of M49, which lies 4\Min2 away, outside the WFC frame to the
East; the outer surface brightness profile is tidally truncated, the
ellipticity is also affected. The surface brightness profile in the
center is brighter (by one to two magnitudes) than that of galaxies in
the same magnitude range. A resolved nucleus is clearly present
in the color images, as well as in the surface brightness
profiles. There is a small blue cluster within 0\Sec1 from the
nucleus, a second about 0\Sec9 to the South-East.

\noindent {\bf 80. VCC 1857}. Very low surface brightness, elongated
system. There are several faint clusters in the frame, and a very
bright one about 6\sec~ West of the center. The position of the center
was determined by block averaging and then contouring the images, and
was held fixed in fitting the isophotes.

\noindent {\bf 81. VCC 1075}. The isophotal center is consistent with
the location of a bright cluster, which is therefore identified as the
nucleus. The central position is held fixed at the nuclear location in
the fitting the isophotes.

\noindent {\bf 82. VCC 1948}. Elongated system. Although there are
some faint clusters around the center, none appears to be at the
isophotal center.

\noindent {\bf 83. VCC 1627}. This galaxy has brighter surface
brightness than galaxies in the same magnitude range. As a
consequence, the nucleus is not as prominent in the surface brightness
profile, although it is clearly visible in the color image. The galaxy
is quite round; there are no signs of disturbance in the isophotes. A
very red cluster is seen about 6\sec~West of the nucleus; otherwise
the galaxy has a small GC system compared to galaxies in the same
magnitude range.

\noindent {\bf 84. VCC 1440}. This galaxy has brighter surface
brightness than galaxies with comparable $B_T$, and a large number  of
bright globular clusters for a galaxy of this magnitude. A bright,
resolved nucleus is clearly present.

\noindent {\bf 85. VCC 230}. This galaxy has a bright
nucleus. The galaxy is quite round, and has what appears to be a
significant globular cluster population. The isophotes are regular.

\noindent {\bf 86. VCC 2050}. Very elongated system with a central,
faint nucleus. The galaxy appears regular. A very red cluster is
seen just over 7\sec~ South of the nucleus.

\noindent {\bf 87. VCC 1993}. The $g$ and $z-$band images show a faint
nucleus, which does not affect significantly the surface brightness
profile. From a visual inspection of the image, there are very few
objects which could be identified as globular clusters.

\noindent {\bf 88. VCC 751}. Fairly elongated galaxy with a bright
nucleus. The isophotes appear regular. There is a very red
background galaxy about 12\sec~East of the nucleus.

\noindent {\bf 89. VCC 1828}. The nucleus is quite prominent.

\noindent {\bf 90. VCC 538 (NGC 4309A)}. Round system, with a bright,
resolved nucleus. There is a bright spiral galaxy about 90\sec~to
the South.

\noindent {\bf 91. VCC 1407}. This galaxy has a bright nucleus;
 the galaxy has  regular isophotes. There seems to be a large
 population of globular clusters. A very blue and small (less than
 0\Sec2 across) elongated feature is seen about 1\Sec5 East of the
 nucleus.

\noindent {\bf 92. VCC 1886}. The brightest cluster/nucleus in the frame
appears to be at the location of the isophotal center; a second,
almost as bright cluster is 4\Sec7 to the North-East.

\noindent {\bf 93. VCC 1199}. The galaxy is a close companion of M49,
which lies only 4\Min5 away, and is a prime candidate for galaxy
harassment.  Its surface brightness is brighter than galaxies in the
same magnitude range, and is tidally truncated in the outer
regions. The nucleus looks elongated (along the major axis of the
galaxy), furthermore, there seems to be, from the $g$ and $z-$band
images, a very thin edge-on disk also aligned with the galaxy major
axis, extending less than 1\sec~on either side of the center. The
residual images also show evidence of a large-scale spiral pattern.

\noindent {\bf 94. VCC 1743}. Very low surface brightness elongated
galaxy, with a very faint nucleus. The nucleus seems to have the same
color as the surrounding galaxy.

\noindent {\bf 95. VCC 1539}. This galaxy has a large number  of
bright globular clusters for a galaxy of this magnitude. The galaxy is
very round, has a bright nucleus, and becomes increasingly
bluer towards the center.

\noindent {\bf 96. VCC 1185}. Small, round system, with a bright
nucleus. The isophotes are regular.

\noindent {\bf 97. VCC 1826}. This galaxy has a very sparse cluster
system.  The nucleus is quite prominent.

\noindent {\bf 98. VCC 1512}. This is a dIrr/dE transition object.
There is a loose association of blue stars at the center. It is not
clear whether this galaxy has dust or whether the darker areas seen
towards the center are simply an illusion created by the presence of
the bright blue associations. There doesn't seem to be a proper
nucleus.

\noindent {\bf 99. VCC 1489}. The light is dominated by a bright,
resolved nucleus.

\noindent {\bf 100. VCC 1661}. This galaxy has a bright, resolved
nucleus. The galaxy appears quite round. The isophotes are regular.

\acknowledgments

It is a pleasure and a privilege to thank Dr. Sidney van den Bergh for
sharing his knowledge and expertise. We are indebted to Dr. Peter
Stetson for help in generating the ACS/WFC PSFs used in this paper.
Special thanks go to Dr. Alister Graham and the anonymous referee for
the helpful comments and suggestions. Support for program GO-9401  was
provided through a grant from the Space Telescope Science Institute,
which is operated by the Association of Universities for Research in
Astronomy, Inc., under NASA contract NAS5-26555. P.C. acknowledges
additional support provided by NASA LTSA grant
NAG5-11714. M.M. acknowledges additional financial support provided by
the Sherman M.Fairchild foundation. D.M. is supported by NSF grant
AST-020631, NASA grant NAG5-9046, and grant HST-AR-09519.01-A from
STScI. M.J.W. acknowledges support through NSF grant AST-0205960. This
research has made use of the NASA/IPAC Extragalactic Database (NED)
which is operated by the Jet Propulsion Laboratory, California
Institute of Technology, under contract with the National Aeronautics
and Space Administration, and the HyperLeda database, {\sf
http://www-obs.univ-lyon1.fr/hypercat/}.

\clearpage

\LongTables
\begin{landscape}
\begin{deluxetable}{rrllrrrrclll}
\tabletypesize{\scriptsize}
\tablecaption{Global Morphological Properties of Program Galaxies.\label{tab1}}
\tablewidth{0pt}
\tablehead{
\colhead{ID} & 
\colhead{VCC} & 
\colhead{RA (J2000)} & 
\colhead{Dec (J2000)} & 
\colhead{$B_T$} &
\colhead{$J$} &
\colhead{$H$} &
\colhead{$K_s$} &
\colhead{$E(B-V)$} &
\colhead{$v_{sys}$} &
\colhead{Type} & 
\colhead{Other} \\
\colhead{} &
\colhead{} &
\colhead{(NED)} &
\colhead{(NED)} &
\colhead{(mag)} &
\colhead{(mag)} &
\colhead{(mag)} &
\colhead{(mag)} &
\colhead{(mag)} &
\colhead{(km s$^{-1}$)} &
\colhead{(NED)} &
\colhead{Names} 
}
\startdata
  1 & 1226 & 12:29:46.79 & +08:00:01.5 &  9.31 &  6.39 $\pm$ 0.02&  5.75 $\pm$ 0.02&  5.51 $\pm$ 0.02& 0.022 &  997$\pm$7  & E2/S0$_1$(2)   & M49,~N4472    \\
  2 & 1316 & 12:30:49.42 & +12:23:28.0 &  9.58 &  6.83 $\pm$ 0.02&  6.16 $\pm$ 0.02&  5.90 $\pm$ 0.02& 0.023 & 1307$\pm$7  & E0             & M87,~N4486    \\
  3 & 1978 & 12:43:39.66 & +11:33:09.4 &  9.81 &  6.76 $\pm$ 0.02&  6.08 $\pm$ 0.02&  5.83 $\pm$ 0.02& 0.026 & 1117$\pm$6  & S0$_1$(2)      & M60,~N4649    \\
  4 &  881 & 12:26:11.74 & +12:56:46.4 & 10.06 &  7.19 $\pm$ 0.02&  6.51 $\pm$ 0.02&  6.28 $\pm$ 0.02& 0.029 & -244$\pm$5  & S0$_1$(3)/E3   & M86,~N4406    \\
  5 &  798 & 12:25:24.04 & +18:11:25.9 & 10.09 &  7.17 $\pm$ 0.02&  6.51 $\pm$ 0.02&  6.26 $\pm$ 0.02& 0.030 &  729$\pm$2  & S0$_1$(3)~pec  & M85,~N4382    \\
  6 &  763 & 12:25:03.72 & +12:53:13.2 & 10.26 &  7.26 $\pm$ 0.02&  6.59 $\pm$ 0.02&  6.35 $\pm$ 0.02& 0.041 & 1060$\pm$6  & E1             & M84,~N4374    \\
  7 &  731 & 12:24:28.20 & +07:19:03.0 & 10.51 &  7.71 $\pm$ 0.02&  7.03 $\pm$ 0.02&  6.80 $\pm$ 0.02& 0.021 & 1243$\pm$6  & E3             & N4365         \\
  8 & 1535 & 12:34:03.10 & +07:41:59.0 & 10.61 &  7.55 $\pm$ 0.02&  6.79 $\pm$ 0.02&  6.55 $\pm$ 0.02& 0.021 &  448$\pm$8  & S0$_3$(6)      & N4526         \\
\enddata
\tablecomments{RA, Dec and $v_{sys}$ are from the
  NASA/IPAC Extragalactic Database; $B_T$ and Hubble Type are from the
  VCC. Infrared magnitudes are from the
  2MASS database. Magnitudes  are not extinction corrected; $E(B-V)$ values are calculated from the
  reddening maps of Schlegel, Finkbeiner, \& Davis (1998)}
\end{deluxetable}
\clearpage
\end{landscape}

\begin{deluxetable}{rrllllll}
\tabletypesize{\scriptsize}
\tablecaption{Morphological Peculiarities.\label{tab1}}
\tablewidth{0pt}
\tablehead{
\colhead{ID} & 
\colhead{VCC} & 
\colhead{Dust} &
\colhead{Small-scale} & 
\colhead{Large-scale} & 
\colhead{Nucleus} &
\colhead{Bar} &
\colhead{Comments}\\
\colhead{} & 
\colhead{} & 
\colhead{} &
\colhead{Disk} & 
\colhead{Disk} & 
\colhead{} &
\colhead{} &
\colhead{}
}
\startdata
1   &  1226    & C1 &    &    & II   &    & Dust lane\\
2   &  1316    & C1 &    &    & 0    &    & Dust filaments. Non thermal nucleus;\\
    &          &    &    &    &      &    & saturated in $g-$band \\
3   &  1978    &    &    &    & II   &    & \\
4   &   881    & C1 &    &    & II   &    & Dust filaments. The surface brightness profile dips\\
    &          &    &    &    &      &    & towards the center\\
5   &   798    & C1:&    & Y: & II   &    & Possible dust patches. Visibly boxy within $r < 1$\sec~\\
6   &   763    & C2 &    &    & 0    &    & Dust lanes; non-thermal nucleus\\
\enddata
\tablecomments{For each galaxy (col 2), the columns report the
  detection of dust (col. 3), a small or large-scale
stellar disk (col. 4 and 5 respectively), a nucleus (col 6) 
and a stellar bar (col. 7). Other peculiarities are noted in column 8. 
A ``Y'' indicates the presence of a bar or stellar disk, a
  ``:'' reflects uncertainty. A disk is defined as having
``small-scale'' if it does not dominate the galaxy morphology at large
radii, viceversa for a large-scale disk. Both a large and small-scale
disk can be detected in the same galaxies when the two disks have
different scale lengths or surface brightnesses, and therefore appear
as distinct from inspection of the images, or in their effect on the
surface brightness profile. The presence of dust is classified according 
to the dust morphology: C1 for faint wisps and patches, C2 for high extinction 
irregular dust lanes, C3 for large, patchy dust disks, and C4 for small, 
regular featureless dust disks (see \S 4.1.1 for detail). 
In the case of the column indicating  whether a nucleus is present, we adopt the
terminology used in Paper VIII: ``0'' means that the presence of a nucleus could
not be investigated either because of the presence of significant dust contamination,
or non-thermal emission from an active nucleus; II refers to galaxies which are
non-nucleated; Ia certainly nucleated; Ib certainly nucleated, but the nucleus is
too faint to derive a reliable fit; Ic likely nucleated; Id possibly nucleated; Ie 
the nucleus is offset from the isophotal center. An ``ITD'' in the last column stands 
for ``Inner Truncated Disk''.}
\end{deluxetable}

\clearpage

\begin{deluxetable}{rrccccccccc}
\tabletypesize{\scriptsize}
\tablecaption{Best Fit parameters to the Luminosity Profile for Program Galaxies.\label{tab3}}
\tablewidth{0pt}
\tablehead{
\colhead{ID} & 
\colhead{VCC} & 
\colhead{Model} & 
\colhead{$\mu_{r_e}$} & 
\colhead{$\gamma$} &
\colhead{$n$} &
\colhead{$r_e$} &
\colhead{$r_b$} &
\colhead{$r_h$} &
\colhead{$m_{nucleus}$}\\
\colhead{} &
\colhead{} &
\colhead{} &
\colhead{(mag arcsec$^{-2}$)} &
\colhead{} &
\colhead{} &
\colhead{(\arcsec)} &
\colhead{(\arcsec)} &
\colhead{(\arcsec)} &
\colhead{(mag)} &
}
\startdata
  1 & 1226 & cS & 23.51 &  0.086 &  5.503 & 208.01 &   1.94 &  \nd  &  \nd \\
    &      &    & 21.98 &  0.093 &  5.626 & 211.38 &   1.86 &  \nd  &  \nd \\
  2 & 1316 & cS & 23.45 &  0.322 &  6.094 & 163.83 &   7.15 &  \nd  &  \nd \\
    &      &    & 21.86 &  0.309 &  6.021 & 164.18 &   6.60 &  \nd  &  \nd \\
  3 & 1978 & cS & 22.52 &  0.206 &  4.766 &  98.16 &   2.56 &  \nd  &  \nd \\
    &      &    & 20.90 &  0.203 &  4.854 &  98.48 &   2.52 &  \nd  &  \nd \\
  4 &  881 & cS & 25.29 &  0.021 &  7.016 & 411.84 &   0.72 &  \nd  &  \nd \\
    &      &    & 23.90 &  0.021 &  7.316 & 440.99 &   0.73 &  \nd  &  \nd \\
\enddata
\tablecomments{For each galaxy, the first row reports parameters
  in the $g-$band, the second in the $z-$band. Surface
  brightnesses are not extinction corrected. Columns 1 and 2 list the
ID and VCC numbers of the galaxies; column 3 shows whether the profile
is best described by a core-S\'ersic (cS) or S\'ersic (S) model (\S 4.2.1). The
next five columns list the parameters of the best fit model, as in
equations 7 and 8; for the core-S\'ersic models, $I_e$ represents the
surface brightness measured at one effective radius, and is related to
$I_b$ (equation 9), as $I_e = I_b~{\rm exp}\{b_n[(r_b/r_e)^{1/n}
  -1]\}$. The last two columns list the half-light radius and total
magnitude of the nuclear King model, when fitted.\newline
$^{(1)}$ Based on the dIrr/dE nature of this object, the irregular
profile shape, and the large discrepancy between the values of $r_e$ measured in
the two filters, $r_e$ and all integrated quantities for this galaxy are deemed unreliable.
}
\end{deluxetable}

\clearpage

\begin{landscape}
\begin{deluxetable}{rrllrcccrrrc}
\tabletypesize{\scriptsize}
\tablecaption{Global Structural Parameters for Program Galaxies.\label{tab4}}
\tablewidth{0pt}
\tablehead{
\colhead{ID} & 
\colhead{VCC} & 
\colhead{RA (J2000)} & 
\colhead{Dec (J2000)} & 
\colhead{$r_e$} & 
\colhead{$\mu(r_e)$} &
\colhead{$\langle \mu(r_e) \rangle$} &
\colhead{$\langle \epsilon \rangle$} &
\colhead{$\langle B4 \rangle$} &
\colhead{$\langle \Theta \rangle$} & 
\colhead{m} &
\colhead{$g-z$}\\ 
\colhead{} & 
\colhead{} & 
\colhead{} & 
\colhead{} & 
\colhead{(arcsec)} & 
\colhead{(mag arcsec$^{-2}$)} &
\colhead{(mag arcsec$^{-2}$)} &
\colhead{} &
\colhead{} &
\colhead{(degrees)} & 
\colhead{(mag)} &
\colhead{(mag)}
}
\startdata
  1 & 1226 & 12:29:46.74 & +8:00:01.463 & 209.15 & 99.99 & 21.96 & 0.09$\pm$0.02 & -0.002$\pm$ 0.002 & -19.9$\pm$  2.8 &  8.37 &  1.60$\pm$0.03\\
    &      &             &              & 212.51 & 99.99 & 20.42 & 0.09$\pm$0.02 & -0.003$\pm$ 0.003 & -18.9$\pm$  2.1 &  6.79 &      \\
  2 & 1316 & 12:30:49.42 & +12:23:28.35 & 171.71 & 99.99 & 21.96 & 0.02$\pm$0.04 &  0.001$\pm$ 0.007 &  46.1$\pm$ 37.1 &  8.79 &  1.60$\pm$0.04\\
    &      &             &              & 170.74 & 99.99 & 20.35 & 0.02$\pm$0.01 &  0.001$\pm$ 0.007 &  15.8$\pm$ 36.7 &  7.19 &      \\
  3 & 1978 & 12:43:40.08 & +11:33:09.23 &  99.34 & 22.83 & 21.06 & 0.11$\pm$0.04 & -0.004$\pm$ 0.003 & -83.6$\pm$  4.4 &  9.08 &  1.62$\pm$0.02\\
    &      &             &              & 100.05 & 21.21 & 19.45 & 0.11$\pm$0.05 & -0.004$\pm$ 0.004 & -83.4$\pm$  3.6 &  7.45 &      \\
  4 &  881 & 12:26:11.76 & +12:56:46.06 & 411.84 & 99.99 & 23.60 & 0.19$\pm$0.02 & -0.005$\pm$ 0.006 & -57.2$\pm$  4.9 &  8.53 &  1.57$\pm$0.05\\
    &      &             &              & 440.99 & 99.99 & 22.19 & 0.19$\pm$0.02 & -0.005$\pm$ 0.006 & -57.2$\pm$  4.9 &  6.97 &      \\
\enddata
\tablecomments{For each galaxy, the first row reports parameters
  in the $g-$band, the second in the $z-$band. Magnitudes and surface
  brightnesses are not extinction corrected.
  Columns 3 and 4 list the RA and Dec of the galaxy center, at J2000, measured directly from
  the images (to be compared to the values given in Table 1). Column 5 
  lists the effective radius of the galaxy; these coincide with the
  values listed in column 7 of Table 3 for galaxies best fit by a
  S\'ersic model, but are slightly larger for the galaxies best fit by a
  core-S\'ersic model (for which column 7 of Table 3 lists the $r_e$ of
  the S\'ersic model fitting the outer parts). Columns 6 and 7 list
  the surface brightness at $r_e$, measured directly from the images
  when possible, and the mean surface brightness within
  $r_e$, integrated from the best fitting model. 
  Columns 8, 9 and 10 list the average ellipticity, B4
  coefficient, and major axis position
  angle measured between  1\sec~and $r_e$. Average values are not
  listed if less than four points are available to compute the mean
  and standard deviation. Finally, the 
  magnitudes listed in the next to last column exclude
  the nucleus, and are obtained by integrating the best fit profile to
infinity. The colors listed in the last column are measured directly
from the surface brightness profile for 1\sec~$< r < r_e$. To avoid
errors associated with the background subtractions, surface brightness values
1 mag below the sky or fainter are excluded.\newline
$^{(1)}$ Based on the dIrr/dE nature of this object, the irregular
profile shape, and the large discrepancy between the values of $r_e$ measured in
the two filters, $r_e$ and all integrated quantities for this galaxy are deemed unreliable.}
\end{deluxetable}
\clearpage
\end{landscape}

\clearpage

\begin{figure}
\epsscale{0.01}
\vskip -0.1in
\caption{(Right) Surface brightness profiles in $g$ (lower points) and $z$
  (upper points) for four galaxies in the ACS Virgo Cluster
  Survey. The data are plotted up to the radius at which the surface
  brightness falls below 10\% of the sky, as in Figure \ref{prof1}.  
  For each galaxy, and in each filter, we show as a
  solid line the best-fit S\'ersic (blue) and core-S\'ersic (red)
  models, each with the addition of a nuclear component when necessary
  (in these cases, the best fits to the nucleus and the galaxy are
  shown separately as dotted and dashed lines respectively). Residuals
  from the best fit are shown in the panels below over the range of
  radii used in constraining the fit (see text for further
  details). The data are plotted against the geometric mean
  radius. The dotted vertical line is placed at r=0\Sec049,
  corresponding to the ACS/WFC pixel size.  
\label{fit1}}
\end{figure}

\clearpage

\begin{figure}
\plottwo{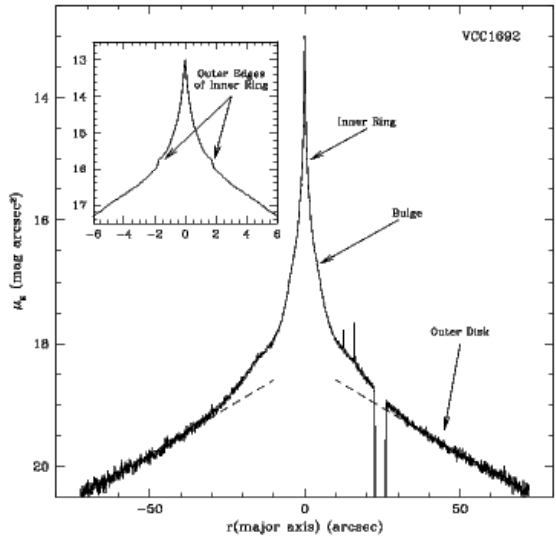}{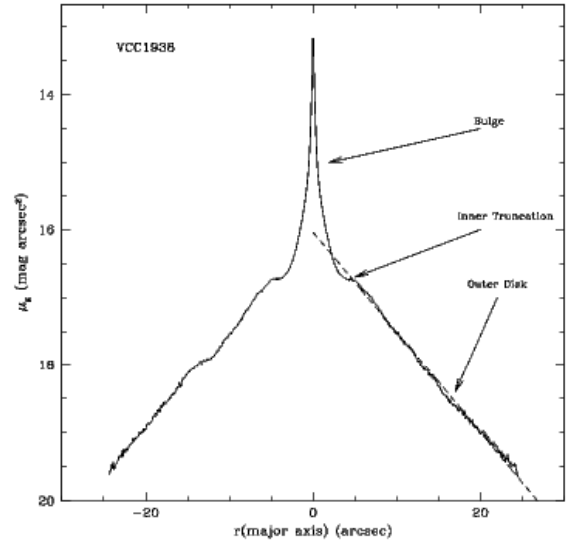}
\caption{(Left) A cut through the $g-$band photometric major axis of VCC 1692
  (NGC 4570), highlighting the several morphological components. As
  shown by the dashed line, the outer profile, dominated by a thin
  stellar disk, is roughly exponential. The bulge contribution is
  visible in the inner $\sim$ 20\sec. The innermost 1\Sec8 are
  dominated by the nuclear ring, the outer boundary of which are
  clearly visible in the expanded view shown in the inset.
\label{vcc1692}}
\end{figure}

\begin{figure}
\vskip -0.1in
\caption{(Right) A cut through the $g-$band photometric major axis of
  VCC 1938 (NGC 4638), clearly showing the inner truncation of he
  outer exponential disk.
\label{vcc1938}}
\end{figure}

\begin{figure}
\vskip 0.8in
\epsscale{1.0}
\plottwo{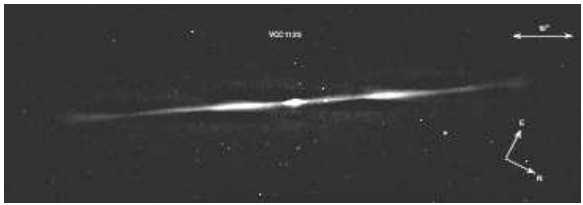}{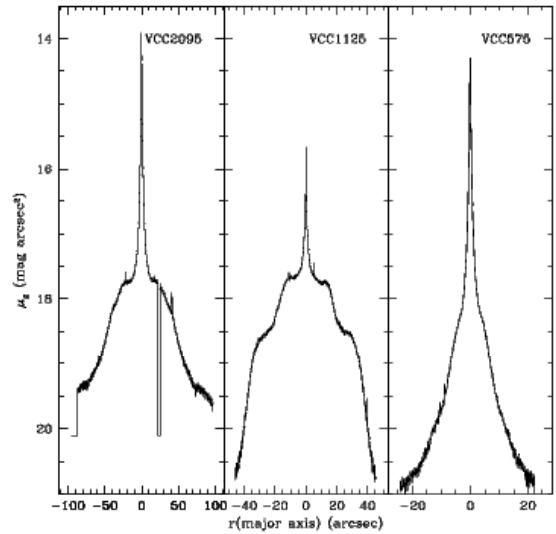}
\caption{(Left) Residual map of VCC 1125, obtained by subtracting the best
  fitting isophotal model from the $g-$band image.
\label{vcc1125}}
\end{figure}

\begin{figure}
\epsscale{1.0}
\vskip -0.1in
\caption{(Right) Cuts through the $g-$band photometric major axes of
  VCC 2095 (NGC 4762), VCC 1125 (NGC 4452) and VCC 575 (NGC 4318).
\label{vcc2095}}
\end{figure}

\clearpage

\begin{figure}
\plottwo{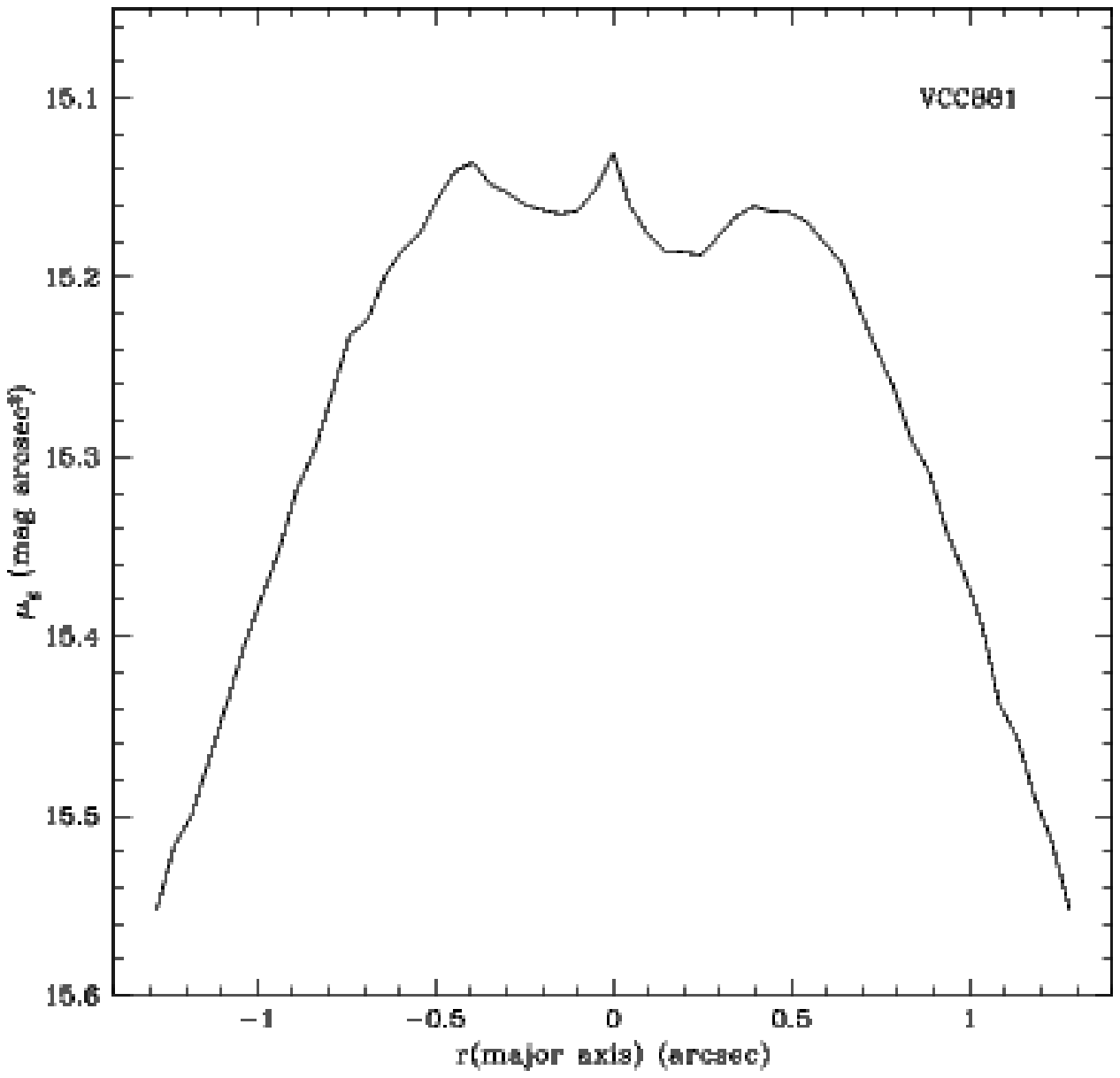}{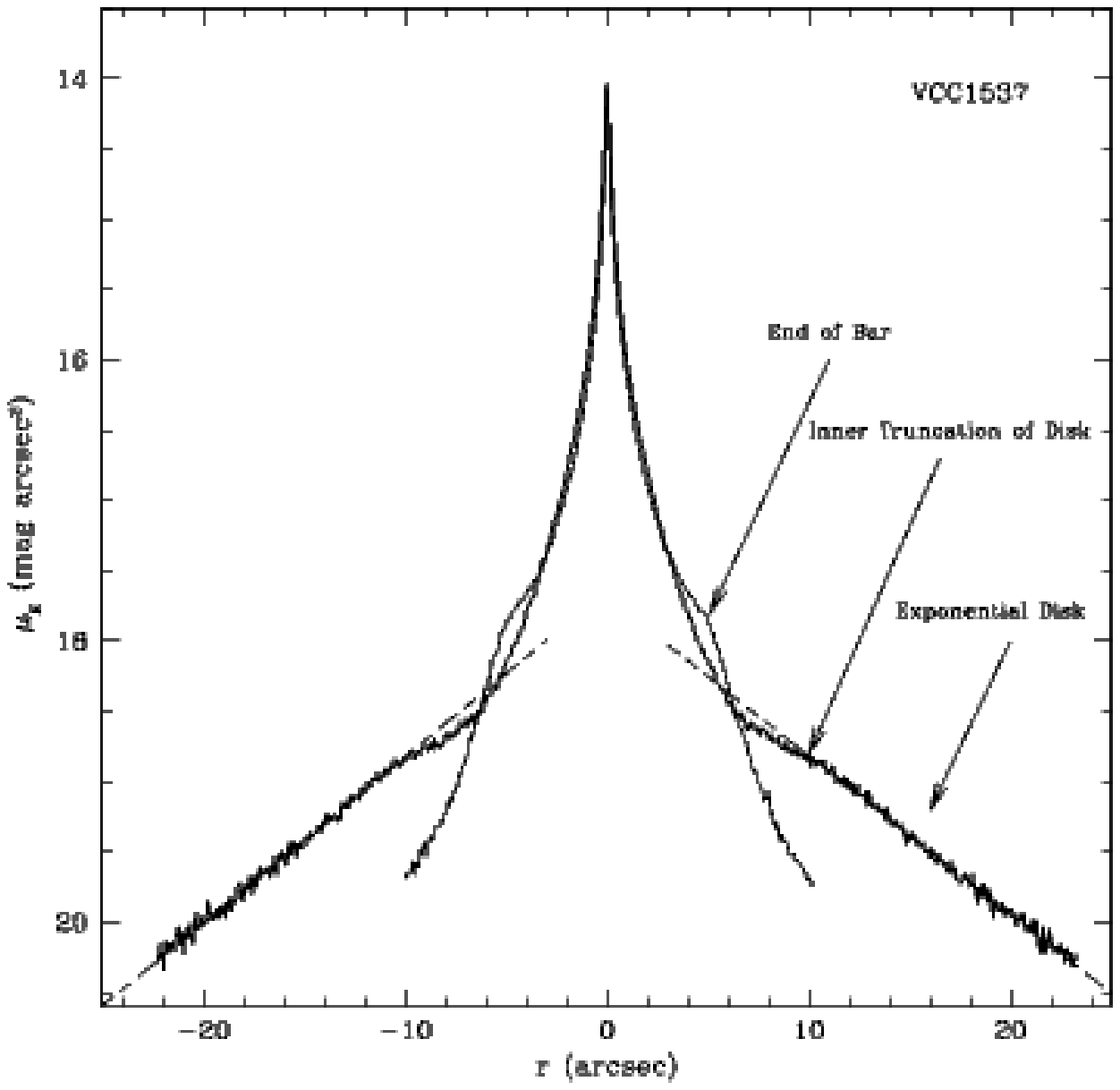}
\caption{(Left) A cut through the $g-$band photometric major axis of
  VCC 881 (NGC 4406), showing the decrease in the surface brightness
  profile in the inner 0\Sec4.
\label{vcc881}}
\end{figure}

\begin{figure}
\vskip -0.1in
\caption{(Right) Cuts through the $g-$band photometric major axis (thick solid
  line) and through the bar axis (thin solid line) of
  VCC 1537 (NGC 4528) $-$ the latter is misaligned by roughly
  30\deg~relative to the minor
  axis of the galaxy. The inner truncation in the exponential disk
  (approximated by the dashed line), as well as the ends of the bar are marked
\label{vcc1537}}
\end{figure}

\begin{figure}
\epsscale{0.7}
\plotone{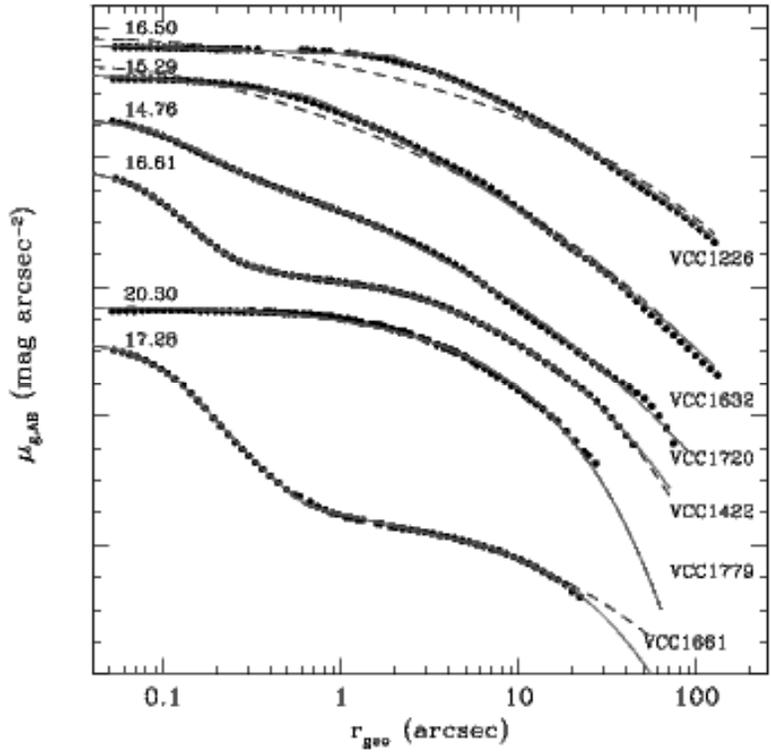}
\caption{Representative $g-$band surface brightness profiles for six
galaxies in the ACS Virgo Cluster Survey, from the brightest (VCC
1226) to the faintest (VCC 1661) galaxy in the sample, plotted as a
function of the geometric mean radius. All galaxies are ``regular''
ellipticals, in the sense of not displaying significant morphological
peculiarities (with the exception of the possible presence of a
nuclear component). The galaxies are arranged, from top to bottom, in
order of increasing total B-band magnitude (decreasing
luminosity). They have been shifted in the vertical direction for
clarity; the value of the $g-$band surface brightness profile at
0\Sec049 is listed on the left-hand-side next to each galaxy profile,
and tickmarks on the ordinate axis are separated by
1~mag~arcsec$^{-2}$. For each galaxy, we overplot the best-fit
S\'ersic (dashed) and core-S\'ersic (solid) model, with addition of a
King model for nucleated galaxies (VCC 1720, VCC 1422 and VCC 1661).
\label{fig07}}
\end{figure}

\clearpage

\begin{figure}
\epsscale{0.7}
\plotone{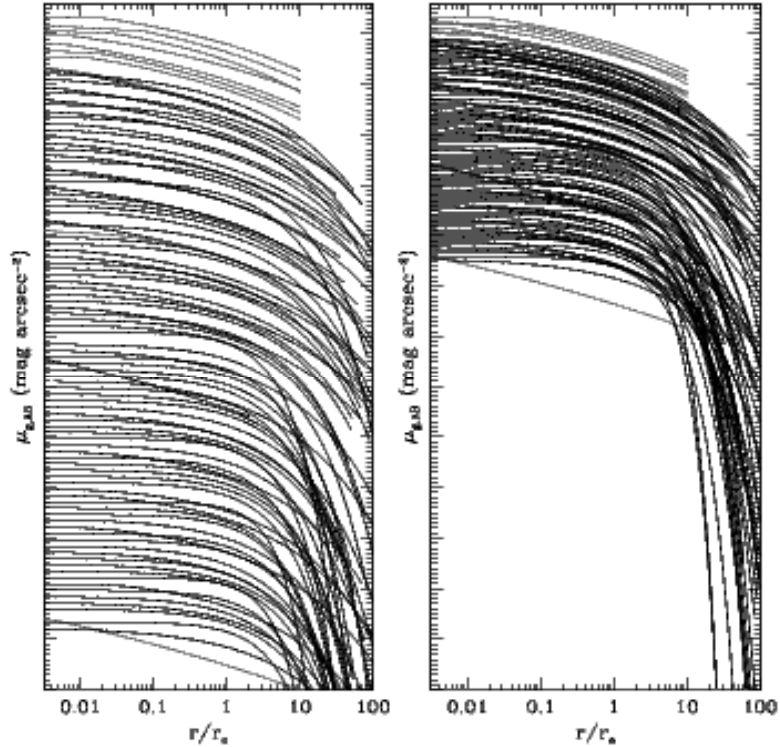}
\caption{A compilation of the best-fit models for the $g-$band surface
  brightness profiles of all ACSVCS galaxies. The models exclude a
  nuclear component when present, and are shown before convolution by
  the ACS/WFC PSF. For each galaxy, the abscissa shows the isophotal
  radius normalized to the effective radius $r_e$, derived from the
  best fit. Along the ordinate, galaxies are ordered according to
  their total B-band magnitudes, with the brightest galaxy (VCC 1226)
  plotted as having the largest central surface brightness, and every
  other galaxy plotted as having central surface brightness shifted
  downwards, by a constant amount, relative to that of VCC 1226.  Each
  small tick on the ordinate corresponds to 1 magnitude
  arcsec$^{-2}$. The two panels differ only in the range of surface
  brightness shown: to the right, the scale is shown to highlight
  difference in the innermost region. In both panels, galaxies best
  fitted by a core-S\'ersic profile are  shown in red.
\label{allprofiles}}
\end{figure}

\begin{figure}
\epsscale{1.0}
\plotone{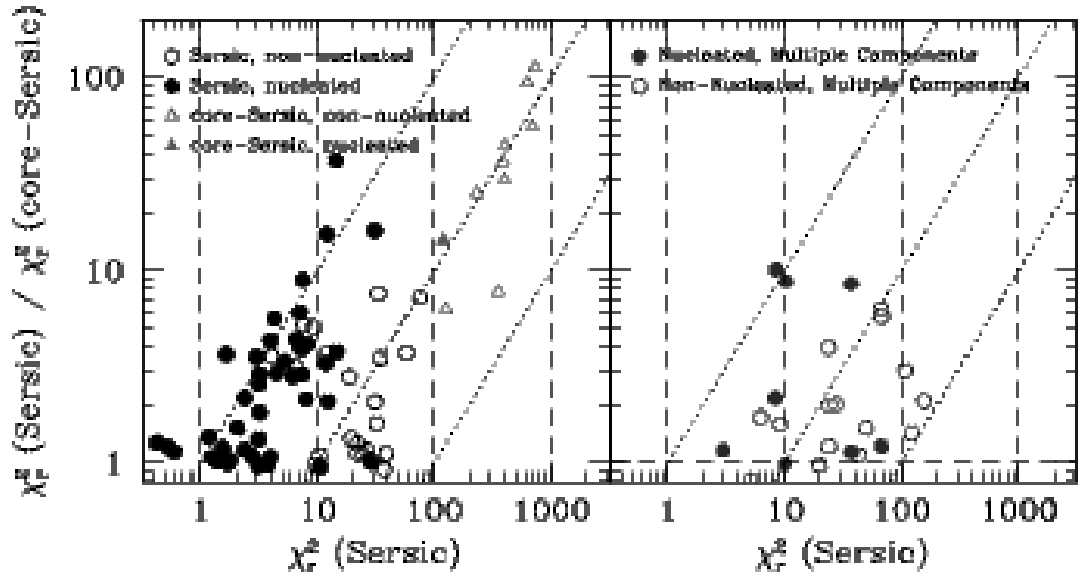}
\caption{The ratio of the reduced $\chi^2$ of the S\'ersic and
  core-S\'ersic fits {\it vs} the reduced $\chi^2$ for the best
  S\'ersic fit  (see text for details). Galaxies with stellar disks
  or bars are shown in the right panel,  `regular' galaxies in the
  left panel. The dotted lines represent locii for which (from left to
  right) $\chi^2_r$(core-S\'ersic)=1,10,100; the dashed lines have the
  same meaning for $\chi^2_r$(S\'ersic). Galaxies plotted as triangles
  are best fit by a core-S\'ersic profile.
\label{chiplot}}
\end{figure}

\clearpage

\begin{figure}
\epsscale{0.85}
\plotone{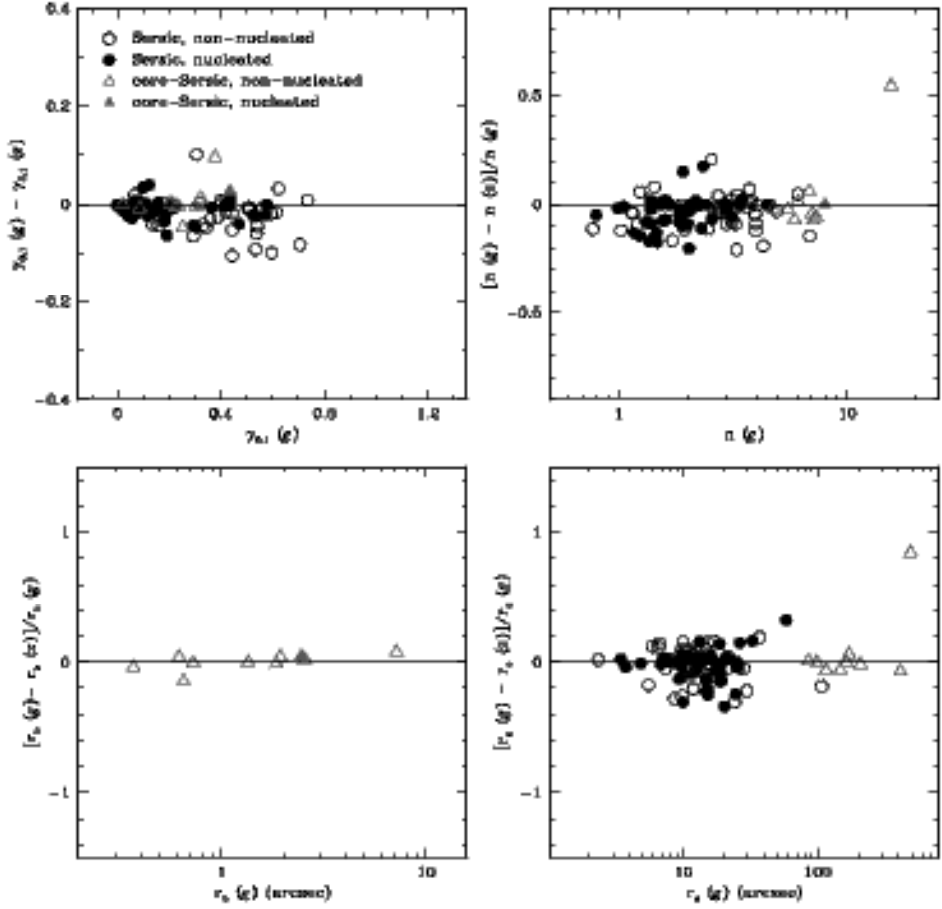}
\caption{Comparison of $\gamma$, $n$, break and effective radii ($r_b$
  and $r_e$) for the best fits to the surface brightness profiles in
  the $g$ and $z-$bands. Galaxies are color-coded as in
  Figure \ref{comparison2}.  In the case of galaxies best fit by a
  S\'ersic profile, the logarithmic slope $\gamma$ is calculated at
  0\Sec1 from the center, and is uniquely defined by $n$ and $r_e$
  (see equation 16).
\label{comparison1}}
\end{figure}

\begin{figure}
\epsscale{0.9}
\plotone{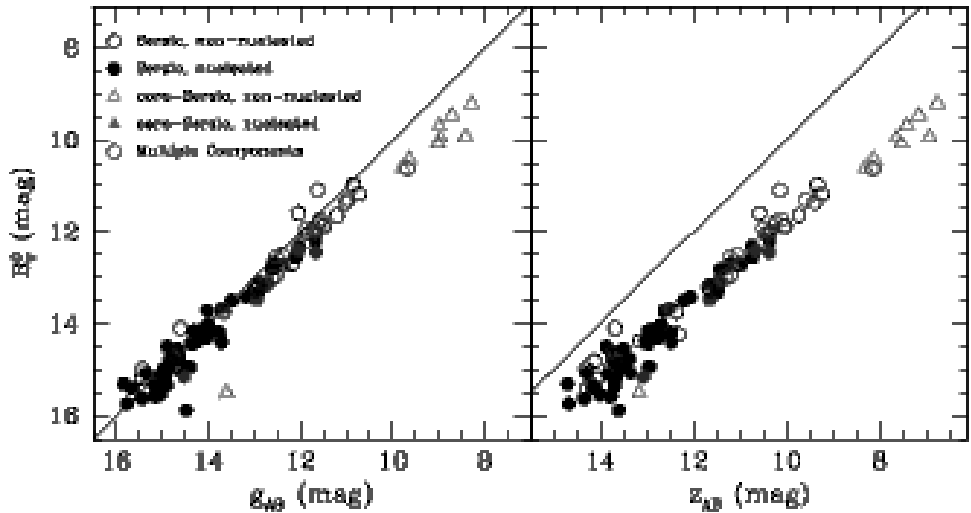}
\caption{Comparison of the $g$ and $z-$band total magnitudes,
  corrected for extinction, derived
  by integrating to infinity the best fitting surface brightness
  profiles, compared to the $B_T^0$ magnitudes from the RC3. The
  diagonal line shows a one-to-one correspondence. Galaxies with
  multiple morphological components, for which the model fits are not
  always satisfactory, are plotted in blue.
\label{magcomp}}
\end{figure}

\clearpage

\begin{figure}
\plotone{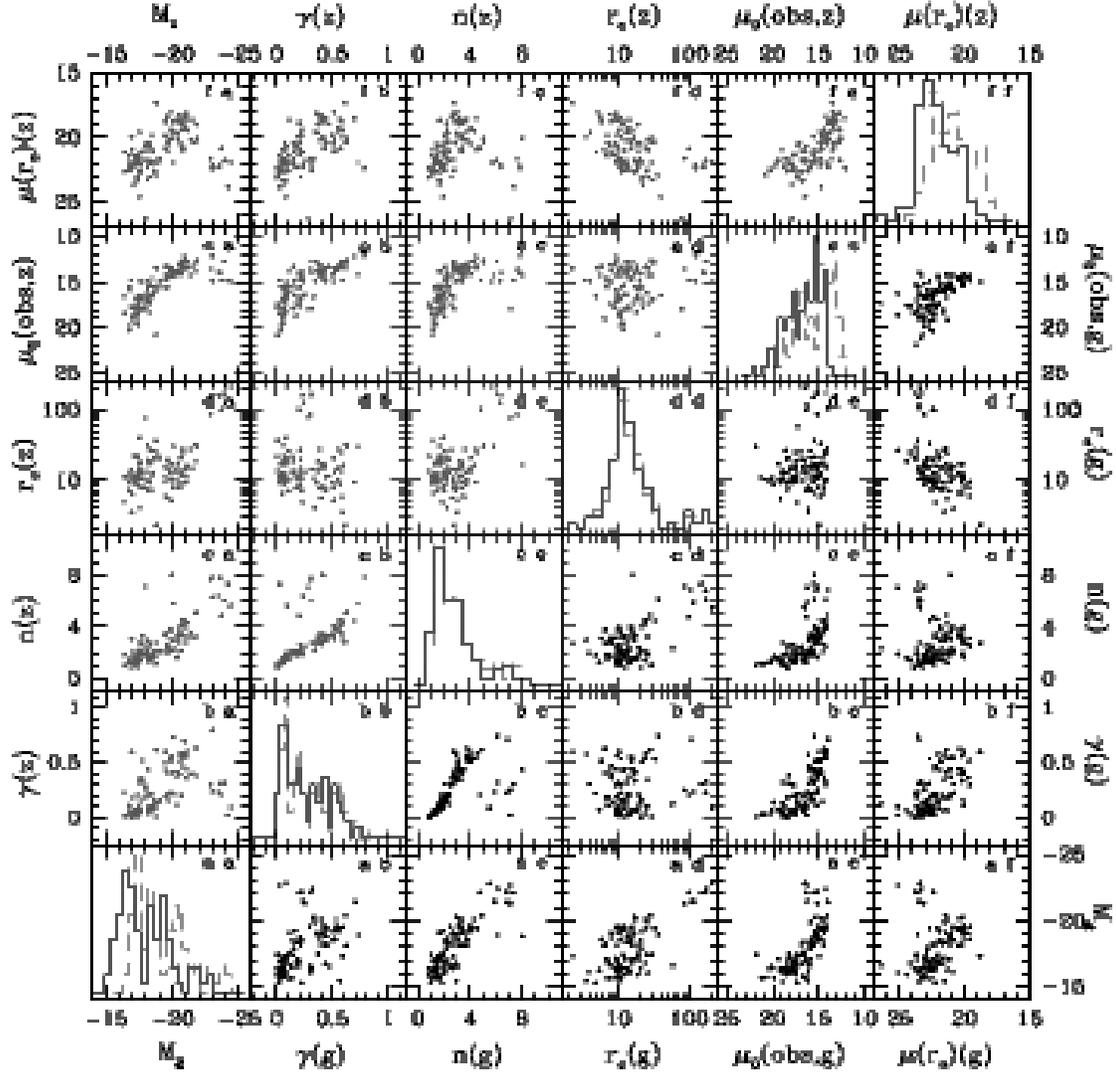}
\caption{Correlations between fit parameters: total absolute magnitude
  from the model profile, integrated to infinity and extinction
  corrected; inner slope $\gamma$ (in the case of galaxies best fit by
  a S\'ersic profile, $\gamma$ is defined as  the logarithmic slope at
  0\Sec1 from the center, see equation 16); S\'ersic index $n$;
  effective radius $r_e$, in arcseconds; extinction corrected central
  surface brightness $\mu_0$ in mag arcsec$^{-2}$, measured directly
  from the images at 0\Sec049, and therefore including a nuclear
  component when present; and extinction corrected surface at $r_e$,
  in mag arcsec$^{-2}$. Parameters recovered for the $g$ and $z-$band
  images are shown in the panels below and above the  diagonal line
  running from bottom-left to top-right, respectively. Histograms of
  each of the parameters (drawn as a solid and dashed line for the $g$
  and $z-$band respectively), are shown along the diagonal (the y-axis
  scale is not shown for the histograms). In this and all subsequent
  figures, histograms are created by binning the data with an
  ``optimal'' bin size given by $2\times IQR\times N^{-1/3}$ (Izenman
  1991), where $N$ is the total number of objects, and $IQR$ is the
  interquartile range (i.e. the range which includes the second and
  third quartile of the ranked distribution of the parameter under
  investigation).
\label{all1}}
\end{figure}

\clearpage

\begin{figure}
\plottwo{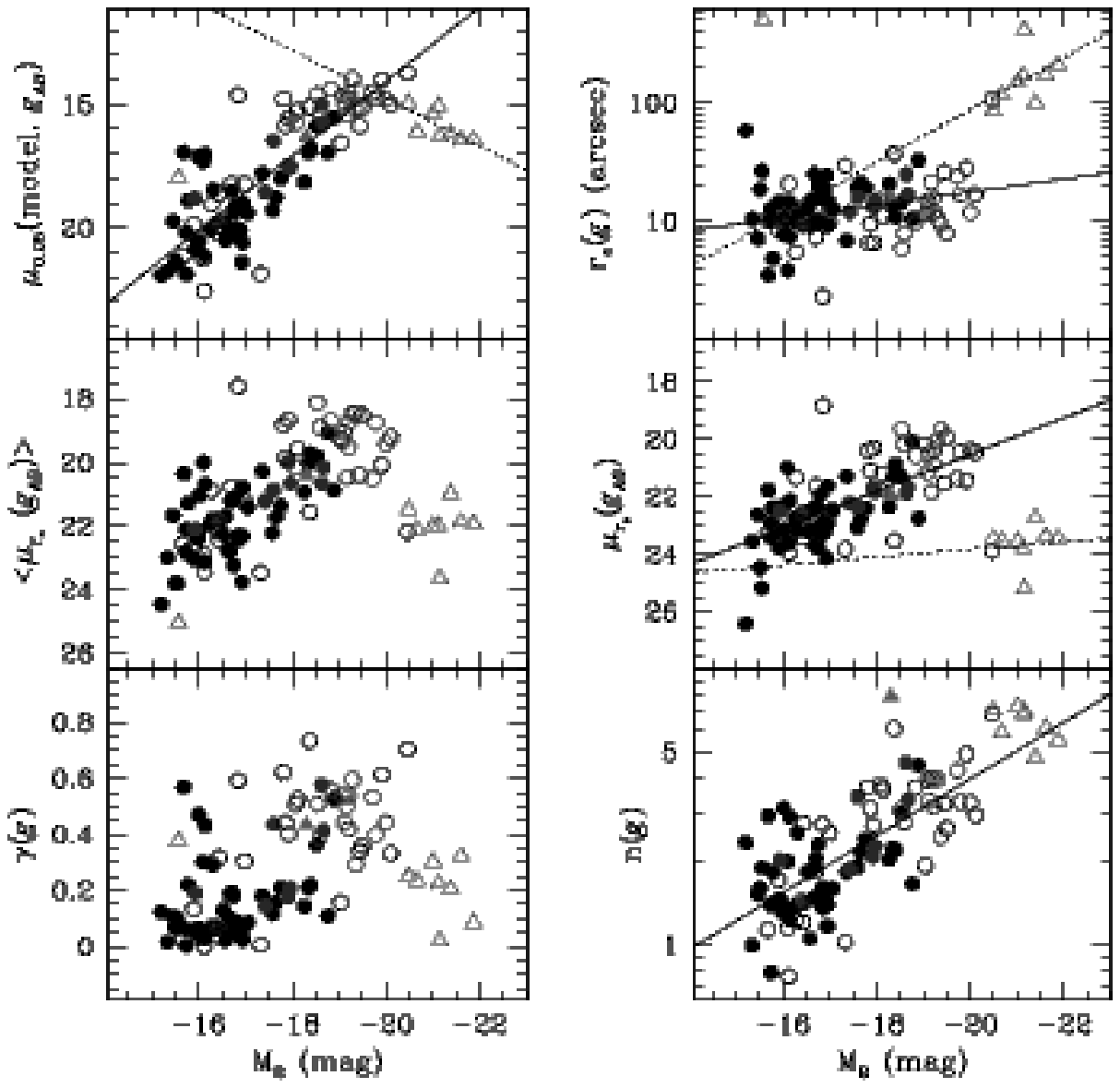}{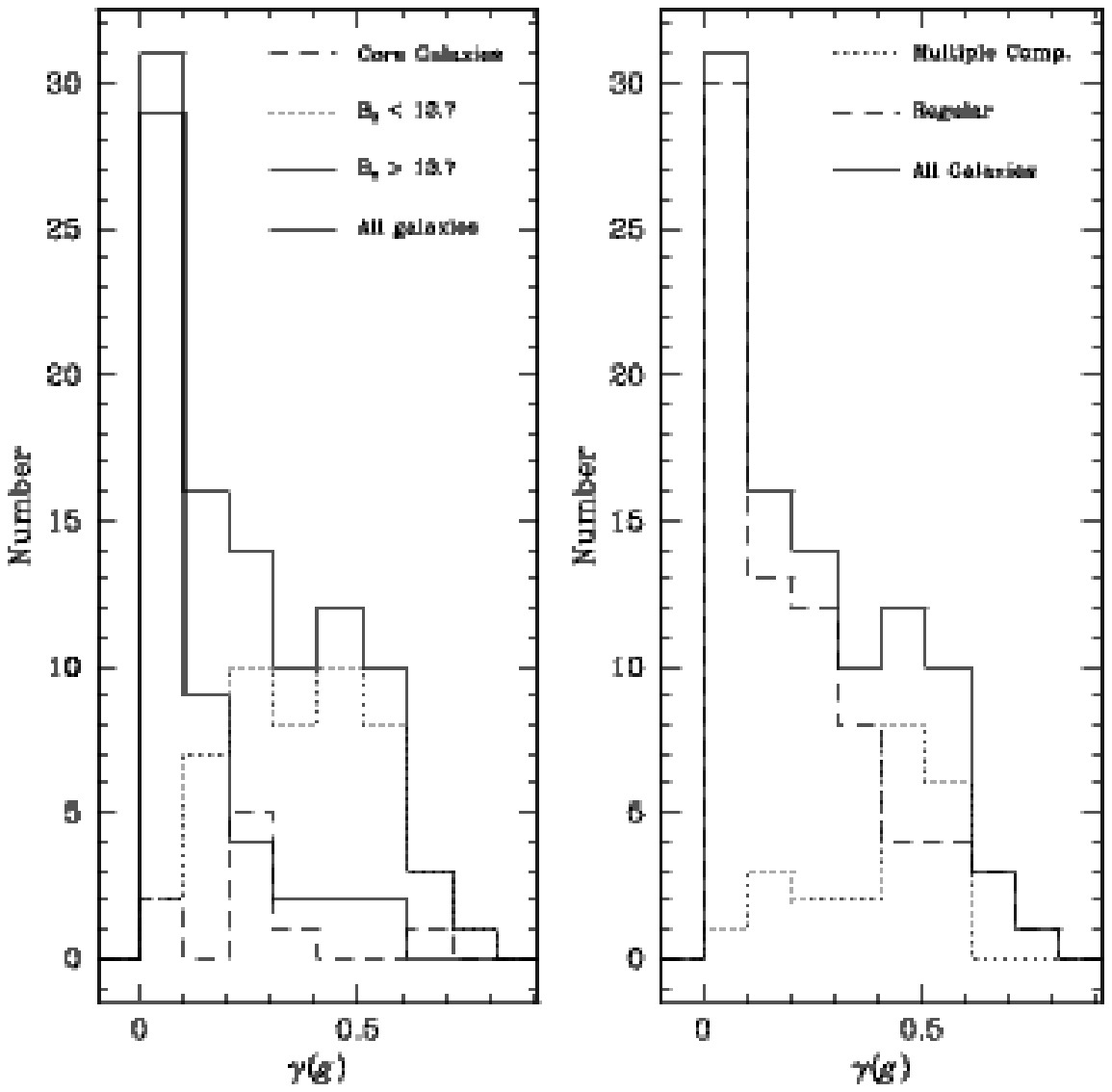}
\caption{(Left) Correlation of parameters characterizing the inner (on the
  left) and outer (on the right) surface brightness profile and the
  total $B-$band magnitude of the host galaxies. On the left panels,
  from top to bottom, are shown the surface brightness measured from
  the best fitting model at 0\Sec049, {\it prior} to PSF convolution
  and excluding a nuclear component, if present; the average surface
  brightness within one effective radius $r_e$, in mag arcsec$^{-2}$;
  and the slope $\gamma$ of the inner profile at $r=$0\Sec1, measured
  prior to PSF convolution (following equation 16 in the case of
  profiles best fit by a S\'ersic model). On the right panels are shown
  the effective radius $r_e$; the surface brightness at $r_e$, in mag
  arcsec$^{-2}$; and the S\'ersic index $n$ characterizing the shape
  of the outer profile.  Galaxies are color-coded as in the legend of
  Figure \ref{magcomp}, blue points refer to galaxies with multiple
  morphological components.
\label{fcorr2}}
\end{figure}

\begin{figure}
\vskip -0.2in
\caption{(Right) The distribution of the $g-$band inner slope $\gamma$ of the
  galaxy profile (i.e excluding a nuclear component) for the entire
  sample (shown by the solid line), and the sample divided according
  to their total $B-$band magnitude (left panel) and the presence of
  multiple morphological components (right panel). $\gamma$ is
  measured from the best fit profiles, and therefore represents the
  slope prior to PSF convolution and excluding a nuclear component. In
  the case of galaxies best fit by a S\'ersic profile, $\gamma$ is 
  measured at 0\Sec1 from the center following equation 16.
\label{hist2}}
\end{figure}

\begin{figure}
\epsscale{0.6}
\vskip 0.2in
\plotone{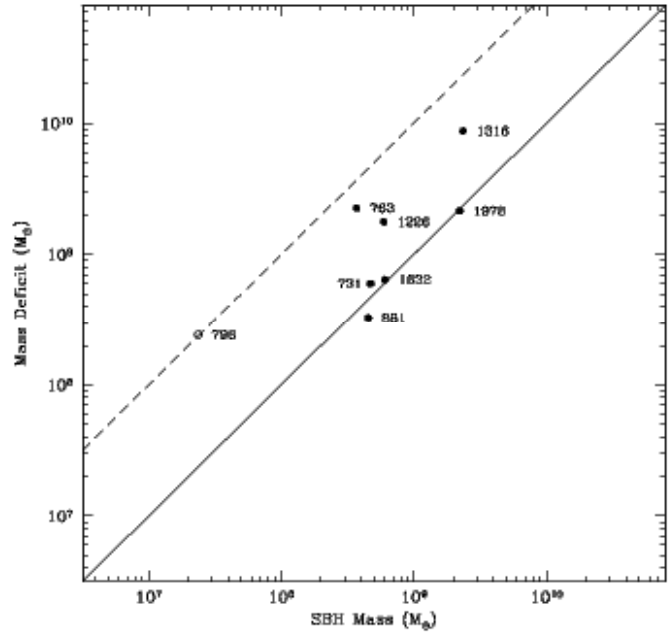}
\caption{The core ``mass deficit'', defined as the  difference between
  the mass enclosed within the break radius by the S\'ersic model that
  best fits the outer surface brightness profile and the best fit
  core-S\'ersic model, plotted against the mass estimated for the
  nuclear SBH (see text for details). The solid and dashed lines are
  plotted for a mass deficit equal to, and ten times larger than, the
  SBH mass. VCC 798 (NGC 4382), shown by the open circle, is likely to
  host a large scale stellar disk.
\label{deficit}}
\end{figure}

\clearpage

\begin{figure}
\plotone{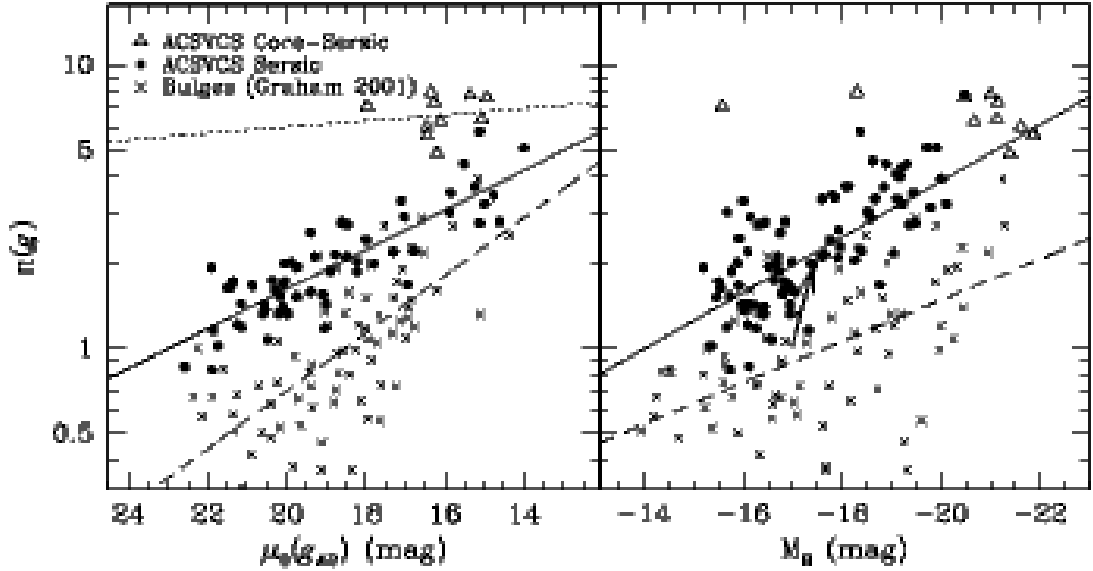}
\caption{Left panel: The S\'ersic index {\it n}, plotted against the
  surface brightness  estimated from the best fitting model at
  0\Sec049, {\it prior to} PSF convolution. Right panel:  {\it n},
  plotted against the total, extinction corrected absolute $B-$band
  magnitude of the host galaxy (from the RC3). The ACSVCS galaxies are
  shown as solid circles and open triangles depending on whether the
  surface brightness profile is best fit by a  S\'ersic or
  core-S\'ersic model, respectively. Crosses represent bulges from
  Graham (2001); $B-$band central surface brightnesses from this paper
  have been converted to $g-$band by assuming $g-B \sim 0.55$
  (Fukugita et al.\ 1995), in addition the bulge data have been
  extinction corrected and total magnitudes have been rescaled to $H_0
  =72$ km/s/Mpc. The solid line  represents the best least square fit
  to the ``S\'ersic'' galaxies (solid circles), the dotted line shows
  the best least square fit to the ``core-S\'ersic'' galaxies, and the
  dashed line shows the best fit to the bulges.  The arrow in the
  bottom right panel shows the change in $n$ and $M_B$ undergone by
  an exponential ($n =1$) bulge after merging with a satellite of half
  its mass, according to N-body simulations by Eliche-Moral et al.\
  (2005).
\label{shape}}
\end{figure}

\begin{figure}
\plottwo{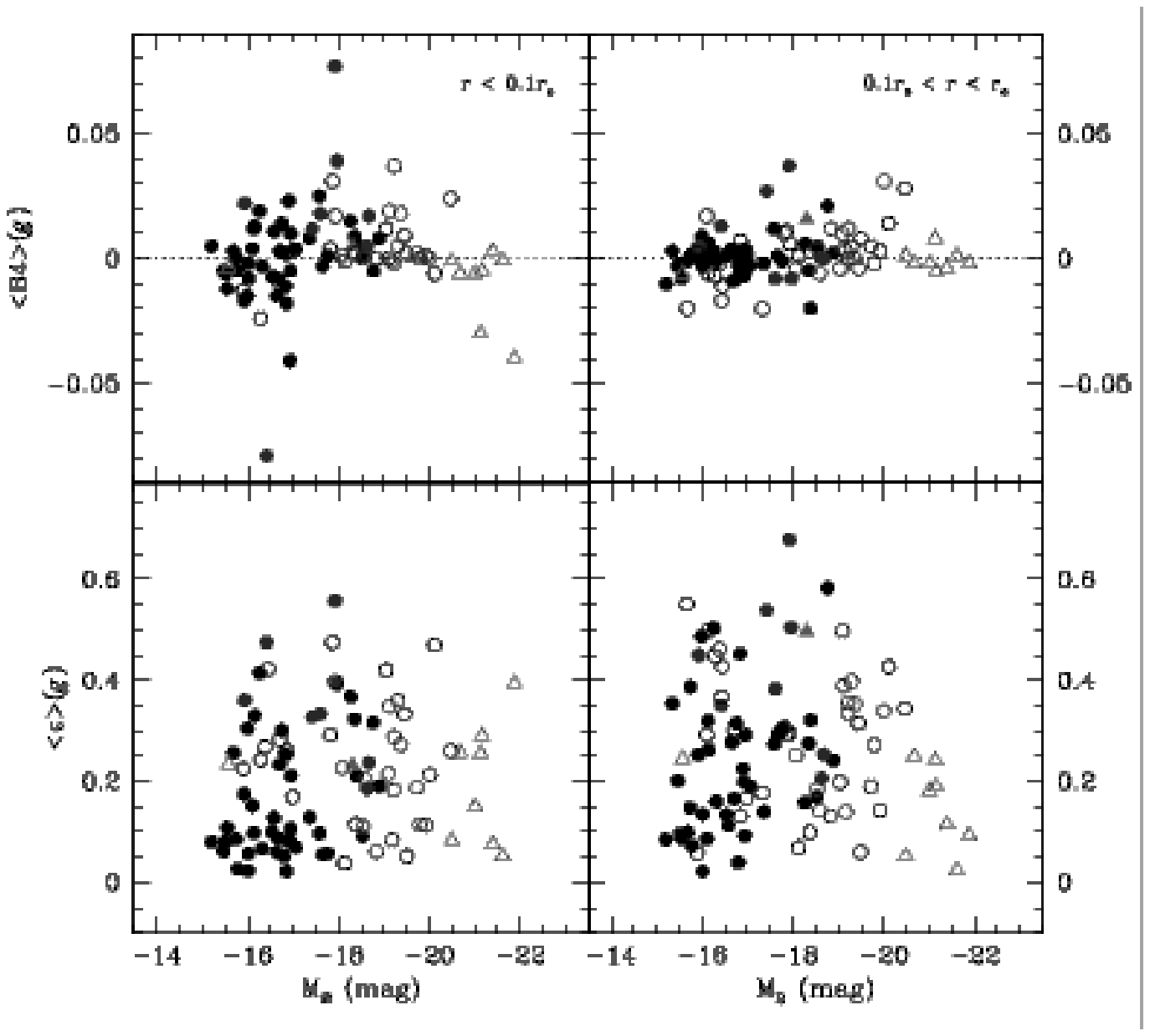}{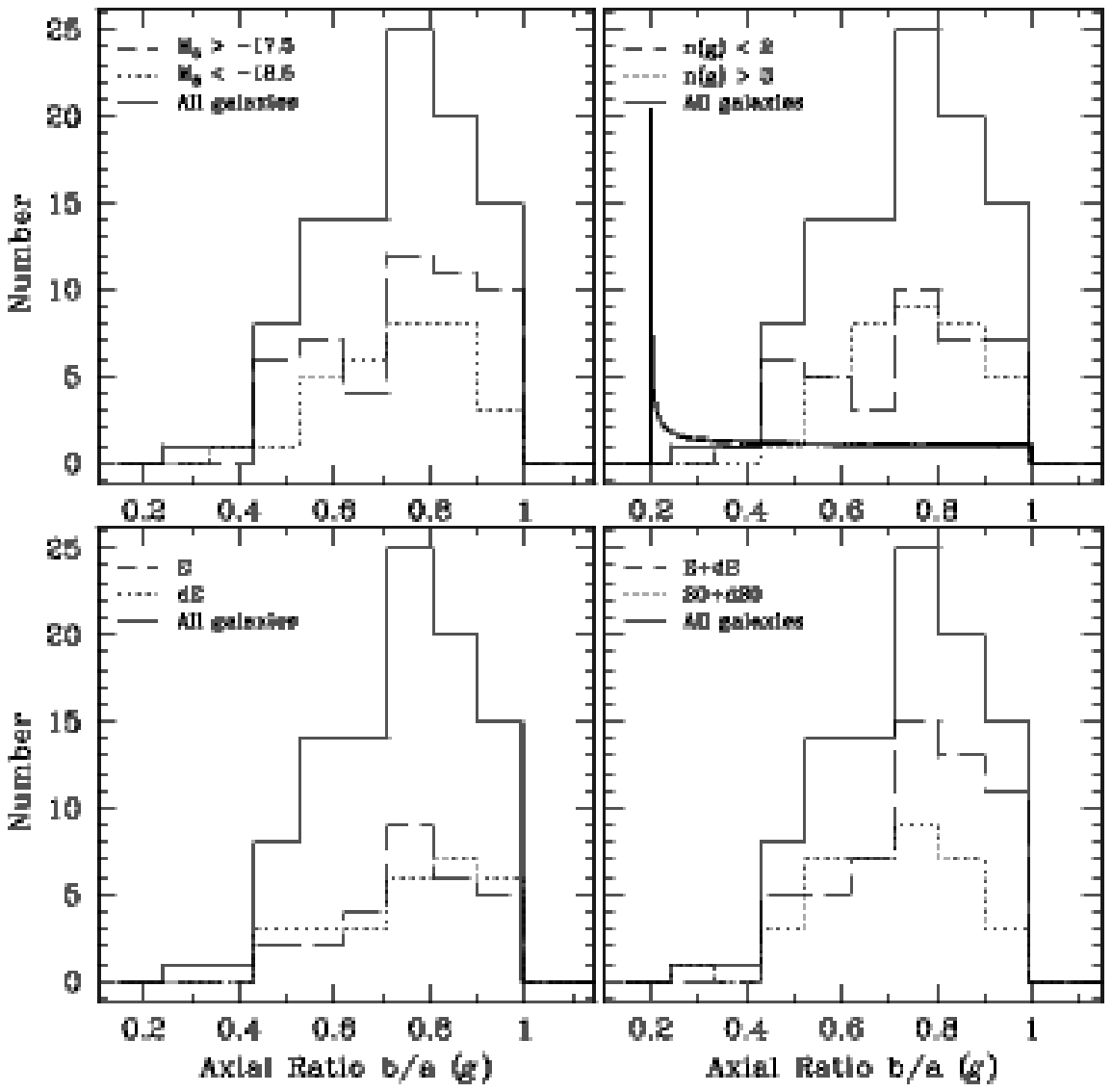}
\caption{(Left) Correlations between the extinction corrected $B-$band
  magnitude and mean ellipticity and B4 coefficients. The latter  are
  calculated within 10\% of the effective radius $r_e$ (excluding the
  nuclear region, if a nucleus is present) in the left panels, and
  between 0.1$r_e$ and $r_e$ in the panels to the right.  Galaxies are
  color-coded as in the legend of Figure \ref{magcomp}, blue points
  refer to galaxies with multiple morphological components.
\label{corr4}}
\end{figure}

\begin{figure}
\vskip -0.2in
\caption{(Right) The distribution of minor to major axes ratios, $b/a$,  for
  the ACSVCS galaxies. In all panels, the thick solid line shows the
  distribution for the entire sample. In the upper left and right
  panels the sample is divided according to absolute $B-$band
  magnitude and S\'ersic index $n$ respectively; the thick continuous line
  in the right panel shows the distribution of axial ratios expected
  for a population of galaxies with intrinsic $b/a=0.2$ and seen at
  random orientations. In the bottom left and right panels, the sample
  is divided according to morphological classification.
\label{ell}}
\end{figure}

\clearpage

\begin{figure}
\plotone{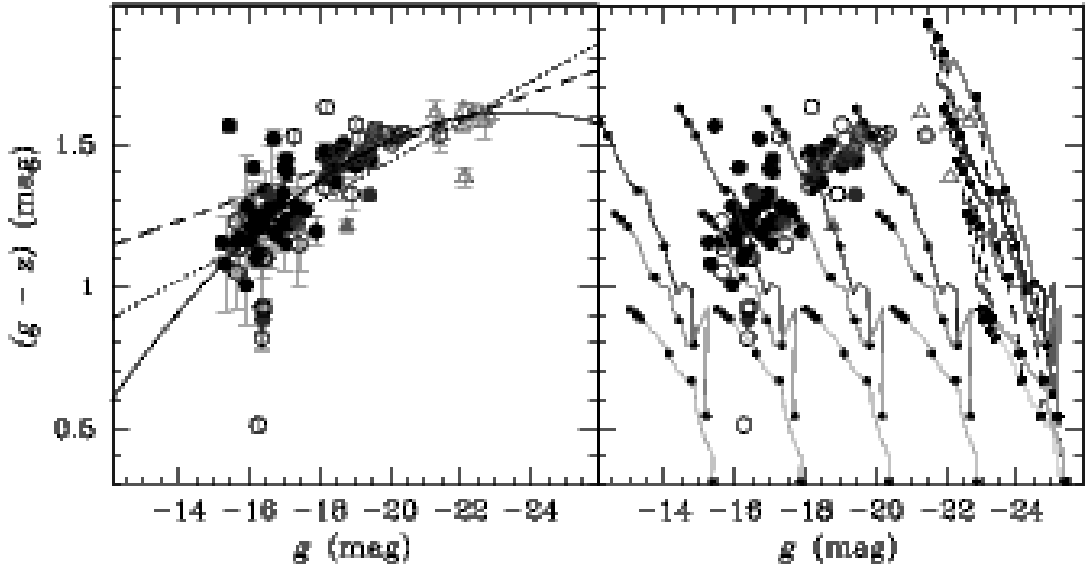}
\caption{Color-Magnitude diagram for the 100 galaxies which comprise
  the ACSVCS. Galaxies are-coded as in Figure \ref{magcomp}. In the
  left panel, the data are plotted with errorbars; the dashed and
  dotted lines corresponds to least-square fits to galaxies fainter
  and brighter than $M_g = -18$ respectively, while the curved line
  represents the best quadratic fit to the entire sample (see equations
  31, 32 and 33). The solid lines in the right panel show the
  prediction of the stellar population synthesis models of Bruzual \&
  Charlot (2003). Models of constant metallicity and varying age are
  shown as solid lines for [Fe/H]=$-$2.25 (yellow), $-$1.65 (cyan),
  $-$0.64 (green), $-$0.33 (red), +0.09 (blue), +0.56
  (magenta). Points along the isometallicity tracks corresponds to
  ages of 1, 2, 4, 8, 12, 15 Gyr (from left to right). Finally, from
  top to bottom, the $g-$band magnitudes have been scaled to a total
  mass $M=10^{12}$, $10^{11}$, $10^{10}$, $10^9$, and $10^8$
  M$_{\odot}$. All magnitudes have been extinction corrected.
\label{cmd1}}
\end{figure}

\begin{figure}
\plotone{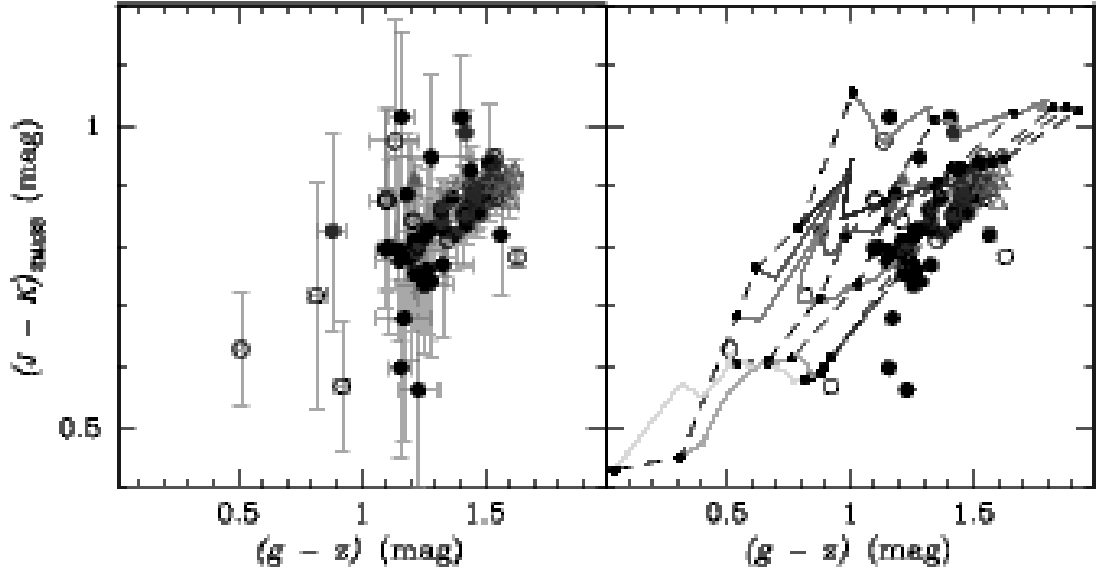}
\caption{$g-z$ vs $J-K$ color-color diagrams. $J$ and $K-$band
  magnitudes are from the 2MASS database. Galaxies are coded as in
  Figure \ref{cmd1}, as are the stellar population synthesis
  models.  All magnitudes have been extinction corrected.
\label{cmd2}}
\end{figure}

\end{document}